\documentclass[botnum, fleqn,reqn]{unmeethesis}

\usepackage[T1]{fontenc}
\usepackage[utf8]{inputenc}
\usepackage[english]{babel}
\usepackage{booktabs} 
\usepackage{graphicx} 
\usepackage{enumitem} 
\usepackage{xcolor} 
\usepackage{amsmath}
\usepackage{amssymb}
\usepackage{mathtools}
\usepackage{amsfonts}
\usepackage{optidef}
\usepackage[square,numbers,sectionbib]{natbib}
\usepackage{subfiles}
\usepackage{lipsum}
\usepackage{url}

\usepackage{caption}
\usepackage{subcaption}
\usepackage{tabularx}
\usepackage{array,ragged2e}
\newcolumntype{P}[1]{>{\RaggedRight\arraybackslash}p{#1}}
\usepackage{multirow}
\usepackage[ruled,vlined]{algorithm2e}
\usepackage{algorithmic}
\usepackage{geometry}
\usepackage{float}

\newcommand{\comment}[1]{}

\usepackage{pdfpages}
\usepackage{subfiles}

\makeatletter
\let\ps@empty\ps@plain
\makeatother

\begin{document}
\pagenumbering{roman}

\vspace{0.8in}
\begin{Large}
 \noindent Phuong Tran
 \end{Large}
  \vspace{1mm}
 \hrule
   \noindent\textit{Candidate} \vspace{0.5in}
 
 \begin{Large}
 \noindent Electrical and Computer Engineering
 \end{Large}
 
\vspace{1mm}
 \hrule
    \noindent\textit{Department} \vspace{0.4in}

   \noindent This thesis is approved, and it is acceptable in quality and form for publication: \vspace{0.2in}\\
\begin{large}
 \noindent \textit{Approved by the Thesis Committee:} \vspace{0.25in}
 \end{large}
 
%
%
%

 
 \noindent  {\large Marios Pattichis}, {\small Chairperson}

  \hrule
 \vspace{9mm}
 \noindent {\large Sylvia Celedon-Pattichis}
   \vspace{1mm}
 \hrule
  \vspace{9mm}
 \noindent {\large Lei Yang}
   \vspace{1mm}
 \hrule
    \thispagestyle{plain}
      \setcounter{page}{1}

	\title {Fast Video-based Face Recognition in Collaborative Learning Environments}
	
	\author{Phuong Tran}

	\degree{Master of Science \\ Computer Engineering}
	
	\documenttype{Thesis}
	
	\previousdegrees{B.S., Computer Science, 2018}
	
	\date{December, \thisyear}
	
	\frontmatter


	\maketitle
	
	\begin{dedication}
		To my parents Xuyen and Tran, grandma Van, grandma Gai, grandpa An, and my late grandfather Mai for believing in me and for always being there for me. \\[3ex]
	\end{dedication}
	
	\begin{acknowledgments}
			\vspace{1.1in} First and foremost,  I would like to thank my advisor Professor Marios
		Pattichis for all his patience in providing me with great advice. I would not have done it without his unconditional support and help. I would also like to thank my thesis committee members: Professor Sylvia Celedón-Pattichis and Professor Lei Yang, not only for being on my committee and letting me present my work, but also for their constructive feedback.
		
		I would like to acknowledge Uncle Huan and Auntie Khanh for supporting me when I first came to the US, and for their continuing unconditional support. I also would like to thank Ms. Phuong Nguyen and Thuy-Hang Cao for being there when I most needed it. Lastly, I'd like to show my appreciation to my ivPCL lab-mates and Miguel Angel Hombrados Herrera for being there through the ups and downs.
		
		This material is based upon work supported by the National Science
		Foundation under Grant No. 1613637, No. 1949230, and No. 1842220. Any opinions or
		findings of this thesis reflect the views of the author.  They do not
		necessarily reflect the views of NSF.
		
	\end{acknowledgments}
	
	\maketitleabstract 
	
	\begin{abstract}
	Face recognition is a classical problem in Computer Vision that has experienced significant progress recently. Yet, face recognition in videos remains challenging. In digital videos, face recognition is complicated by occlusion, pose and lighting variations, and persons entering and leaving the scene. The goal of the thesis is to develop a fast method for face recognition in digital videos that is applicable to large datasets. Instead of the standard video-based methods that are tested on short videos, the goal of the approach is to be applicable to long educational videos of several minutes to hours, with the ultimate goal of testing over a thousand hours of videos.
		
		The thesis introduces several methods to address the problems associated with video face recognition. First, to address issues associated with pose and lighting variations, a collection of face prototypes is associated with each student. Second, to speed up the process, sampling, K-means Clustering, and a combination of both are used to reduce the number of face prototypes associated with each student. Third, to further speed up the method, the videos are processed at different frame rates. Fourth, the thesis proposes the use of active sets to address occlusion and also to eliminate the need to apply face recognition on video frames with slow face motions. Fifth, the thesis develops a group face detector that recognizes students within a collaborative learning group, while rejecting out-of-group face detections. Sixth, the thesis introduces a face DeID for protecting the identities of the students. Seventh, the thesis uses data augmentation to increase the size of the training set. The different methods are combined using multi-objective optimization to guarantee that the full method remains fast without sacrificing accuracy.
		
		To test the approach, the thesis develops the AOLME dataset that consists of 138 student faces with 81 boys and 57 girls of ages 10 to 14, which were predominantly Latina/o students. The video dataset consists of 3 Cohorts, 3 Levels from two schools (Urban and Rural) throughout the course of 3 years. Each Cohort and Level contain multiple sessions and an average of 5 small groups of 4 students per school. Each session has from 4 to 9 videos that average 20 minutes each. The thesis trained on different video clips for recognizing 32 different students from both schools. The training and validation datasets consisted of 22 different sessions, whereas the test set contained videos from seven other sessions. Different sessions were used for training, validation, and testing. The video face recognition was tested on 13 video clips extracted from different groups, with a duration that ranges from 10 seconds to 10 minutes. Compared to the baseline method, the final optimized method resulted in very fast recognition times with significant improvements in face recognition accuracy. Using face prototype sampling only, the proposed method achieved an accuracy of 71.8\% compared to 62.3\% for the baseline system, while running 11.6 times faster.
		\clearpage 
	\end{abstract}
	
	\tableofcontents
	\listoffigures
	\listoftables
	
		\begin{glossary}{Longest  string}
		\item[IoU Ratio]Intersection over Union is an evaluation metric used to measure the accuracy of an object detector on a particular dataset. The IoU is the ratio of the overlapping area of ground truth and predicted area to the total area.
		\item[AOLME] The Advancing Out-of-School Learning in Mathematics and Engineering research study
		\item[SOTA] state-of-the-art
		\item[centroid] The center point in a countour
		\item[Ground Truth] A set of labelled data that serves as a point of comparison 
		\item[CNN] Convolutional Neural Network
		\item [DCNN] DEEP Convolutional Neural Network
		\item [DeID] De-Identification
	\end{glossary} 
	
	\mainmatter
	
	\chapter{Introduction}
	The front face recognition topic has been tackled by many researchers with approaches ranging from traditional classification methods to deep learning methods, and the results have been excellent. This is shown through the Face Recognition Homepage \cite{FRHomePage} where the new databases and algorithms of multiple approaches are updated from the early 2000s to the current year. These algorithms include simple image-based face classification using SVM to video-based face recognition using CNN. However, when it comes to face recognition from different poses, there are still significant challenges. The goal of this thesis is to develop video face recognition methods for recognizing students from different poses. The hope is that the methods will support educational researchers in assessing student participation in collaborative learning groups. A collaborative learning group is represented by the group of students closest to the camera. Background groups are not considered part of the collaborative group that the thesis is analyzing. There is a possibility that students or facilitators move between groups.
	Thus, the algorithms need to recognize the current members of the group from a larger group of students. Video face recognition in collaborative learning environments requires that we address occlusion, dynamic presence of participants, lighting, and pose variations. Our goal is to develop fast and accurate methods that can be used to quantify student participation as measured by their presence in their learning groups.
	
	As mentioned above, a fundamental challenge of this thesis's dataset is that 
	face recognition needs to work
	on a large variety of poses.
	As long as a sufficiently small part 
	of the face is visible,
	the algorithm needs to identify the student.
	This requirement covers occlusion.
	
	Furthermore, students may disappear or reappear
	because the camera moves, or the students take
	a break, or because they have to leave the group.
	Hence, there are significant challenges for video face recognition that is unique to this thesis and are not present in standard face recognition.

	\section{Motivation}
	
	As part of a collaboration project between engineering 
	and education, the goal is to assist educational researchers to analyze how students who join the program learn and/or facilitate the learning of other students. Therefore,
	the problem of identifying who's who is crucial for assessing student participation in the project.

	This thesis's main motivation is to develop robust methods to track and recognize the participants in the AOLME program. Furthermore, the developed methods need to be fast. Eventually, this thesis will need to apply the methods to approximately one thousand hours of videos that need to be analyzed. This thesis focuses on detecting all the students in the collaborative groups along with tracking and recognizing them over the entire session. This will contribute to the analyses needed for educational researchers to keep track of whether a student is newly joined or such student is a returning participant, how often students show up during each session, and over the entire program.
	
\begin{figure*}[!t]
		\centering
		(a)~\includegraphics[width=0.27\textwidth]{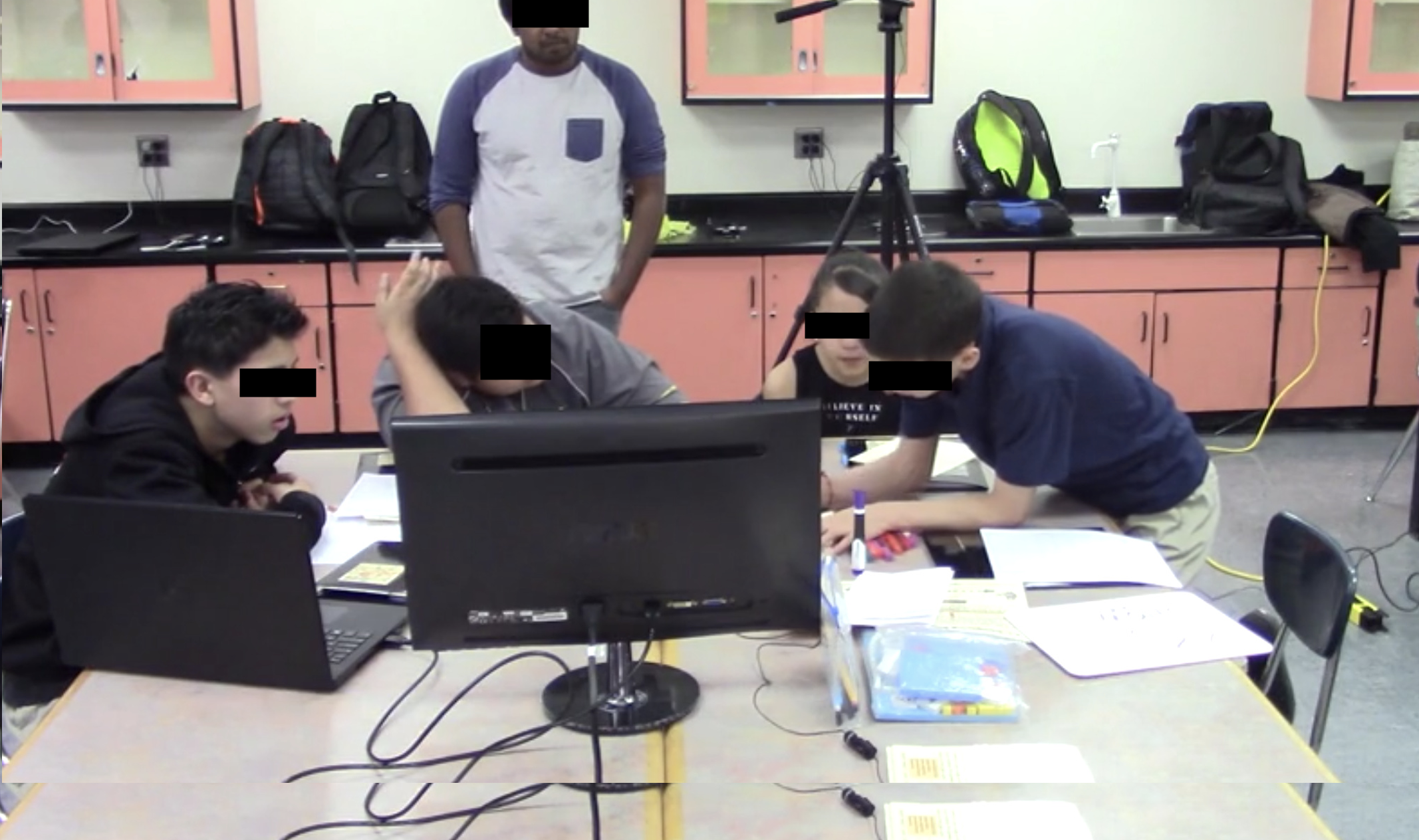}~
		(b)~\includegraphics[width=0.27\textwidth]{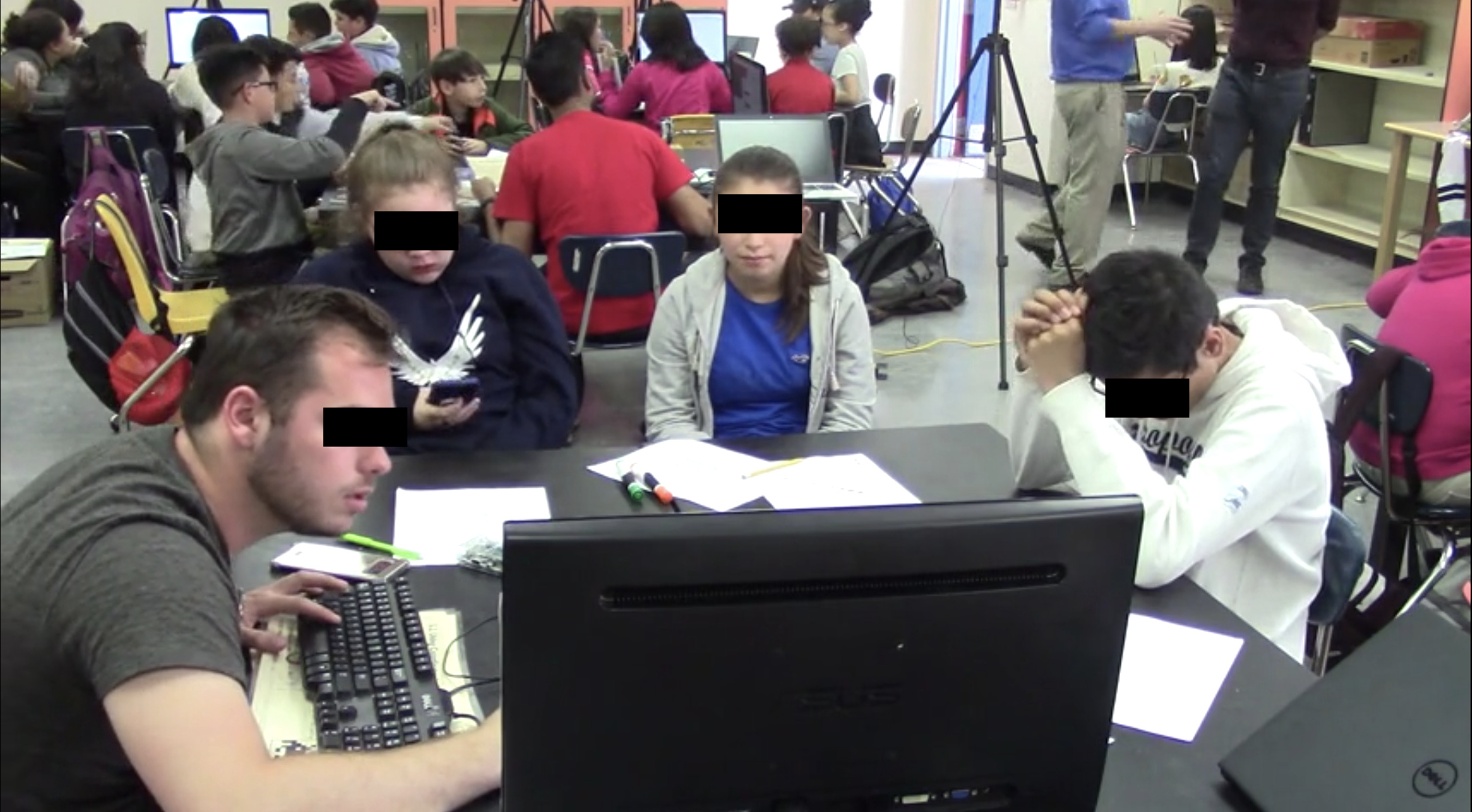}~
		(c)~\includegraphics[width=0.27\textwidth]{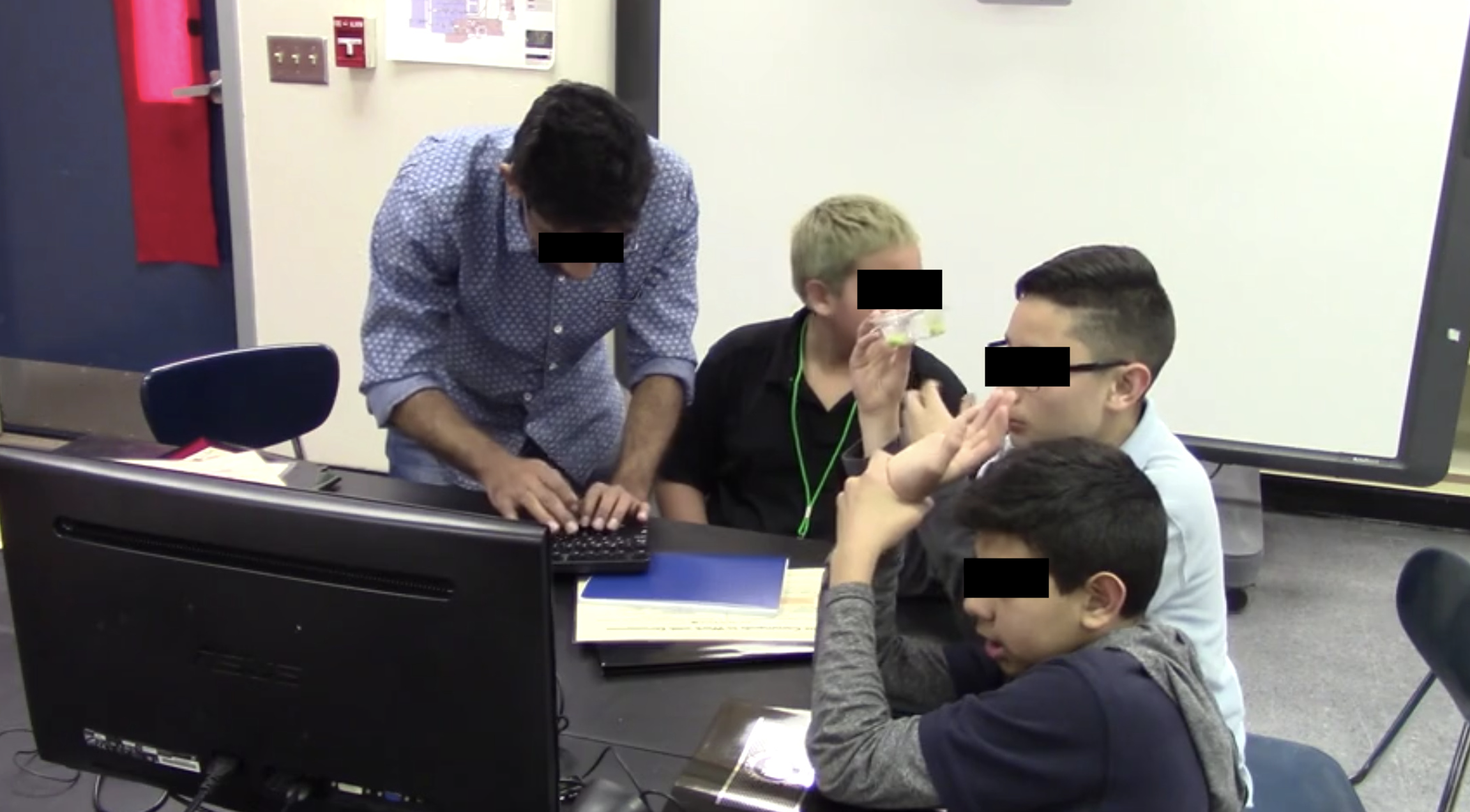}~\\[0.05 true in]
		(d)~\includegraphics[width=0.27\textwidth]{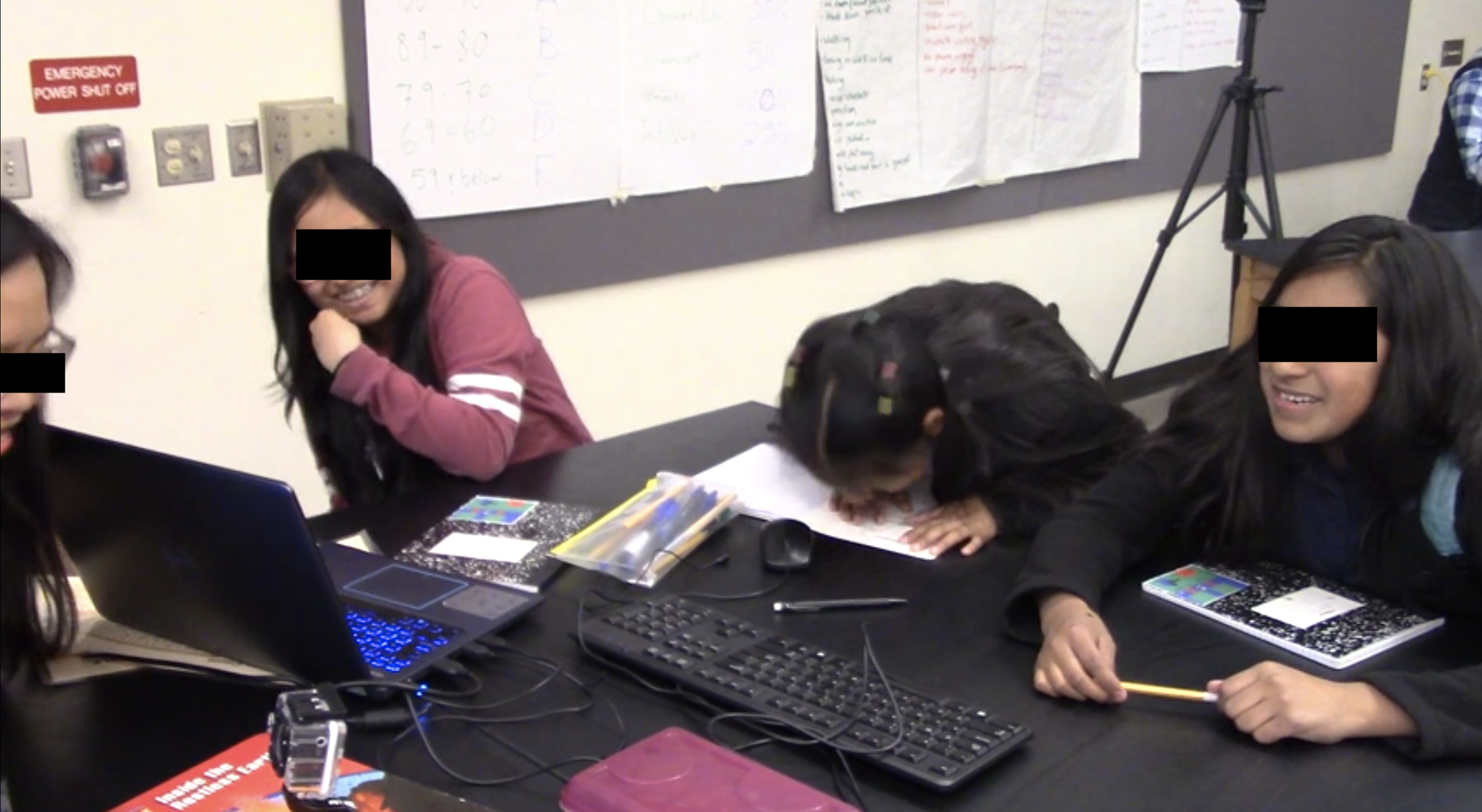}~
		(e)~\includegraphics[width=0.27\textwidth]{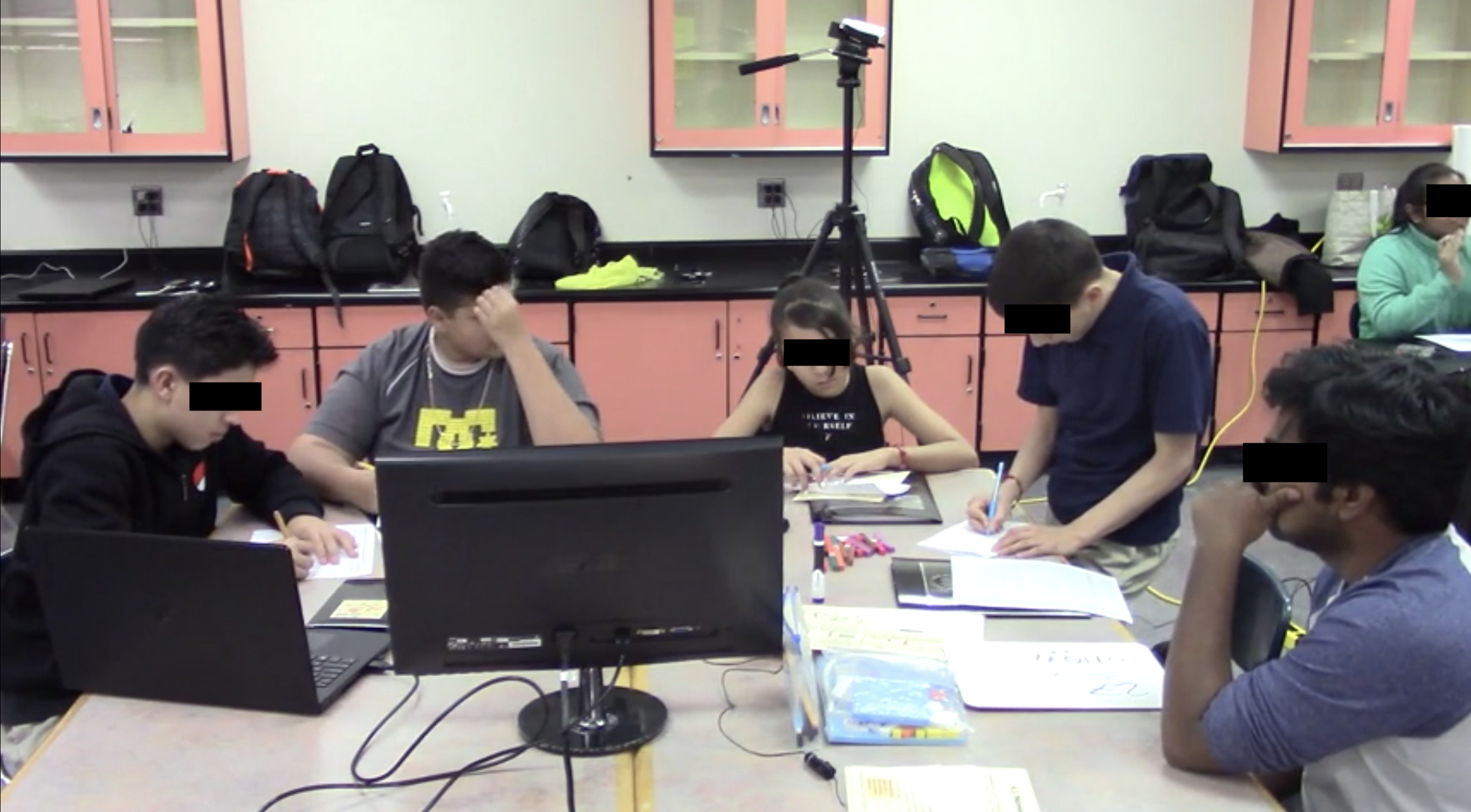}~
		(f)~\includegraphics[width=0.27\textwidth]{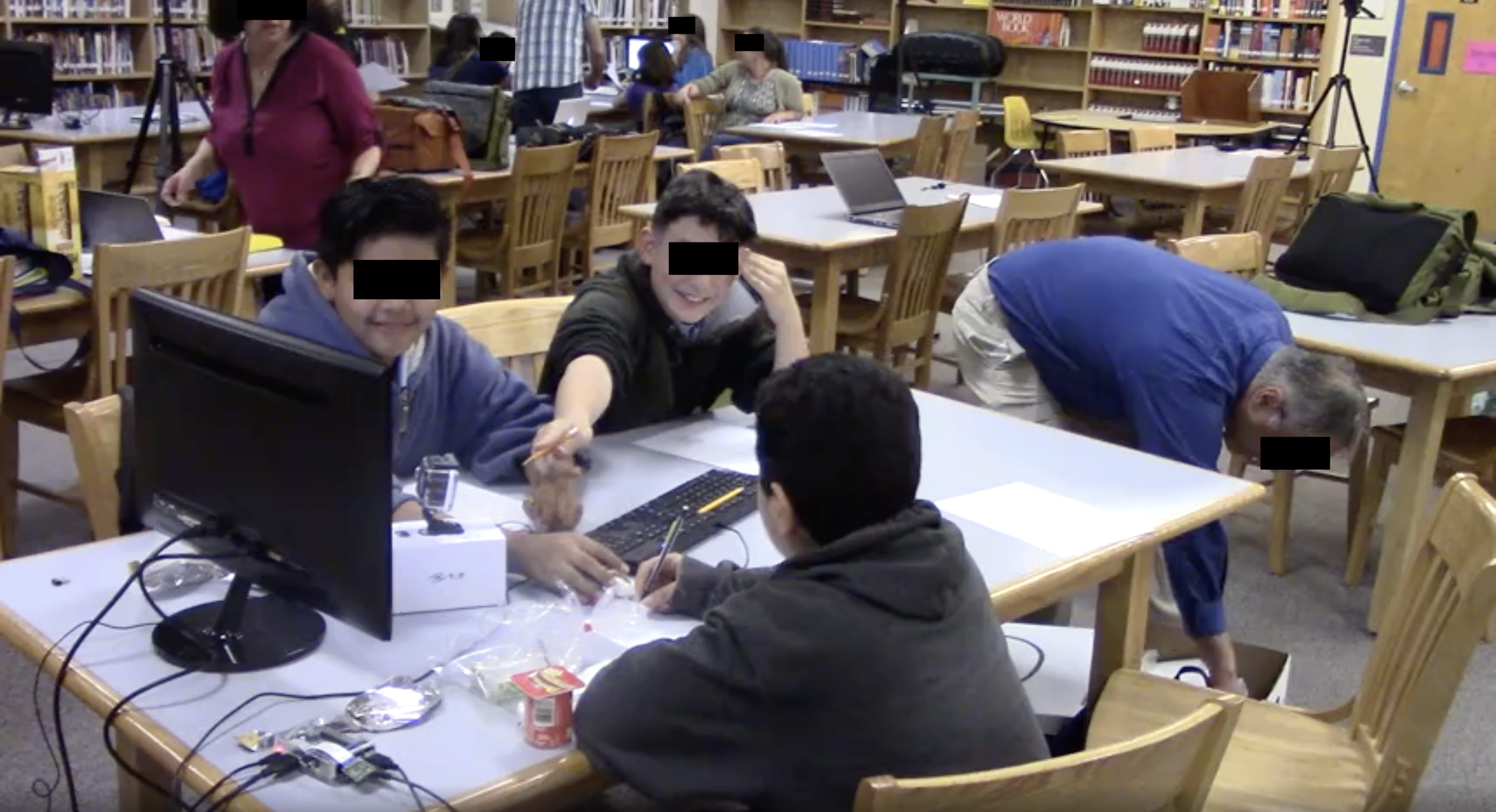}~\\[0.05 true in]
		(g)~\includegraphics[width=0.27\textwidth]{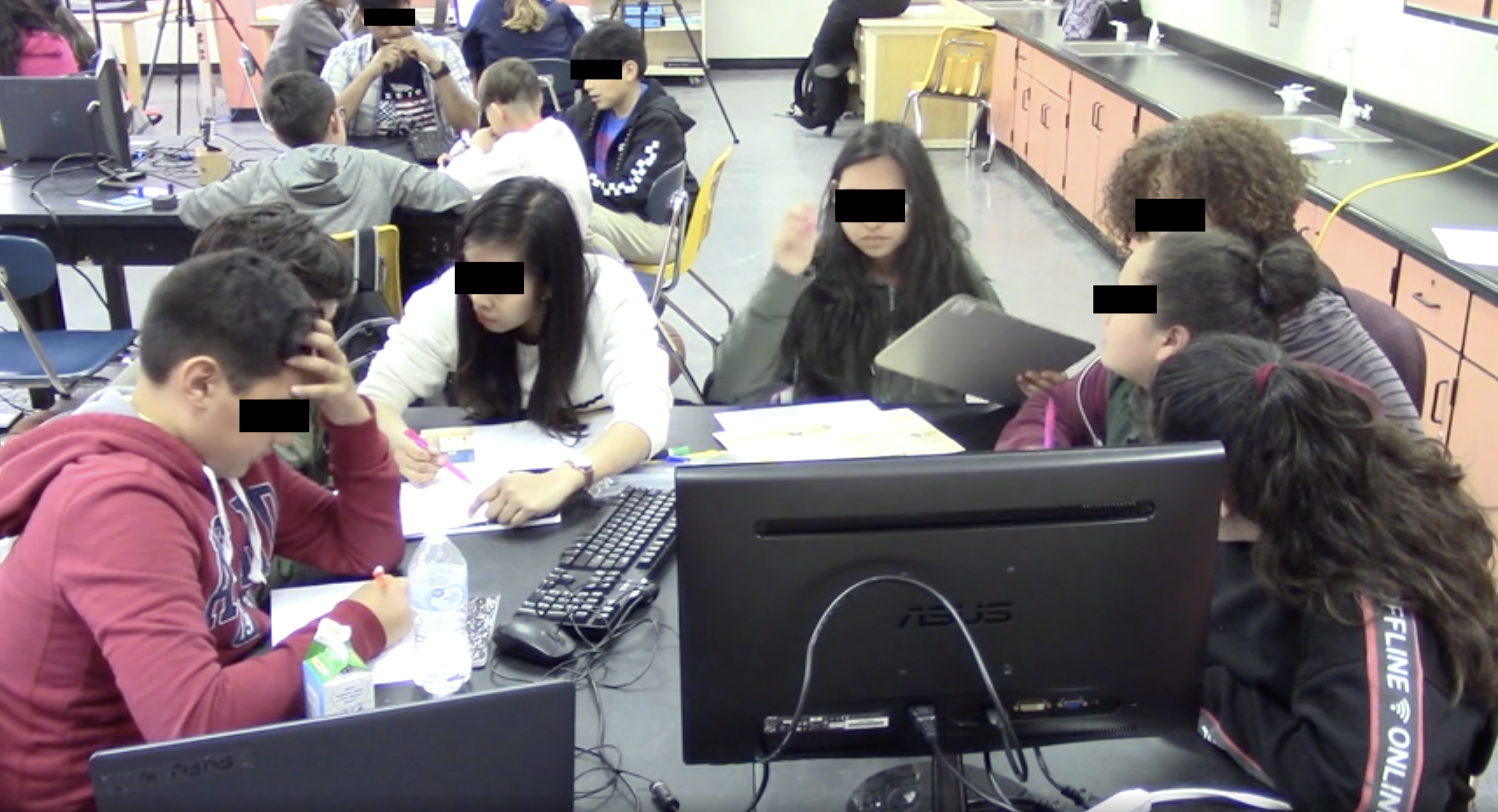}~
		(h)~\includegraphics[width=0.27\textwidth]{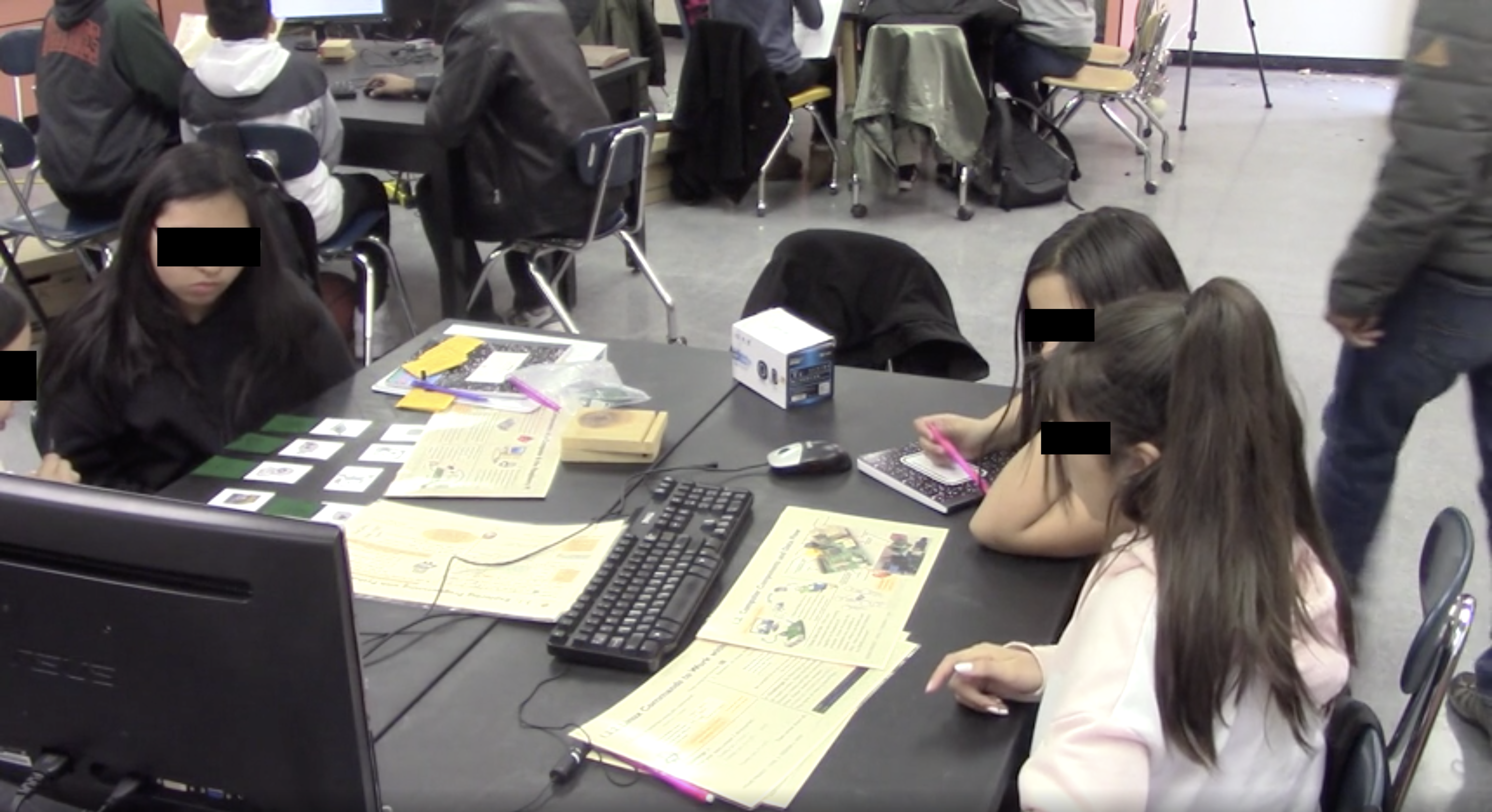}~
		(i)~\includegraphics[width=0.27\textwidth]{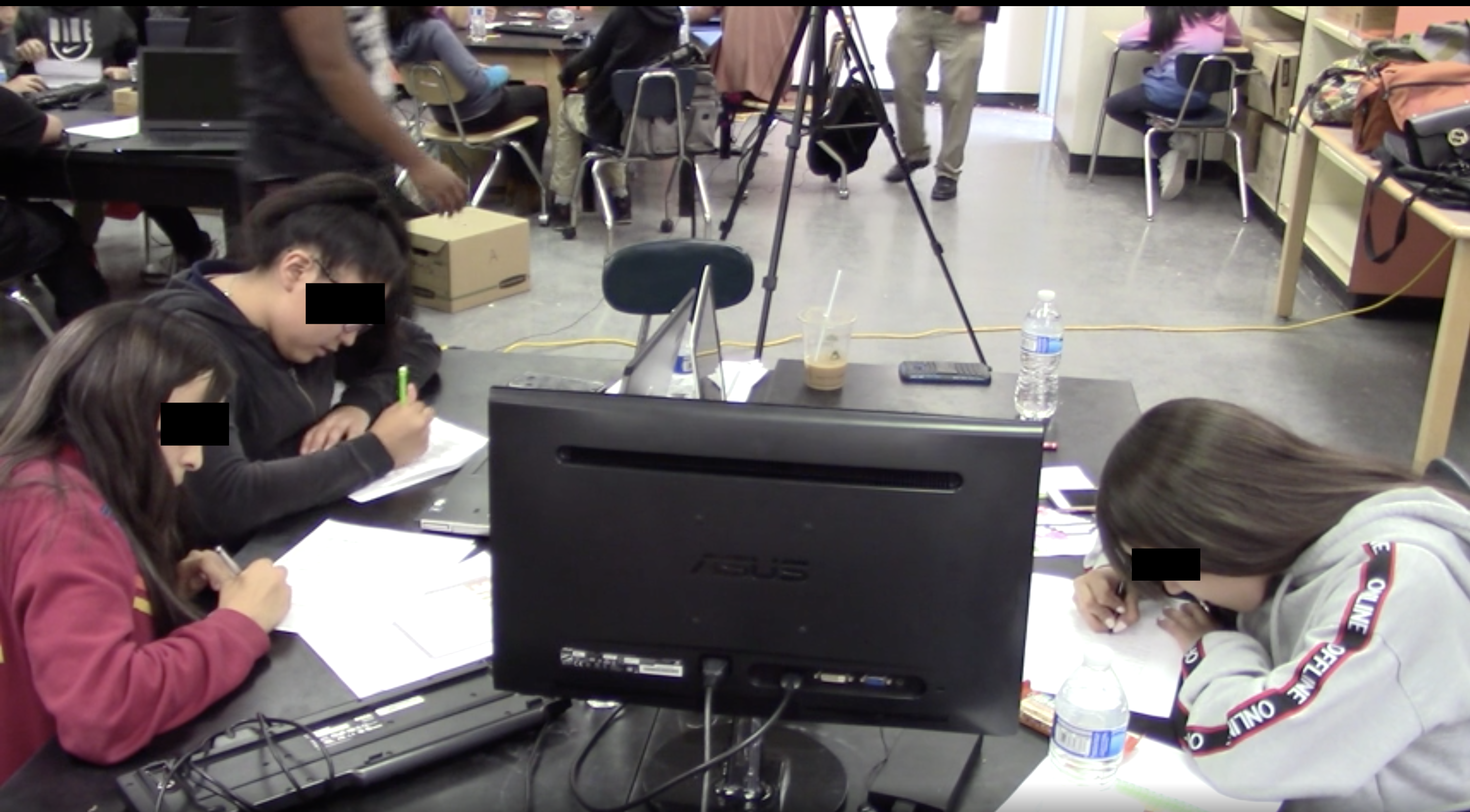}~\\[0.05 true in]
		(j)~\includegraphics[width=0.27\textwidth]{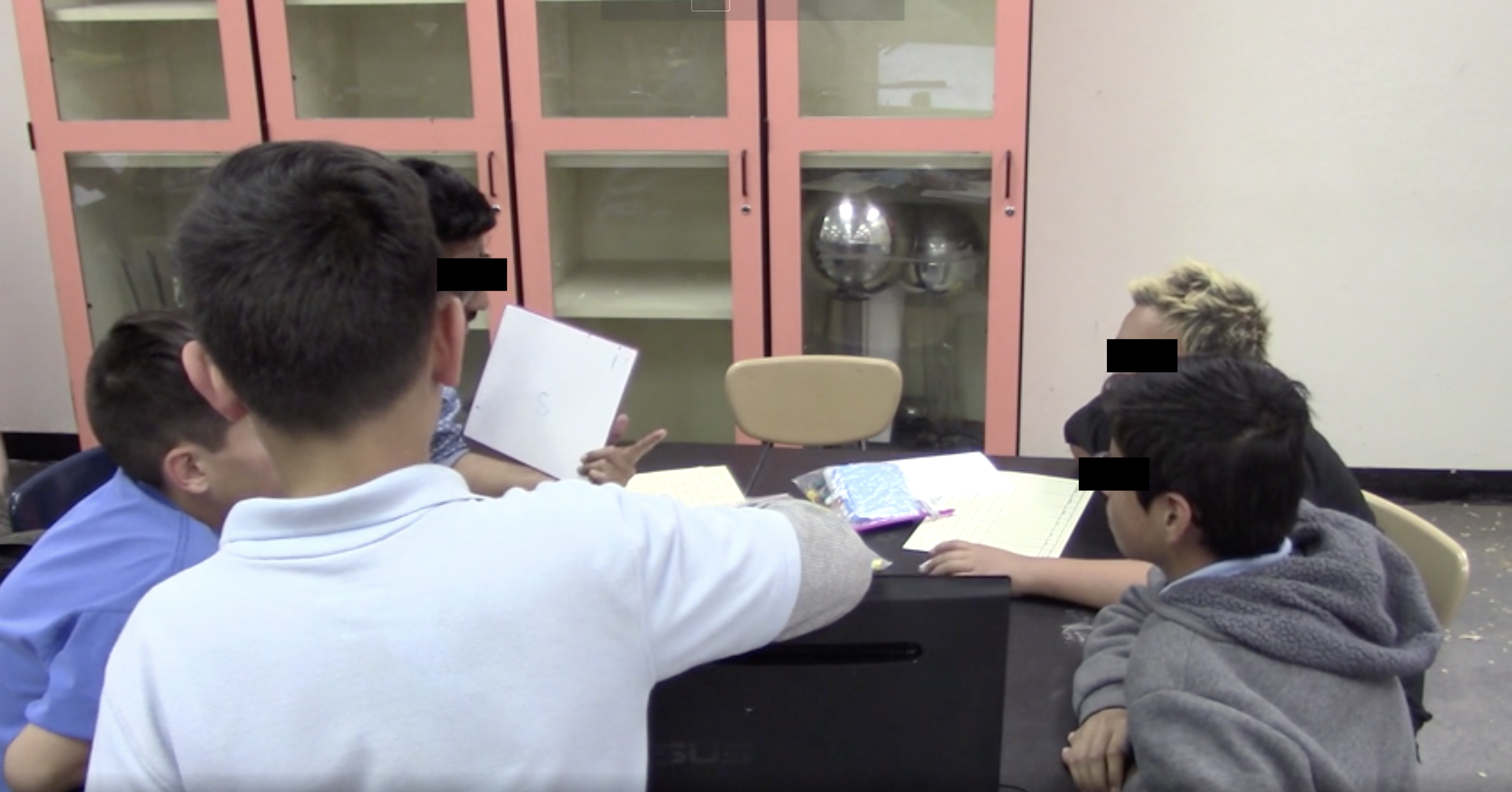}~
		(k)~\includegraphics[width=0.27\textwidth]{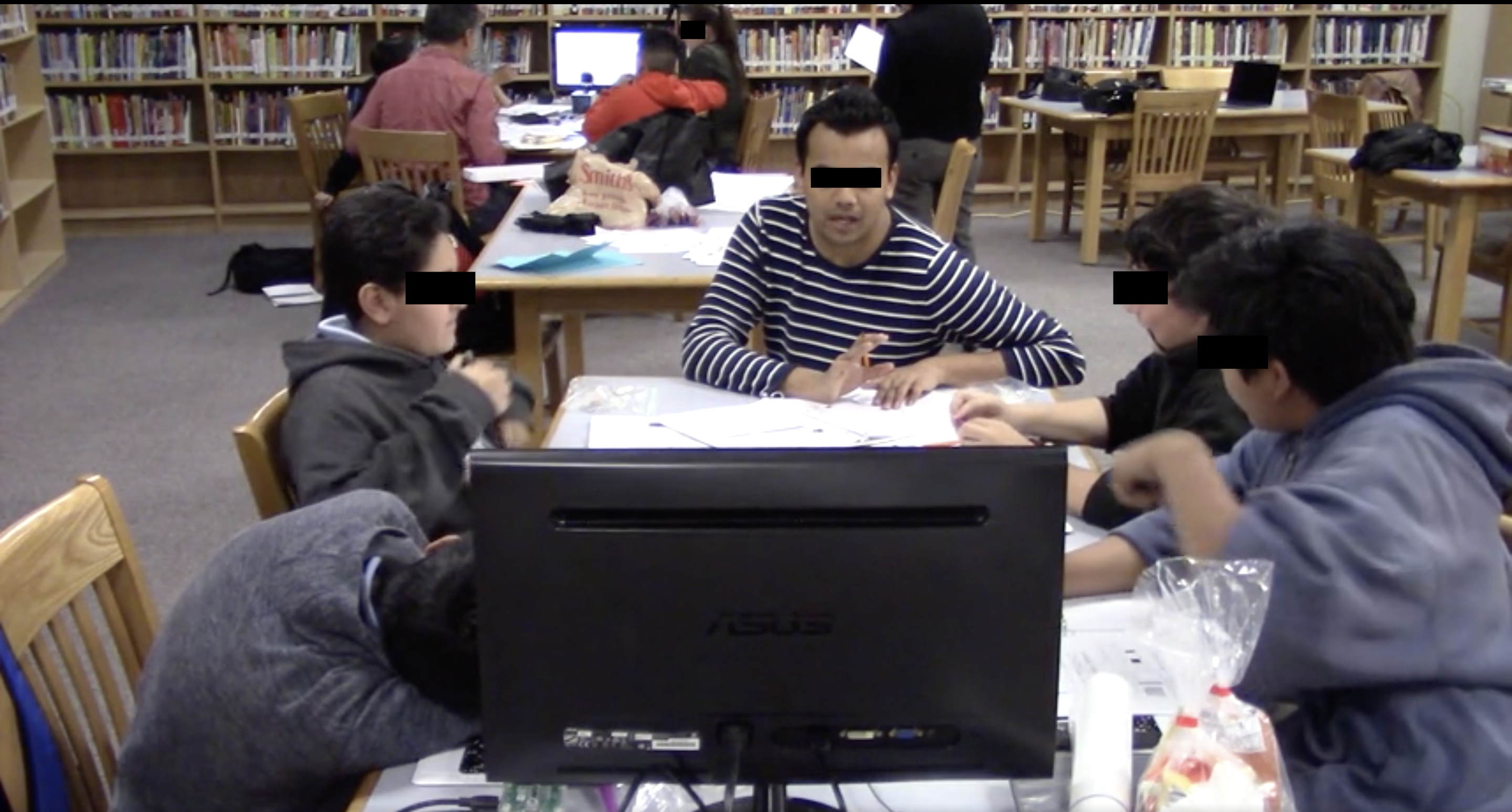}~
		(l)~\includegraphics[width=0.27\textwidth]{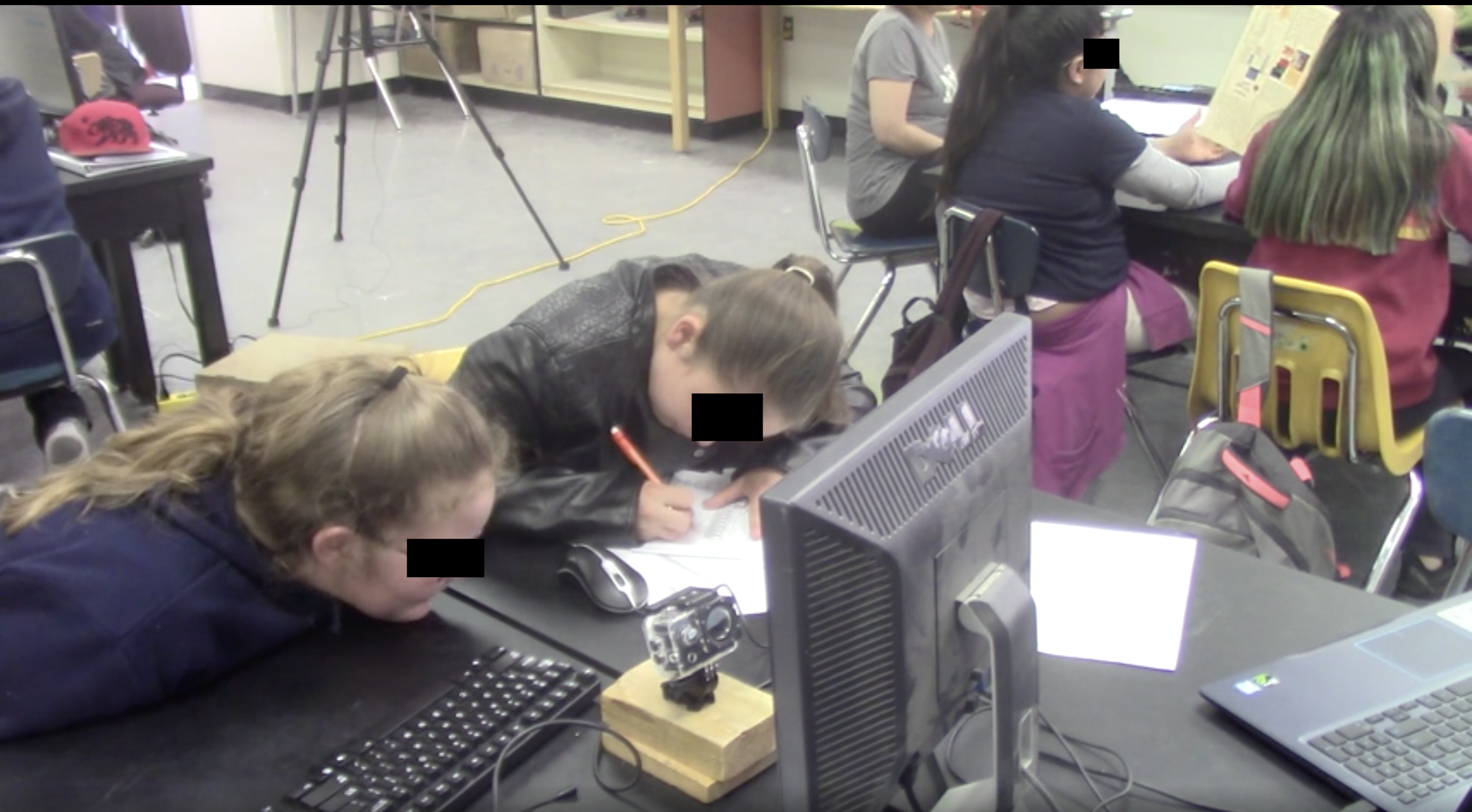}~\\[0.05 true in]
		(m)~\includegraphics[width=0.27\textwidth]{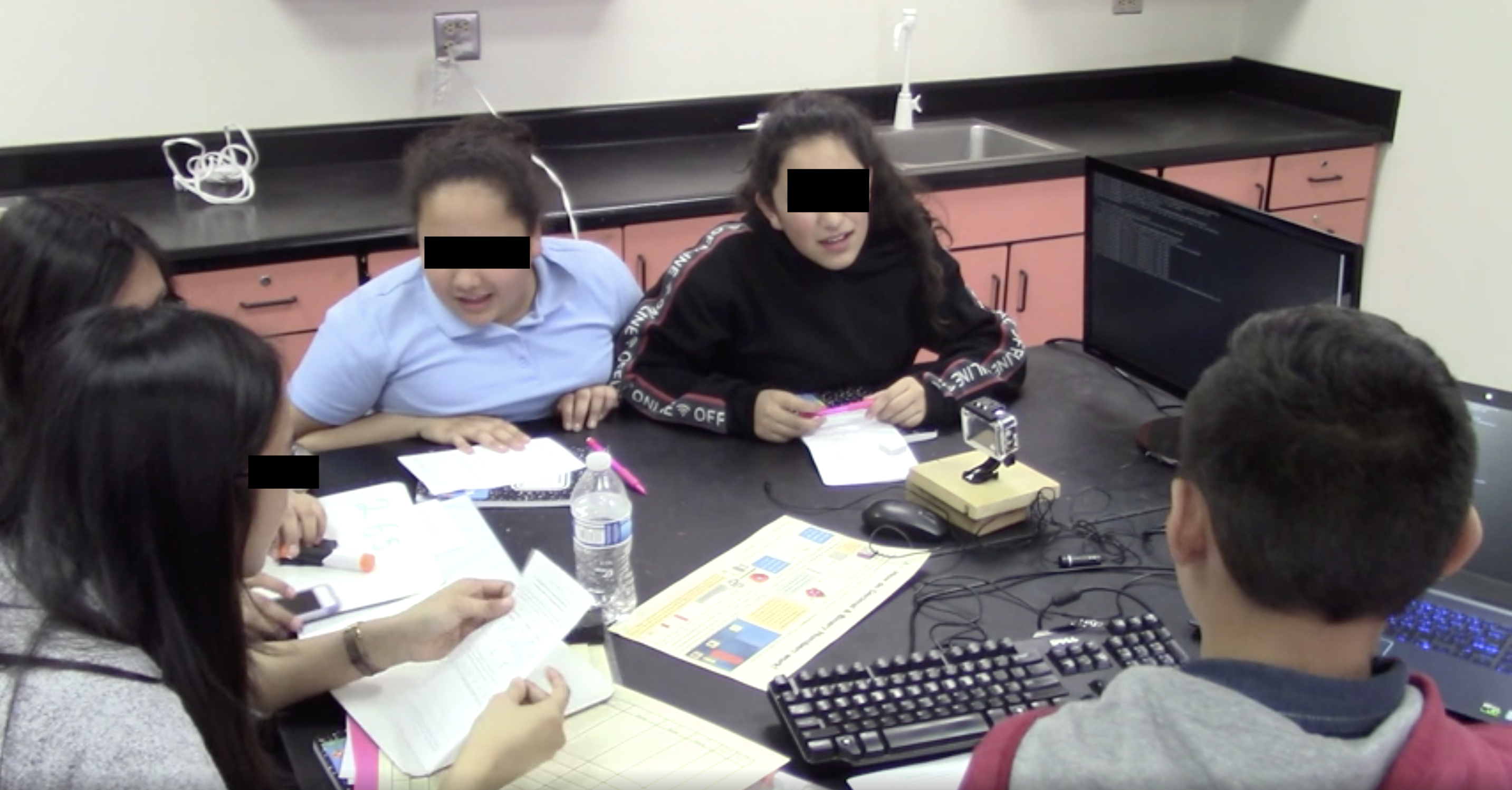}~
		(n)~\includegraphics[width=0.27\textwidth]{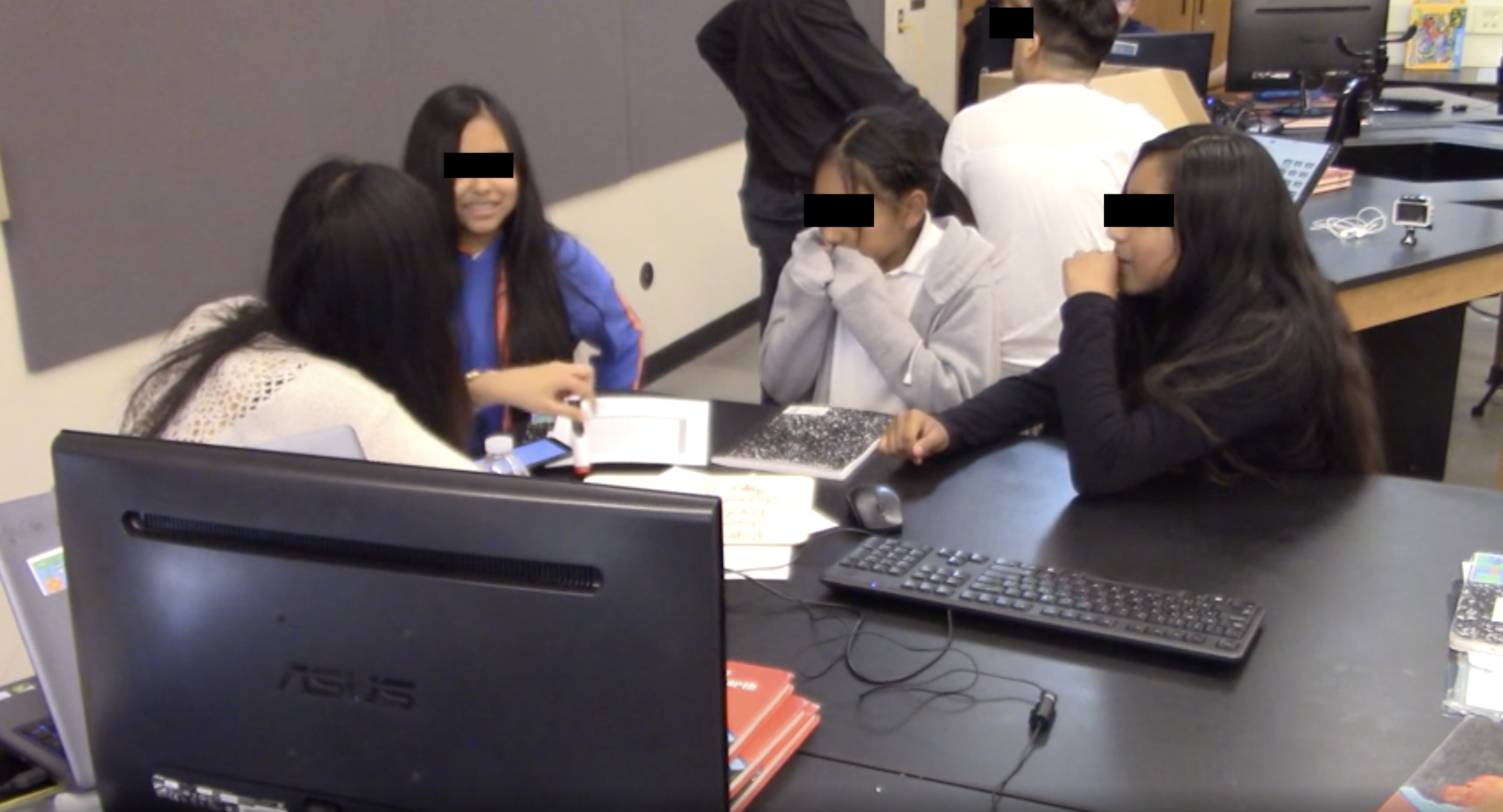}~
		(o)~\includegraphics[width=0.27\textwidth]{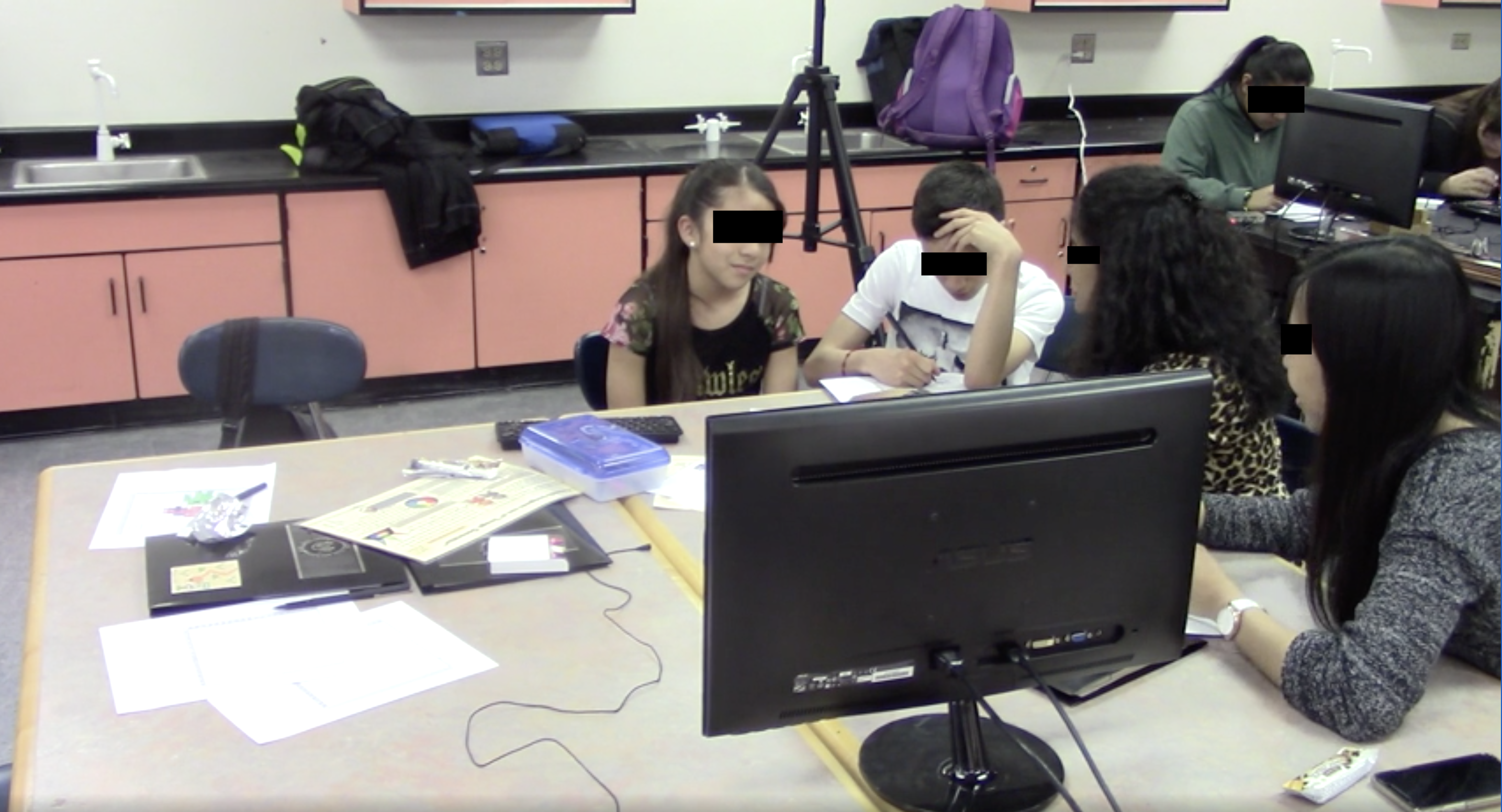}~\\[0.05 true in]
		\caption[Examples of frames extracted from multiple videos] {
			Examples of frames extracted from multiple videos showing the challenges of video recognition from the AOLME video datasets.} 
		\label{fig:Problems}
	\end{figure*}

	A summary of the challenges associated with recognizing faces from the AOLME video datasets is shown in Figure \ref{fig:Problems}. The thesis encountered occlusion which resulted in recognition failures as the students were occluded by another student (see a, g, j, and m in Figure \ref{fig:Problems}), by the camera angle (see d, f, i, j, h, and n in Figure \ref{fig:Problems}), or by their positions (see d, e, and k in Figure \ref{fig:Problems}). Pose variation is also another problem when most of the faces are covered when they turn (see b, c, l, and o in Figure \ref{fig:Problems}). In addition, there are 5 to 6 groups in an AOLME session, but the algorithm only focuses on the collaborative group, which is the one that is closest to the camera. Unlike other common datasets that many other methods choose to test on (e.g., VoxCeleb1, CelebA, etc.), where different camera lenses are focused on only the celebrities and blur the background, the AOLME dataset keeps everyone in focus within each video frame. The issue occurs as there exist too many background groups with multiple out-of-group faces: i.e, (b), (e), (f), and (g).
	
	The thesis focuses on only part of the set of all the collaborative learning groups. In the future, the proposed method can be integrated to identify all the students who joined the program.

\section{Thesis Statement}
	
	The thesis is focused on the development of fast and robust methods for face recognition in the AOLME video datasets. First, the method uses a K-means approach to identify image clusters for recognizing faces from different poses. Second, the method applies multi-objective optimization to study the inter-dependency between recognition rate, number of clusters, and recognition accuracy. Along the Pareto front of optimal combinations, the proposed approach selects an optimal number of face clusters that provide for a fast approach without sacrificing recognition accuracy. Third, the thesis applies frame rate skipping to boost recognition rate when participants do not move within a small number of frames. Fourth, the thesis uses the past recognition history to deal with occlusions and hence support consistent recognition throughout the video. Lastly, the thesis combines all previous approaches with data augmentation with transformation to increase the size of the training dataset. Compared to InsightFace, the proposed system provides for significantly faster recognition rates and higher accuracy.
	
\section{Contributions}
	The contributions of this thesis include:
	\begin{enumerate}
		\item Clustering methods to identify image clusters
		for recognizing faces from different poses.
		\item Multi-objective optimization and frame rate skipping to reduce recognition time.
		\item Robust tracking of the participants with multi-frame/video processing to deal with occlusions for consistent recognition.
		\item DeID faces in  digital videos to protect individuals' identities.
		\item Data augmentation to increase the size of the training dataset.
	\end{enumerate}
	
		\section{\label{section:overview}Overview}
	The remainder of the thesis is organized into 5 chapters:
	\begin{itemize}
		\item \textbf{Chapter 2: Background.} This chapter describes prior work.
		\item \textbf{Chapter 3: Dataset.} This chapter describes Dataset Organization and Ground Truth.
		\item \textbf{Chapter 3: Methods.} This chapter provides a description of the video face recognition methods.
		\item \textbf{Chapter 4: Results.} This chapter provides a summary for a baseline image and the proposed video face recognition method.
		\item \textbf{Chapter 5: Conclusion and future work.} This chapter provides a summary of the thesis and future work proposal.
	\end{itemize}

	\chapter{Background}
	This chapter provides a summary of prior research in ivPCL lab, face detection, and face recognition methods along with some commonly used datasets for face detection and recognition. Face recognition is a classic problem; thus, there has been extensive research on the subject (\cite{FRHomePage}, \cite{Deng}, \cite{PCA}...). This thesis will present the recent methods that involve video-based methods for face detection and face recognition. 
	
	\section{Face Detection}
	In order to recognize a face, the algorithm first needs to detect if a face exists. A table of common face detection datasets will be shown in Table \ref{detectDataset}. A short summary of relevant methods will also be provided.
	
	\begin{table}[ht!]
		\caption[Summary of commonly used face detection datasets.]{Summary of commonly used face detection datasets. These
			datasets contain a large number of classes with images extracted
			from multiple sources in which YTF and  iQIYI-VID are the two datasets that contain videos, whereas the rest are only image-based datasets.}
 		\label{detectDataset}
 		\scalebox{0.99}
  		{
  			\renewcommand{\arraystretch}{1.75}
  			\begin{tabular}{P{1.8cm} P{5cm} P{6.5cm}}
  			\hline
 				\hline
 				\textbf{Dataset} & \textbf{Summary} & \textbf{URL}\\
 				\hline
 				\hline
 				AFW & 
 				$\bullet$ 205 images with 473 labeled face using Flickr images. \newline
 				$\bullet$ annotations include a rectangular bounding box, 6 landmarks and the pose angles. & \url{https://vision.ics.uci.edu/papers/ZhuR_CVPR_2012/ZhuR_CVPR_2012.pdf} \\
				
 				FDDB & 
 				$\bullet$ annotations for 5,171 faces in a set of 2,845 images.	 & \url{http://vis-www.cs.umass.edu/fddb/index.html} \\
				
 				PASCAL FACE & 
 				$\bullet$ 851 images and 1,341 annotated faces. & \url{http://host.robots.ox.ac.uk/pascal/VOC/databases.html} \\
				
 				WIDER FACE &
 				$\bullet$ 32, 203 images with 393, 703 labeled faces.   \newline
 				$\bullet$ large variations in appearance, pose, and scale. & \url{https://www.tensorflow.org/datasets/catalog/wider_face} \\
				
 				IJB-A & 
 				$\bullet$ 24,327 images and 49,759 faces for both face detection and recognition & \url{https://www.nist.gov/itl/iad/image-group/ijb-dataset-request-form} \\
				
 				MALF &
 				$\bullet$ 5,250 images and 11,931 faces.\newline
 				$\bullet$ first face detection dataset that supports fine-gained evaluation. & \url{http://www.cbsr.ia.ac.cn/faceevaluation} \\
 				\hline
  			\end{tabular}
  		}
	\end{table}
	
\textbf{Haar Cascade \cite{Haars}:}\\
	Haar Cascade is a machine learning-based method with a simple model.  Although this approach gives lots of false predictions and does not work with occlusion and non-frontal faces, the fast run time, simple architecture, and the fact that it can perform face detection at different scales make the approach an interesting baseline method.
	
\begin{figure}[!b]
		\centering
		\includegraphics[width=1\textwidth]{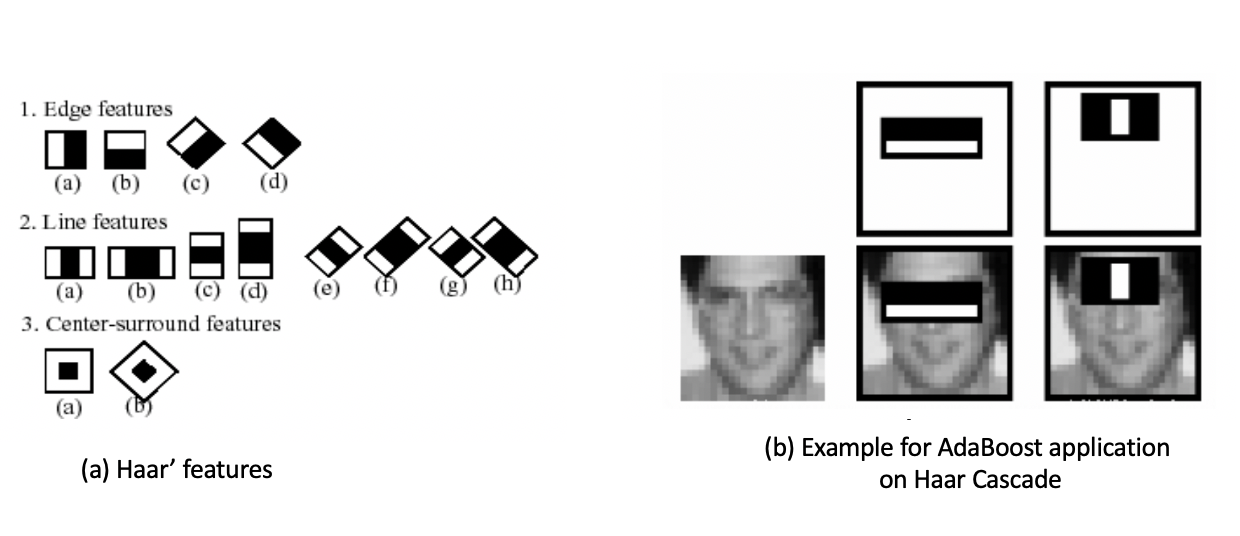}
		\caption{Haar Cascade Features \cite{Haars}.} 
		\label{Haars}
	\end{figure}

	The method uses Integral Image which is calculated by the sum of all the pixels that are above and on its left of the original image. The integral image allows us to compute the sums of pixel intensities within any given rectangular region of the given image. AdaBoost is applied to reject features that are irrelevant and only keep the best-representing features. The feature representation and AdaBoost application to the method example are shown in Figure \ref{Haars} (a) and (b), respectively. In addition, Attentional Cascade is proposed to reduce training time as it processes a group of features in stages; thus, if it fails on the first stage, the window is discarded. The next stage is processed and, whichever window passes all stages, is assigned to be a face.
	
	OpenCV has provided a library for implementing Haar Cascade for easy accessibility with trained XML files to help with the process of acquiring face (positive) and no face (negative) training images.

	\textbf{YOLO \cite{YOLO}: }  \\
	YOLO (You Only Look Once) is a SOTA deep learning object detection method, including face detection, that produces results in real-time. YOLO uses a single neural network to train the whole image, which makes it very robust. It takes an image as input and outputs the bounding boxes, which include the height, width, center, and label, and each region's prediction confidence. The image is divided into multiple blocks or cells, with each block or cell responsible for detecting the objects that appear inside if the center of the object belongs to this cell. YOLO attempts to remove false positives by removing predictions that give low IoU between the predicted and the actual detection boxes. An example of object detection using YOLO is shown in Figure \cite{YOLO}.
	\begin{figure}[!b]
		\centering
		\includegraphics[width=1\textwidth]{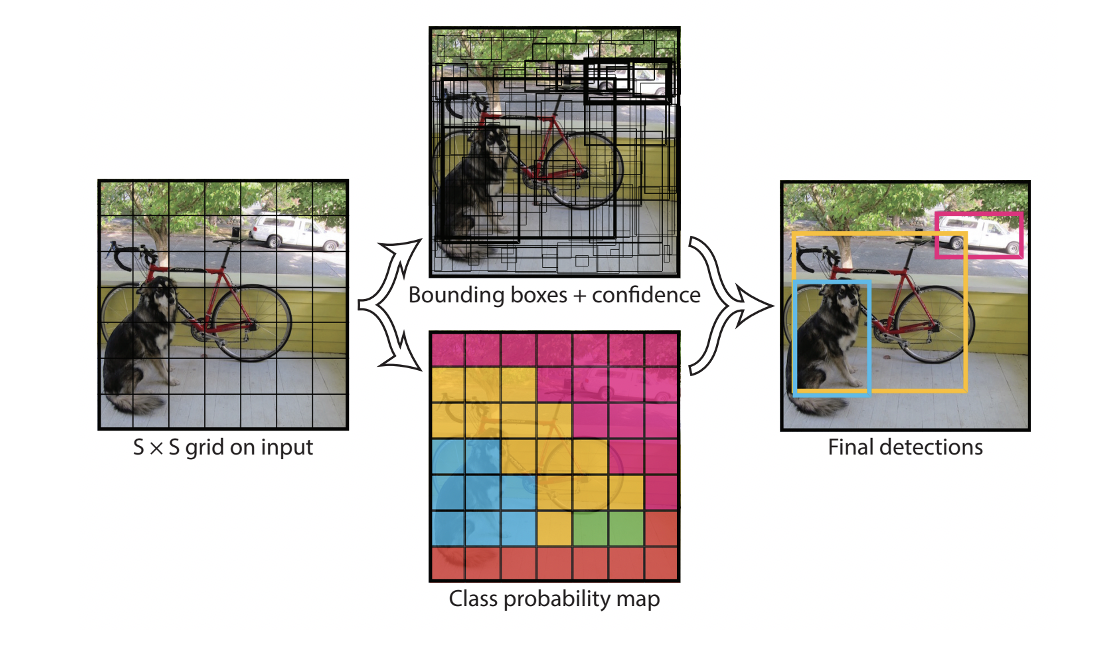}
		\caption{Object detection using YOLO \cite{YOLO}.} 
		\label{YOLO}
	\end{figure} 
	
	\textbf{Dlib \cite{Dlib}:} \\
	Dlib is a C++-based toolkit for machine learning and data analysis applications; Dlib is Python-friendly as it is quite simple to use Python bindings. Dlib is commonly used in face detection and face landmark detection as it can estimate 68 pairs of facial landmarks. Figure \cite{Dlib} (a) provides a visualization of Dlib landmarks. There are two commonly used methods with Dlib: Histogram of Oriented Gradients (HOG) and CNN-based.
	
	\begin{enumerate}
		\item 
		Dlib-HOG \\
		Dlib is a feature extraction method using Histogram of Oriented Gradients (HOG) and SVM. HOG does not deal with face detection associated with different poses; thus, the method only works for frontal faces and/or slightly rotated frontal faces. Dlib-HOG does not do well with large occlusions and because the images require a minimum size of 80x80, it does not work well with detecting small area faces. Figure \cite{Dlib}(b) \cite{leoDiCap} shows each cell histogram computation. These cells are combined into one final cell that represents the entire face.
		
		\item
		Dlib-CNN\\
		This method uses a Maximum-Margin Object Detector (MMOD) with CNN-based features. With a simple training process, a large dataset for training is not required.
		As a CNN-based method, Dlib with CNN can detect faces at multiple poses, but with the downside of not being able to run in real-time video with a CPU.
		
	\end{enumerate}

	\begin{figure}[!t]
		\centering
		\includegraphics[width=1\textwidth]{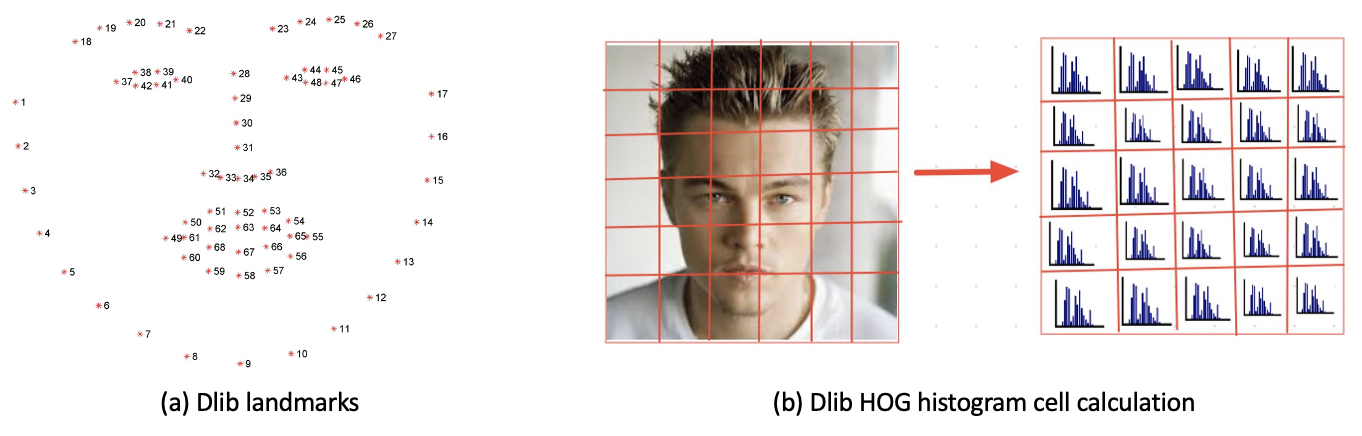}
		\caption{Dlib landmarks and Dlib-HOG histogram for each cell calculation \cite{leoDiCap}.} 
		\label{Dlib}
	\end{figure} 
	
	\textbf{Multi-Task Cascaded Convolutional Neural Networks (MTCNN) \cite{MTCNN}:}
	\begin{figure}[!b]
		\centering
		\includegraphics[width=1\textwidth]{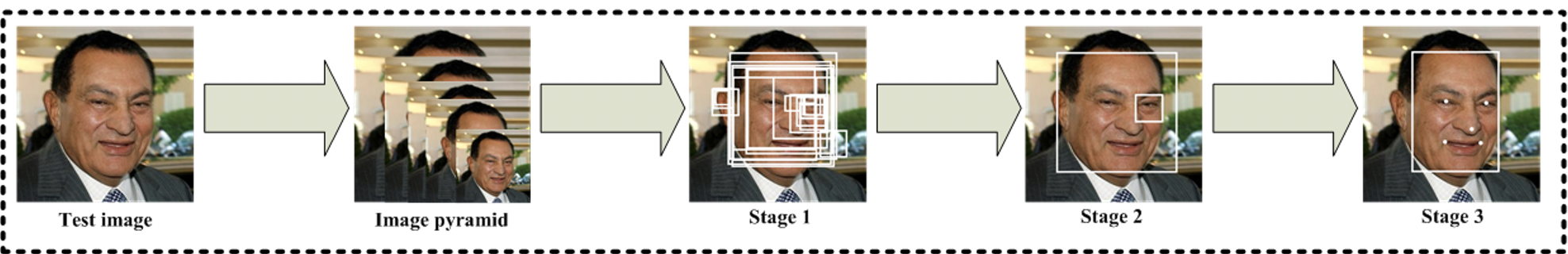}
		\caption{MTCNN \cite{MTCNN}.} 
		\label{MTCNN}
	\end{figure} 
	MTCNN is a deep learning-based face detection method that returns not only the detected bounding boxes with their coordinates but also 5 landmark points. MTCNN uses cascade structures with three stages that consisted of the three connected neural networks including Proposal Network (P-Net), Refinement Network (R-Net), and Output Network (O-Net), in that order.  As Figure \ref{MTCNN} shows, first, the image is resized multiple times for face detection as varied scales and detected bounding boxes are found through P-Net. The output of this network is fitted to the next stage (R-Net). Applying Non-Maximum Suppression (NMS), which chooses one bounding box out of many other overlapping ones (Figure \ref{MTCNN} shows the transition from stage 1 to stage 2), further eliminates false positives. The last stage (O-Net) produces a bounding box on the detected face along with the five landmarks, which is used for alignment. The thesis adopts MTCNN for face detection and alignment.
	
	\section{Face Recognition}
	
	In this section, the thesis will provide a summary of commonly used face recognition methods and datasets. The goal of video face recognition is to associate a label for each face. These faces can come from a still image or video. A table of common face recognition datasets and common face recognition methods are given in Tables \ref{faceDataset} and \ref{faceMethods}, respectively. 
	
	\begin{table}[ht!]
		\caption[Summary of commonly used face recognition datasets.]{Summary of commonly used face recognition datasets. These
			datasets contain a large number of classes with images extracted
			from multiple sources in which YTF and iQIYI-VID are the two datasets that contain videos, whereas the rest are only image-based datasets.}
		\label{faceDataset}
		\scalebox{0.99}
		{
			\renewcommand{\arraystretch}{1.75}
			\begin{tabular}{ P{3.7cm} P{10cm}}
				\hline
				\textbf{Dataset} & \textbf{Summary} \\
				\hline
				\hline
				
				YTF  (Videos) \cite{YTF} & 
				$\bullet$ 3,425 videos with 1595 different people labels (averaging 2.15 videos per subject). \newline
				$\bullet$ All videos are downloaded from YouTube. \\
				
				iQIYI-VID (Videos) \cite{iQIYI-VID}
				& 
				$\bullet$ 600,000 video clips with 5,000 celebrities labels, with duration ranges from 1 to 39 seconds.	\newline
				$\bullet$ Human annotation for Ground truth. All videos are from iQIYI variety shows, films, and television dramas (mostly Asians).
				\\
				CASIA  (Images)\cite{CASIA}& 
				$\bullet$ 7,491 authentic and 5,123 spliced images with JPEG, TIFF, BMP image types, averaging 37 images/person. \newline
				$\bullet$ Controlled environments, single person recognition.\newline
				$\bullet$ Manual source of ground truth.
				\\
				
				LFW (Images) \cite{LFWTech} &
				$\bullet$ 13,000 images of faces, averaging 2.3 images/person \newline
				$\bullet$ Same face scale.\newline
				$\bullet$ Single person recognition. \newline
				$\bullet$ Ground truth source collected from the Internet.\\
				VGG Face (Images) \cite{VGG} 
				& 
				$\bullet$ 2.6 million images with 2,622 identities, averaging of 362 images/person. \newline
				$\bullet$ Same face scale. \newline
				$\bullet$ Ground truth source collected from the Internet. \\
				
				{\begin{tabular}[c]{@{}l@{}}Megaface (Images)\\ (Some children)\cite{Megaface}\end{tabular}}
				
				& 
				$\bullet$ 1,000,000 faces with their respective bounding boxes, 672K identities.\newline
				$\bullet$ All images obtained from Flickr (Yahoo's dataset) and licensed under Creative Commons.	    \\
				{\begin{tabular}[c]{@{}l@{}}\textbf{AOLME} \textbf{(Videos)}\\ \textbf{(Mostly children)}\end{tabular}}
				
				& 
				$\bullet$ $>$150 faces with 2,200 hours of videos.\newline
				$\bullet$ $>$ 1h/session/single sessions. \newline
				$\bullet$ $>$5h/groups of sessions, $>$50 hours/Cohort.\newline
				$\bullet$ Collaborative learning environments with multiple person per frame and various face scales. \newline
				$\bullet$ Ground truth was collected manually.\\
				\hline
			\end{tabular}
		}
	\end{table}

	\begin{table}[ht!]
		\caption{Summary of the common face recognition methods.}
		\label{faceMethods}
		\scalebox{0.79}
		{
			\renewcommand{\arraystretch}{1.75}
			\begin{tabular}{P{3cm} P{5cm} P{3cm}  P{6cm} P{7.5cm}}
				\hline
				\textbf{Author} & \textbf{Summary} & \textbf{Datasets} & \textbf{Results} \\
				\hline 
				\hline
				\textbf Choi et al. (2020)  Gabor Face Representations with DCNN  &
				$\bullet$ different Gabor face representations in training and DCNN phases in testing. \linebreak
				$\bullet$ ensemble of Gabor DCNN base models \linebreak
				$\bullet$ combine FR outputs of individual Gabor DCNN members.  & $\bullet$ FERET \newline $\bullet$ CAS-PEAL-R1 \newline $\bullet$ LFW & 
				$\bullet$  outperforms all hand-crafted FR approaches by a large margin. \linebreak
				$\bullet$ 93.6\% on close- \& 77.1\% on open-set identification protocols; outperforms other SOTA methods, except for DeepID2+
				
				\\
				Deng et al. (2019) Additive Angular Margin Loss for Deep Face Recognition & 
				$\bullet$ solves Softmax’s problem of discriminating in face recognition model to high degree for open-set classification \newline $\bullet$ adds an additive angular margin to the angle between features and target weight to maximize face class separability. &  $\bullet$ Megaface\newline $\bullet$ IJB-B \newline $\bullet$ IJB-C \newline $\bullet$ Trillion-Paris \newline $\bullet$ iQIYI-VID & 
				$\bullet$ surpasses FaceNet; comparable results on ID\&better results on verification (Ver) than CosFace under large protocol\linebreak
				$\bullet$ ID result of 84.840\% (@FPR=1e-3) and comparable verification performance to the most recent submission(CIGIT IRSEC) from the lead-board\\
				Liu et al. (2018) SphereFace: Deep Hypersphere Embedding for Face Recognition & 
				$\bullet$ addresses deep face recognition (FR) problem under open-set protocol \linebreak
				$\bullet$ proposes angular softmax(A-Softmax) loss that enables CNNs to learn angular discriminative features  & $\bullet$ LFW \newline $\bullet$ YTF &
				$\bullet$ 95\% accuracy \linebreak
				$\bullet$ best performance trained on WebFace in Jan 2018\\
				\hline
			\end{tabular} 
		}
	\end{table}	
	
	\noindent\textbf{Face Recognition from Multi-Pose Image Sequence} \cite{eigen}.\\
	This paper applied projections to Eigenfaces to recognize faces from different poses. Pattern vectors created the trajectory in Eigenspace in which each trajectory corresponded to a unique face. The recognition process was done by comparing each need-to-recognize face with the prototype trajectories calculated during training time with different poses.

	\noindent\textbf{Face Recognition using Hidden Markov Models} \cite{Markov}.\\
	The paper proposed adaptive Hidden Markov Models (HMM) for face recognition in videos. The training data of each face is of similar background (video sequences). The testing sets were composed of less than 500 frames. HMM learned the temporal changes of the training video sequences and applied that to the recognition process. The label is chosen based on the best likelihood score provided by HMM.
	
	\noindent\textbf{Attention-aware Deep Reinforcement Learning for Video Face Recognition} \cite{attentions}. \\
	This paper proposed a method towards attention-aware deep reinforcement learning (ADRL) for face recognition in video. Instead of processing all frames, it rejected ambiguous and/or misleading frames by using a Markov decision process (MDP). This paper combined feature learning, which took an input as a whole video processed by a DCNN model for temporal representations per frame, and attention learning, which was an evaluation network that took both image and feature spaces as inputs and produced valued and relevant frames (attention frames). The algorithm used video pairs as MDP and trained the evaluation network by using reinforcement learning.

	\noindent\textbf{Ensemble of Deep Convolutional Neural Networks With Gabor Face Representations for Face Recognition} \cite{Choi}.\\
	This paper introduced the "Gabor DCNN ensemble" method that combined different Gabor face representations instead of just one fixed Gabor filter as inputs to DCNNs in training and testing. The paper focused on testing different combinations to find different patterns that could go beyond gray-scale or RGB image inputs and to improve the accuracy on different environment scales (different light settings, poses, facial expressions).
	
    \noindent\textbf{SphereFace: Deep Hypersphere Embedding for Face Recognition} \cite{Liu}. \\
	This paper introduced the angular softmax (A-Softmax) loss function with the aim to get a smaller maximal intra-class distance than the minimal inter-class distance by learning angular discriminative features. The paper normalized the classifier weights to 1 to improve the accuracy when dealing with open-set face recognition datasets, which have untrained faces in the testing sets.
	
	\noindent\textbf{Insightface} \cite{Deng}.\\
	The InsightFace system developed the use of 
	Additive Angular Margin Loss for Deep Face Recognition 
	(ArcFace) on a large-scale image database with trillions of pairs 
	and a large-scale video dataset,
	and tested on multiple datasets with different loss function models (ArcFace, Softmax, CosFace, etc.). The authors’ goal was to replace the traditional Softmax because the linear transformation matrix's size increases linearly with the identity counts. Also, Softmax's learning features did not work very well on open-set classification. ArcFace improved the discrimination ability by removing instability during training time. ArcFace took the dot product between the features and the last fully
	connected layer, which was the cosine distance between the normalized features and weight vectors. The angle was calculated using the arc-cosine
	function, and then the additive angular margin value was added to
	the angle, which allowed the target logit to be returned by the cosine function.
	
\begin{figure}[!t]
		\centering
		\includegraphics[width=1\textwidth]{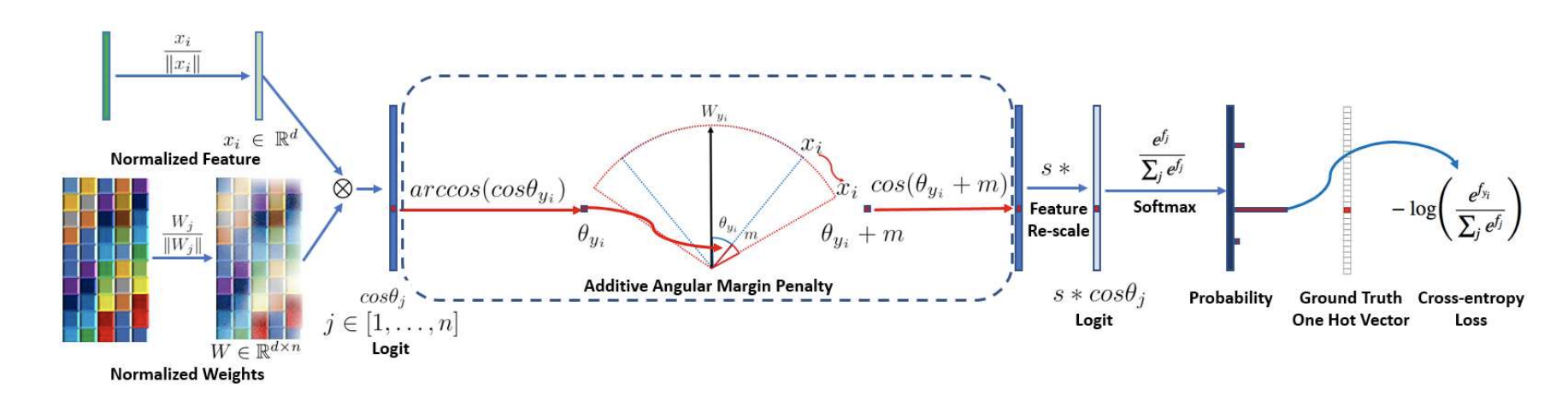}
		\caption{ArcFace loss function \cite{Deng}.} 
		\label{ArcFace}
	\end{figure}

	This system got all the training faces to proper locations and aligned to size 112x112. The system chose one anchor face representation per person (known face). Using feature extractions and Euclidean distances, the paper computed the distances between the detected face with the known face, and the label was taken from the known face with the minimum distance to the detected face. This thesis adopts InsightFace as the baseline face recognition system because of its SOTA performance.
	Figure \ref{ArcFace} shows the ArcFace loss function used in training a DCNN for face recognition.  
	
	\section{ivPCL projects}
	Besides the work on face recognition, as mentioned above, ivPCL introduced multiple fast and dependable methods for video activity and object detection (see Table \ref{tab:ivpcl_prior_research}). The summary covers the development of various methods applied to different problems associated with AOLME video datasets. The research included video activity detection and object detection. The current thesis with the motivation of assisting educational researchers with fast identification provides methods to recognize such students using clustering methods to include multiple poses alongside robust tracking of participants to deal with occlusions. In addition, multi-objective optimization is applied to reduce recognition time and improve accuracy. Lastly, face DeID is integrated into the thesis to ensure the students' privacy.
	
	\newpage
	\begin{table}[]
		\caption[Projects from image and video processing and communications lab ]{Projects from image and video processing and communications lab (ivpcl) that focus on video analysis research involving activity and object detection.}
		\label{tab:ivpcl_prior_research}
		\scalebox{0.79}
		{
			\renewcommand{\arraystretch}{1.5}
			\begin{tabular}{P{2 cm} P{7.5 cm}  P{7.8 cm}}
				\hline
				\textbf{Author} & \textbf{Title} & \textbf{Summary}\\
				\hline
				\hline

				Tapia, L.S.,  et al. (2021) \cite{Luis2021}  & Bilingual Speech Recognition by Estimating Speaker Geometry from Video Data & Applying interactive video analysis system to estimate the 3D speaker geometry for realistic audio simulations. The paper attempted to create simulated audio  dataset with complicated background noise that is similar to real-life classroom recordings.  \\
				
				Shi, W., et al. (2021) \cite{Shi2021}	& Talking Detection in Collaborative Learning Environments & The paper presented a new method to detect talking combined with head detection in collaborative learning environment videos. The method used a projection of motion vectors and majority voting classification system.\\
				
				Tran, P., et al. (2021) \cite{Tran2021} & Facial Recognition in Collaborative Learning Videos & The paper presented fast approach to face recognition in collaborative learning environment. The method showed improvement on dealing with multi-poses and occlusions in addition to using face prototypes with K-means and Sparse Sampling to boost accuracy and reduce recognition time. \\ 
				
				Teeparthi, S., et al. (2021) \cite{Sravani2021} & Fast Hand Detection in Collaborative Learning Environments &  A new method was applied for detecting hands
				in AOLME videos. \\
				
				Shi, W., et al. (2021) \cite{Shi2021b} & Person Detection in Collaborative Group Learning Environments Using Multiple Representations & The paper introduced problem with detecting groups of students from classroom videos. The problem requires students' detection with different angles and groups separation in long videos (one to two hours each). The paper proposed an approach of using multiple image representations with AM-FM components to solve the problem. \\
				
				Jatla, V., et al. (2021) \cite{VJ2021} & Long-term Human Video Activity Quantification of Student Participation & Proposed long video testing on student participation in collaborative learning environments using the small training data. The methods matched the accuracy with SOTA methods with 1500x fewer parameters.
				\\
			\end{tabular} 
		}
	\end{table}	     
	\begin{table}[hbt]
		\scalebox{0.79}
		{
			\renewcommand{\arraystretch}{1.5}
			\begin{tabular}{P{2 cm} P{7.5 cm}  P{7.8 cm}} 
				
				Ulloa, A., et, al. 2021 \cite{Alvaro} & Deep-learning-assisted analysis of echocardiographic videos improves predictions of all-cause mortality & Applied machine learning to aid doctors and cardiologist with mortality predictions in near future based on health records and historical data. The paper improved the sensitivity by 13\% while keeping specificity unchanged for one-year predictions. \\

				Jatla, V., et, al. 2020 \cite{VJ2020} & Image processing methods for coronal hole segmentation, matching, and map classification & Introduced methods to choose optimal physical models to automatically detect coronal holes from the image subject to solar image observations. \\
				
				Tapia, L., et, al. 2020 \cite{Luis2019} & The Importance of the Instantaneous Phase for Face Detection using Simple Convolutional Neural Networks & The authors demonstrated that FM images with low-complexity neural networks 
				can provide face detection results that
				can only be achieved with much more complex deep learning systems.\\

				Shi, W., et, al. 2018 \cite{shi2018dynamic} & Dynamic group interactions in collaborative learning videos. & The authors developed methods
				to detect where participants
				were looking at, and also classified different types of interactions.  \\
				
				Shi, W., et, al. 2018 \cite{Shi2018} &  Robust head detection in collaborative learning environments using am-fm representations. & Using AM-FM decomposition, the authors introduced methods for face detection including front face and back-of-the-heads. \\
				
				Jacoby, A., et, al. 2018 \cite{Abby2018} & Context-Sensitive Human Activity Classification in Collaborative Learning Environments & the authors introduced methods for detecting
				writing, typing, and talking activities using motion vectors and deep learning. \\
				\bottomrule
			\end{tabular} 
		}
	\end{table}
		
	\section{Uniqueness of AOLME Dataset} 
	AOLME is different from other typical datasets in the way that it primarily includes long videos which are over an hour. Each video has multiple activities but is not limited to typing, talking, eating, and writing. Many challenges impact the results like occlusion, multiple camera angles, illumination issues, multiple people performing the same activity, fast and random movements, people moving across the videos, and activities in the background. In addition, the AOLME dataset is different from standard datasets which mostly consist of celebrities. These datasets' images focus on the celebrities and blur out the backgrounds. The AOLME's video frames contain both the students in the collaborative group (in front of the camera) and the background, non-analyzing groups. Because of the fact that the AOLME videos include faces that do not appear in the training set, it is considered an open-set dataset.

	\chapter{Dataset}
	
	The AOLME project stands for the Advancing Out-of-school Learning in Mathematics and Engineering. It is a collaboration between the Electrical and Computer Engineering Department and the Department of Language, Literacy, and Sociocultural Studies. AOLME implemented an after-school program that used an integrated curriculum of mathematics and computer programming to support middle school students to learn to code in Python using the Raspberry Pi. To assess the teaching and learning of the curriculum, AOLME collected video data of the students and their facilitators while learning the material in the curriculum.
	AOLME generated a large amount of data, including group interactions, monitor and screen recordings that make up to more than 2,000 hours. This thesis focuses only on group interaction data that total about 1,000 hours. The project lasted two years, from 2017 to 2019, in two different middle schools that the thesis will refer to as Urban and Rural middle schools. There is one Cohort per academic year and two to three Levels per Cohort. Each video has $1920\times1080$ resolution at $30$ or $60$ frames per second. This section summarizes how to select the training and testing data, perform the ground truth, and generate the entire AOLME recognition dataset.

	\section{AOLME Dataset Overview}
	
	The AOLME dataset consists of three Cohorts, each with one to three Levels, where each level consists of different sessions. Overall, AOLME had approximately 50 separate groups of students from two middle schools. The three Cohorts, implemented in three years from 2017-2019, are denoted by CxLy format, representing Cohort x Level y with x and y ranging from one to three. The training and testing data are entirely separate and come from different sessions, even different Levels, and Cohorts.
	
	\begin{figure}[!t]
		\centering
		\includegraphics        [width=1\textwidth]{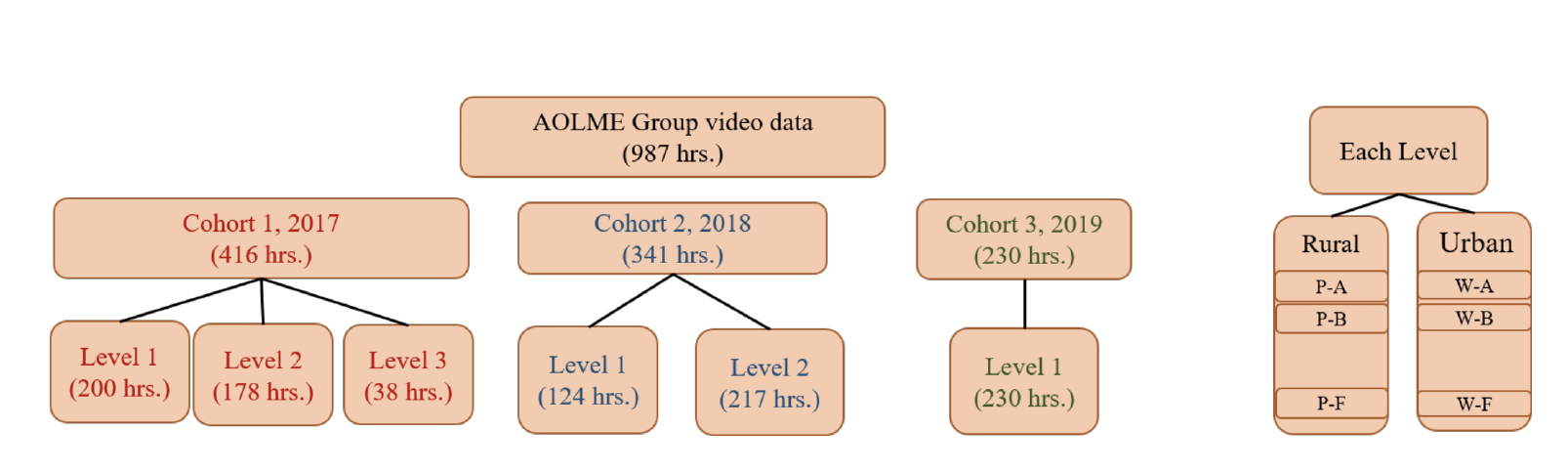}
		\caption{AOLME Hierarchy \cite{sravaniThesis}. }
		\label{aolmeDataset}
	\end{figure}

  	\section{AOLME Student Dataset}
	To ensure the students' privacy, researchers from the College of Education and Human Science and I worked together to develop a way to rename each instead of using their real names. The pseudo-names were created based on the format of someName+uniqueID+schoolInitial (W or P). A dataset of 138 students was generated, excluding Cohort 1 Level 3.  The following sections provide these students' faces (covered) and their pseudo-names in the AOLME program. These sections include trained students for this thesis, ready-to-train students (with available ground truth), and unprocessed students.
	
	\subsection{AOLME32}
	This subsection includes 32 trained and tested students. Figures \ref{fig:trainedDatabase} and \ref{fig:trainedCont} provide individuals’ pseudo-names with their identities protected. Students are split based on their group in each school that they participated. As the program encouraged students who joined a previous Cohort and Levels to come back and become co-facilitators (Co-fac), there exist students who appear in multiple groups.

	Each video session has four to nine videos with a duration of either twenty-three or sixteen minutes; the thesis randomly sampled hundreds of five to ten second long clips from these session videos as training data. Multiple students repeatedly appear throughout the years. For example, in Figures \ref{fig:trainedDatabase} and \ref{fig:trainedCont}, notice that Alvaro70P and Jesus69P appeared in Cohort 2 Level 1 (C2L1) as students and Cohort 3 Level 1 (C3L1) as co-facilitators. Figure \ref{aolmeDataset} \cite{sravaniThesis} demonstrates the hierarchy of the AOLME program between 2017 and 2019.
	
	As the thesis tried out multiple approaches for facial recognition, there are multiple trained datasets for each. Table \ref{aolme32} shows the dataset used for each approach.
      
    \begin{table}[]
		\caption{Datasets for K-means Clustering, Sparse Sampling, and Data Augmentation.}
		\label{aolme32}
		\begin{tabular}{cl}
			\hline
			Approach & \multicolumn{1}{c}{Group} \\ \hline
			\multirow{5}{*}{K-means Clustering}  & G-C1L1W-C-Kushal          \\
			& G-C2L1P-D-Chaitu          \\
			& G-C3L1P-E-Joaquin         \\
			& G-C3L1-A-Ankit            \\
			& G-C1L1P-F-Carlos     \\\hline
			\multirow{7}{*}{Sparse Sampling}                       & G-C1L1P-C-Kelly          \\ & G-C2L1P-D-Chaitu          \\
			& G-C2L1P-E-Krithika     \\ & G-C3L1W-D-Phuong          \\
			& G-C3L1P-C-Phuong       \\ & G-C3L1P-D-Ivonne          \\
			& G-C3L1P-E-Joaquin         \\ \hline
			\multicolumn{1}{l}{\multirow{3}{*}{Data Augmentation}} & G-C2L1P-E-Krithika        \\
			\multicolumn{1}{l}{} & G-C3L1P-D-Ivonne          \\
			\multicolumn{1}{l}{} & G-C3L1W-D-Phuong     \\
			\hline
		\end{tabular}
	\end{table}

	\subsection{K-means Clustering Dataset}
	The dataset to train and test for this approach has 18 participants (16 males and two females), which come from the AOLME32.
	
	\subsection{Sparse Clustering Dataset}
	The dataset to train and test for this approach has 24 participants, which come from the
	AOLME32. These 24 consist of 11 males and 13 females, of which 20 are between ten and fourteen.
	
	\subsection{Data Augmentation Dataset}
	The dataset for this approach has 11 participants, a subset of AOLME32. These students are hand-picked from the Sparse Clustering Dataset because the algorithm failed to recognize them or wrongly labeled them as others within this dataset. The dataset comprises three males and eight females.
	
	\subsection{AOLME41}
	This subsection contains the students with accessible ground truth; yet, these students have not been used for training and tested. The ground truth video clips come from multiple sessions throughout different Cohorts. These students and the students from AOLME32 are from the dataset described in Table \ref{proposedTest}, which are chosen based on the priority for analysis purposes from educational researchers. AOLME41 is demonstrated in Figures \ref{wait} and \ref{waitCont} in the Appendix.
	
	\subsection{AOLME83}
	
	This subsection contains the rest of the students that are yet to be processed. These students are not yet trained/tested and do not have available ground truth. AOLME83 is given in Figures \ref{noKids},  \ref{noCont} and \ref{noCont2} in the Appendix.
	
	\section{Video Face Recognition Dataset}
	\subsection{Ground Truth for Video Face Recognition}
	\subsubsection{Ground Truth Process}
	The thesis used Matlab built-in tool Video Labeler to generate Ground Truth. The sessions to generate Ground Truth were chosen based on faces appearing in the proposed test dataset (Figure \ref{proposedTest}). Thus, the entire ground truth contains 38 different sessions and 695 short clips for training and 13 sessions. Each video clip is 16-23 minutes and is used for testing and is chosen from 73 faces from the AOLME32 and AOLME 41. Each face contains an average of 9,500 images, which makes the total training data images for these 73 faces almost 700,000 images. The process to generate ground truth for training and testing are slightly different.
	
	For training data Ground Truth, this thesis chose multiple dates for faces' variability.  Each clip segment lasts around five to ten seconds, and these clip segments are chosen randomly throughout each session, where multiple faces appear in each frame.
	
	After the segments are selected, using the Video Labeler tool in Matlab, spatial bounding boxes were put manually around each face in the collaborative group with their corresponding label from the naming database. This thesis then tracked the face throughout the clips with the aid of the Video Labeler. The outputs are .mat files containing the time of each frame along with the faces' locations and labels appearing in said frame. 
	\begin{figure}[!t]
		\centering
		\includegraphics[width=1\textwidth]{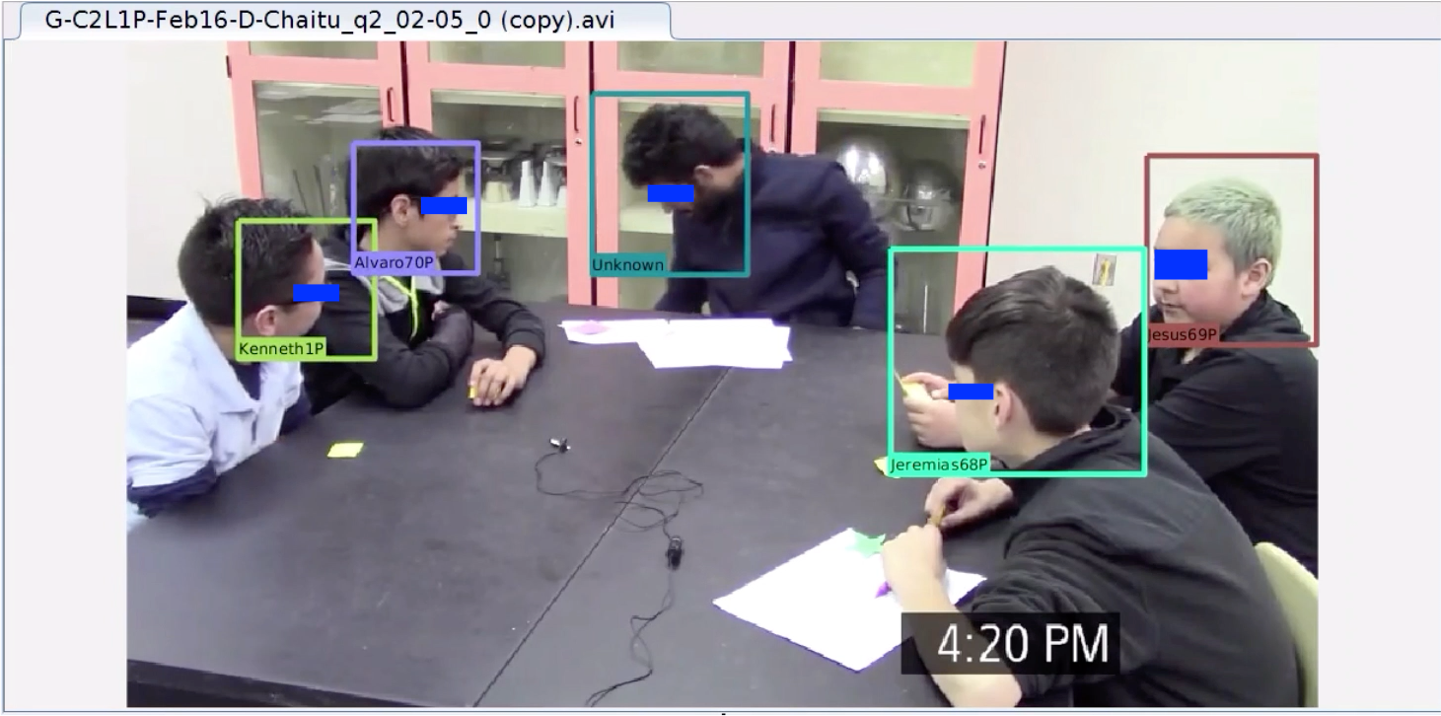}
		\caption{Ground Truth Example.} 
		\label{fig:Ground Truth Example}
	\end{figure}
	
	\begin{figure}[!b]
		\centering
		\includegraphics[width=1\textwidth]{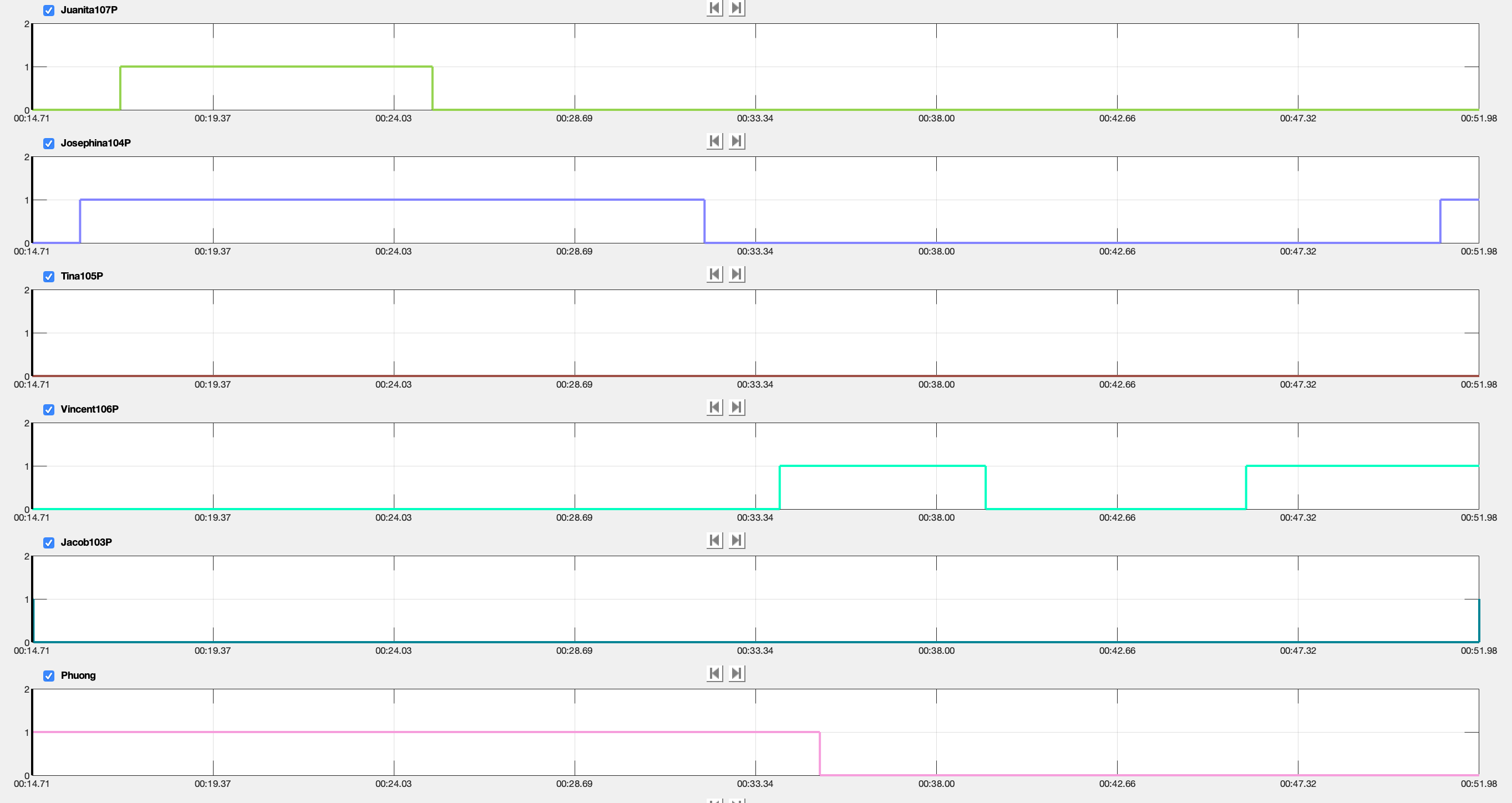}
		\caption[An example of ground truth visualization in Matlab.]{An example of ground truth visualization in Matlab. The plot shows the time-stamps of when the students were visible, including partially occluded. If the students were completely occluded, they were marked as absent.} 
		\label{fig:gtVidLabel}
	\end{figure}
	After getting the locations and labels of each face, the thesis loaded the .mat files to a Matlab script to extract frames and get the short clips of each individual for each clip segment. For example, if a clip is ten seconds long and there are four people in the clips, the outputs for training data would be four ten-second long clips where each clip belongs to each person in such clips.
	
\begin{figure}[]
		\centering
		\includegraphics[width=1\textwidth]{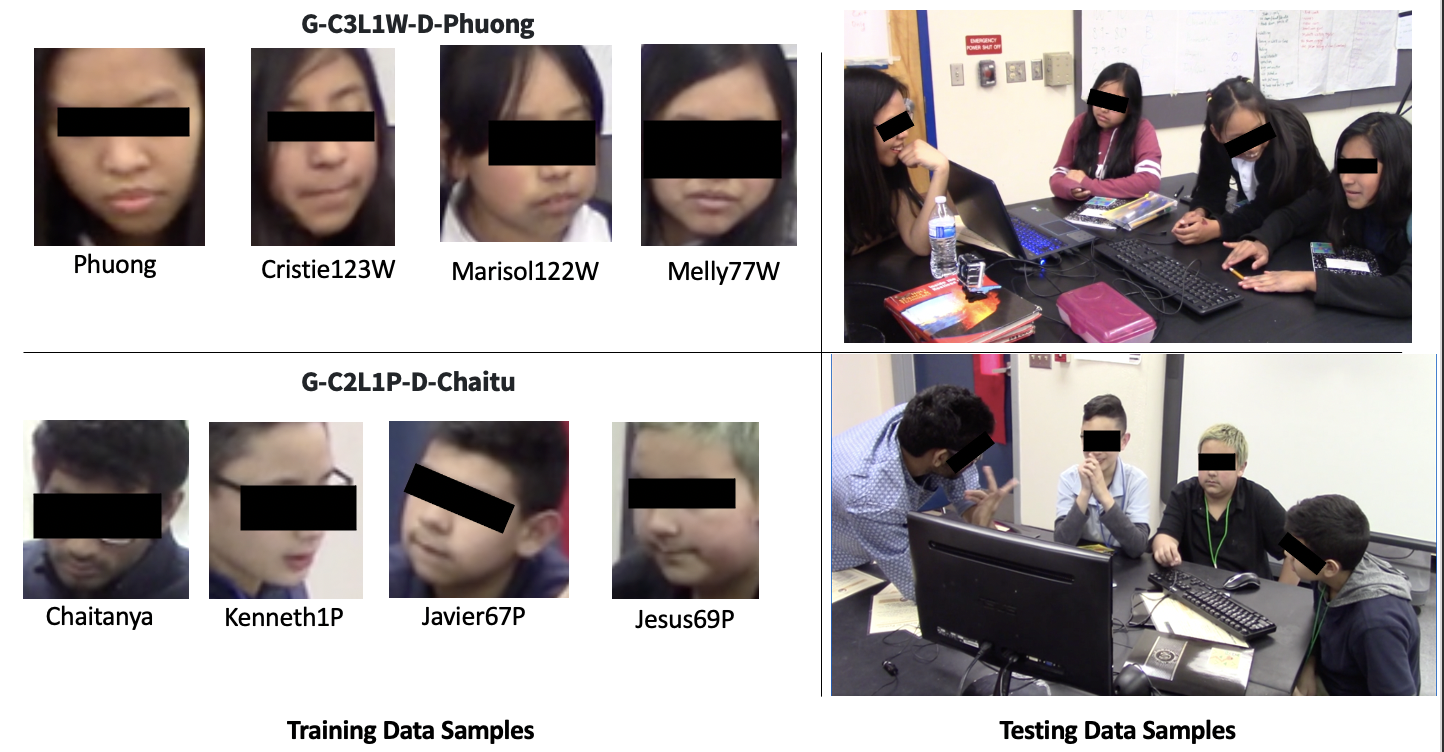}
		\caption{Training and Testing Samples.} 
		\label{testSamples}
	\end{figure}

	The videos are sixteen to twenty-three minutes long for testing ground truth as each session has four to nine videos. The thesis used the Matlab Video Labeler tool for this process. Instead of dividing each clip based on each face, the thesis divided each clip based on frames. After getting the .mat files, the thesis loaded them, got the students' locations and labels appearing in each frame, and saved them as text files. Thus, if '0.txt' is loaded, it will provide all the faces' labels and locations that belong to frame 0. Figure \ref{testSamples} shows examples of train and test video samples.

    To easily understand the Ground Truth process, Figures \ref{fig:Ground Truth Example} and \ref{fig:gtVidLabel} are shown for visualization purposes. Figure \ref{fig:Ground Truth Example} demonstrates how to track faces' locations. Each face is bounded by a rectangular box with the label shown on top (or on the left-hand side). Figure \ref{fig:gtVidLabel} shows starting and ending times of a video for all the individuals in the video. If a student is in the frame, it will show a bump at that time. Otherwise, a flat line implies that the student moved out of the frame or is fully occluded.
	
	As this project aims to assist educational researchers with data analysis for the AOLME project, the thesis does not want to provide information that could reveal the identity of the students. Thus, to protect the students' identity and privacy, the thesis came up with pseudo-names for identification and analysis instead of real names. The naming convention was an agreement between the School of Engineering and the College of Education and Human Sciences in which the syntax is "Pseudo-name + Unique ID + School\_Initial (i.e., Kenneth1P). 
	\begin{table}[!h]
		\caption[Complete Training ground truth for faces from AOLME data.]{Complete Training ground truth for faces from AOLME data. Each clip is five to ten seconds at 30 or 60 fps. The dates in bold are the ones that are in AOLME32.}
		\label{TrainingSet}
		\scalebox{0.61}
		{
			\begin{tabular}{l l l l l l l}
				\multicolumn{7}{c}{\textbf{Cohort 1 Level 1}}\\
				\multicolumn{7}{c}{\textbf{Rural}}\\
				Group & Dates  & \# Faces & \# Clips & Total Videos Proc. & Total Time Proc.(sec) & Total Hours \\
				C & \textbf{Feb16, Feb25}        & 7  & 27  & 8 & 170 & 4.3 \\
				B & Feb25, Mar30        & 4  & 60  & 8 & 504 & 4.4  \\
				E & Feb25, Mar09, Apr06 & 11 & 57  & 5 & 438 & 6.1  \\
				F & Mar09               & 4  & 13  & 1 & 114 & 1.7 \\
				A & Feb25               & 5  & 17  & 3 & 101 & 2.8\\
				
				\multicolumn{7}{c}{\textbf{Urban}}\\
				Group & Dates & \# Faces & \# Clips & Total Videos Proc. & Total Time Proc. (sec) & Total Hours \\
				C   & \textbf{Feb21, May02} & 5 & 27 & 2 & 232 &  2.8 \\
				~\\
				\hline
				\textbf{Total}          &    &36 & 191 & 27 & 1,327 & 21.1\\
				\hline
				\\
				\\
				\multicolumn{7}{c}{\textbf{Cohort 2 Level 1}}\\
				\multicolumn{7}{c}{\textbf{Rural}}\\
				Group & Dates  & \# Faces & \# Clips & Total Videos Proc. & Total Time Proc.(sec) & Total Hours \\
				D & \textbf{Feb23, Mar22, Apr12, Apr19}  & 6  & 78  & 8  & 656 & 7.3 \\
				B & Feb16, Mar08, Apr05         & 12 & 42  & 11 & 321 & 4.9 \\
				C & Feb23, Apr05, Apr26         & 9  & 68  & 7  & 534 & 5.4 \\
				E & \textbf{Feb16, Feb23}                & 9  & 52  & 6  & 444 & 3.0 \\
				
				\multicolumn{7}{c}{\textbf{Urban}}\\
				Group & Dates & \# Faces & \# Clips & Total Videos Proc. & Total Time Proc. (sec) & Total Hours \\
				B   & Mar20, Apr03, May15 & 13 & 70 & 8 & 545  &  4.2 \\
				~\\
				\hline
				\textbf{Total}          &    & 49 & 310 & 40 & 2,500 & 24.8\\
				\hline
				\\
				\\
				\multicolumn{7}{c}{\textbf{Cohort 3 Level 1}}\\
				\multicolumn{7}{c}{\textbf{Rural}}\\
				Group & Dates  & \# Faces & \# Clips & Total Videos Proc. & Total Time Proc.(sec) & Total Hours \\
				D &  \textbf{Feb14, Feb28, Mar21, Apr25} & 7 & 39  & 4  & 327 & 6.1 \\
				C &  \textbf{Feb14, Mar21, Mar28}        & 8 & 52  & 3  & 430 & 4.6 \\
				E &  \textbf{Feb14, Mar21}               & 5 & 34  & 2  & 279 & 2.9 \\
				
				\multicolumn{7}{c}{\textbf{Urban}}\\
				Group & Dates & \# Faces & \# Clips & Total Videos Proc. & Total Time Proc. (sec) & Total Hours \\
				D   & \textbf{Feb12, Feb26, Mar05} & 9 & 49 & 4 & 362 & 3.5  \\
				~\\
				\hline
				\textbf{Total}          &    & 29 & 174 & 13 & 1,388 & 17.1\\
				\hline
			\end{tabular}
		}
	\end{table}

	\subsection{Training set}
	The training dataset came from different sessions and even Cohorts and Levels (if the students returned). A summary of the training set broken down by levels is in Table \ref{TrainingSet}.
	
	\subsection{Testing set}
	The testing data selection comes from educational researchers and prior choices. The mutually exclusive test dataset includes 13 video clips from seven different groups from both Urban and Rural middle schools. 
	
	\begin{table}[]
		\caption[Proposed test dataset that was agreed upon with the UNM  College of Education and Human Sciences.]{\label{proposedTest} Proposed test dataset that was agreed upon with the UNM College of Education and Human Sciences. Data was collected from 13 different groups from both Urban and Rural middle schools. Time range varies from C1L1 (2017) to C3L1 (2019). The sessions are chosen based on prioritized analysis purposes.}  
		\centering
		\begin{tabular}{|c|c|c|l|c|}
			\hline
			{\bfseries GroupID} & {\bfseries Cohort} & {\bfseries Group} & \multicolumn{1}{c}{\bfseries{Date}}         & {\bfseries Urban/Rural}  \\ \hline
			1&	C1L1   & B     & Mar02     & Rural       \\
			2&	C1L1   & C     & Mar30     & Rural       \\
			3&	C1L1   & C     & Apr06     & Rural       \\
			4&	C1L1   & C     & Apr13     & Rural       \\
			5&	C1L1   & E     & Mar02     & Rural       \\
			6&	C2L1   & B     & Feb23     & Urban      \\
			7&	C2L1   & C     & Apr12     & Rural \\ 
			8&	C2L1   & D     & Mar08     & Rural \\
			9&	C2L1   & E     & Apr12     & Rural \\
			10&	C2L1   & B     & Feb27     & Urban \\
			11&	C3L1   & C     & Apr11     & Rural \\
			12&	C3L1   & D     & Feb21     & Rural \\
			13&	C3L1   & D     & Mar19     & Urban \\\hline
		\end{tabular}
	\end{table}

	\begin{table}[]
		\caption[Test dataset used for face recognition.]{\label{TestSet} Test dataset used for face recognition. Data collected from 7 different groups from both Urban and Rural middle schools. Time range varies from C1L1 (2017) to C3L1 (2019). If a test video clip comes from the same group session, the thesis differentiates with a suffix A,B,C (i.e: 2A, 2B,..). Even if the test clips are from same sessions, they are at different times. }  
		\centering
		\begin{tabular}{|c|c|c|l|c|}
			\hline
			{\bfseries GroupID} & {\bfseries Cohort} & {\bfseries Group} & \multicolumn{1}{c}{\bfseries{Date}}          & {\bfseries Urban/Rural}  \\ \hline
			1&	C1L1   & C     & May9      & Urban       \\
			2&	C2L1   & D     & Mar08     & Rural       \\
			3&	C3L1   & D     & Mar19     & Urban   \\
			4&	C3L1   & E     & Mar28     & Rural       \\
			5&	C2L1   & E     & Apr12     & Rural        \\
			6&	C3L1   & D     & Feb21     & Rural         \\
			7&	C3L1   & C     & Apr11     & Rural \\ \hline
			
		\end{tabular}
	\end{table}
	
	\chapter{Methodology}
	This chapter describes the methodology for face recognition. The chapter describes a fast method for video-based face recognition, face DeID, data augmentation, and clustering methods for minimizing the number of required face prototypes.
	\section{Video Faces Recognizer}\label{sec:videorec}
\begin{figure}[!t]
		\centering
		\includegraphics[width=.7\textwidth]{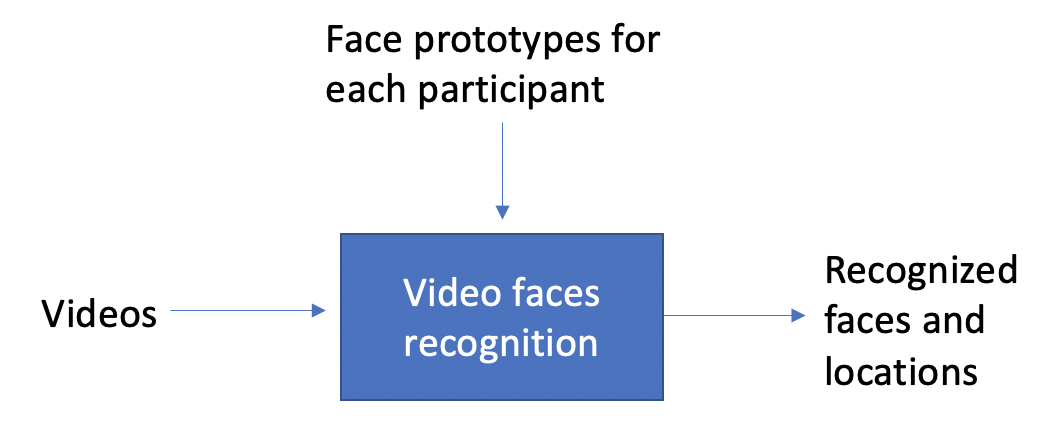}
		\begin{algorithm}[H]
			\SetAlgoLined
			\textbf{Input:}\\
			$\quad$ video clips associated with each participant.\newline
			\textbf{Output:}\\
			$\quad$ facePrototypes associated with each participant.\newline
			\textbf{for} \text{each participant} \\
			\quad$\textbf{Apply} \text{ K-means Clustering}$ \\
			\quad$\textbf{Select} \text{ cluster means}  $ \\
			\quad$\textbf{Find} \text{ nearest images from cluster centroids}$\\
			\quad$\textbf{Align} \text{ faces to 112x112}$\\
			\textbf{end}
			\label{k-means}
			\caption{Computing Face prototypes using K-means.}
		\end{algorithm}
		\caption[Block diagram for recognizing faces from videos.]{
			Block diagram for recognizing faces from videos.
			Each face is associated with a collection of face prototypes.
			Face prototypes are computed using video sampling, K-means Clustering, or both.
			K-means Clustering is described here.}
		\label{fig:systemDiag}
	\end{figure}

	Figure \ref{fig:systemDiag} presents a block diagram of the overall system. This thesis adopts InsightFace as the baseline face recognition system because of its SOTA performance. The video recognition system requires a set of face prototypes associated with each participant. The video face recognition algorithm detects the faces in the input video and computes minimum distances to the face prototypes to identify each participant. The system uses past detection history to handle occlusion issues.
	
	Table \ref{Methods} gives a summary of the video face recognition methods. First, the thesis uses data augmentation to increase the training data size. Then, the thesis uses either Sparse Sampling, K-means Clustering, or combining both to extract an optimal number of face prototypes. After processing each video frame, the thesis uses frame skipping to speed up processing further.   
	
	\begin{table}[]
		\caption{\label{Methods}
			Video face recognition methods.}
		\begin{center}
			\centering
			\begin{tabular}{p{0.24\textwidth}p{0.20\textwidth}p{0.13\textwidth}p{0.13\textwidth}p{0.16\textwidth}}\toprule
				{\begin{tabular}[c]{@{}l@{}}\textbf{Method}\\ \textbf{}\end{tabular}}    &
				{\begin{tabular}[c]{@{}l@{}}\textbf{Augmentation}\\ \textbf{}\end{tabular}} &
				{\begin{tabular}[c]{@{}l@{}}\textbf{Sparse}\\ \textbf{Sampling}\end{tabular}} &
				{\begin{tabular}[c]{@{}l@{}}\textbf{Frames}\\ \textbf{ Skipping}\end{tabular}} &
				{\begin{tabular}[c]{@{}l@{}}\textbf{K-means}\\ \textbf{ Clustering}\end{tabular}}    \\
				\toprule
				{\begin{tabular}[c]{@{}l@{}} Baseline  \\ \textbf{}\end{tabular}}         & x  & x  & x      & x \\~\\
				
				{\begin{tabular}[c]{@{}l@{}}Proposed Method 1 \\ (Face Prototypes \\ with K-means)\end{tabular}}         & x & x  & x      & \checkmark  \\~\\
				
				{\begin{tabular}[c]{@{}l@{}}Proposed Method 2 \\ (Face Prototypes \\ with Sampling)\end{tabular}}     & x    & \checkmark    & x      & x \\~\\
				
				{\begin{tabular}[c]{@{}l@{}}Proposed Method 3  \\ (Augmented)\end{tabular}}     & \checkmark    & \checkmark    & x    & x \\~\\
				
				{\begin{tabular}[c]{@{}l@{}}Proposed Method 4  \\ (Fast) \end{tabular}}     & \checkmark    & \checkmark    & \checkmark      & \checkmark \\
				\\\bottomrule
			\end{tabular}
		\end{center}
	\end{table}	
	
		\subsection{Computation of Face Prototypes}\label{sec:clusters}
	
\begin{figure}[!t]
		\centering
		\includegraphics[width=1\textwidth]{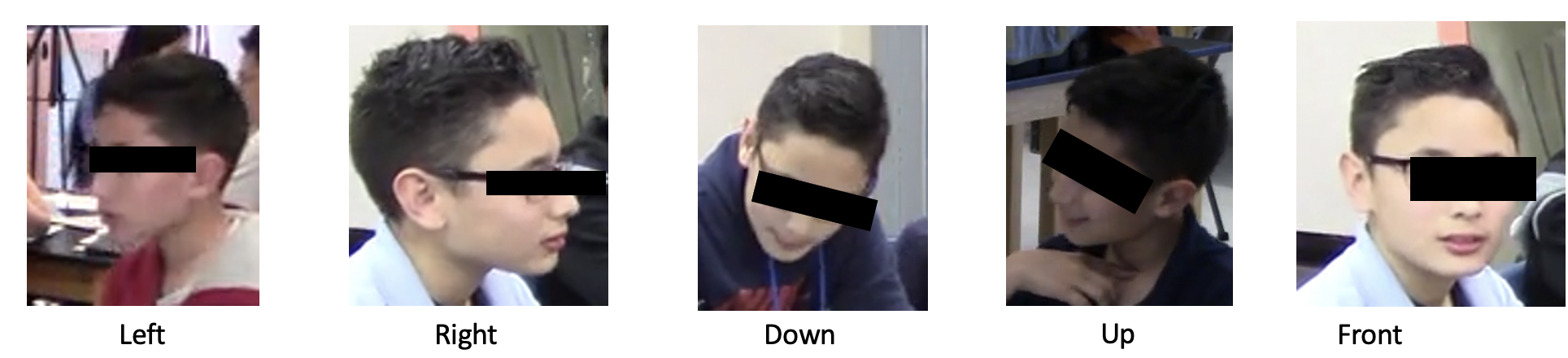}
		\caption{Face pose variability in the AOLME dataset.} 
		\label{fig:angleVar}
	\end{figure}

	The baseline method chose one image as an anchor (known) image to decide the label for the detected face using minimum distance. However, using only one image per face did not work at all on the AOLME dataset. The reason is that the variability in poses of each face in the videos when they move around. Following that, the thesis sets the entire dataset as known images. However, with the total training dataset images of almost 700,000, the time taken to recognize each face was significantly long. In addition, there are many near-identical images because if an individual does not move much during a video, the frame image will not change. Thus, the thesis comes up with an approach to compute a set of face prototypes for each participant, a subset of the training data to represent these known images. This approach reduces the number of images needed for comparison to get face labels and emphasizes the pose variability in the AOLME dataset. Figure \ref{fig:angleVar} illustrates the varied poses of a participant. The thesis uses two methods to compute the face prototypes: K-means Clustering and Sparse Sampling for training videos. Each sampled face is aligned and resized to $112\times 112$.
	
	\noindent\textbf{K-means Clustering:} \\
		
\begin{figure}[]
		\centering
		\includegraphics[width=1\textwidth]{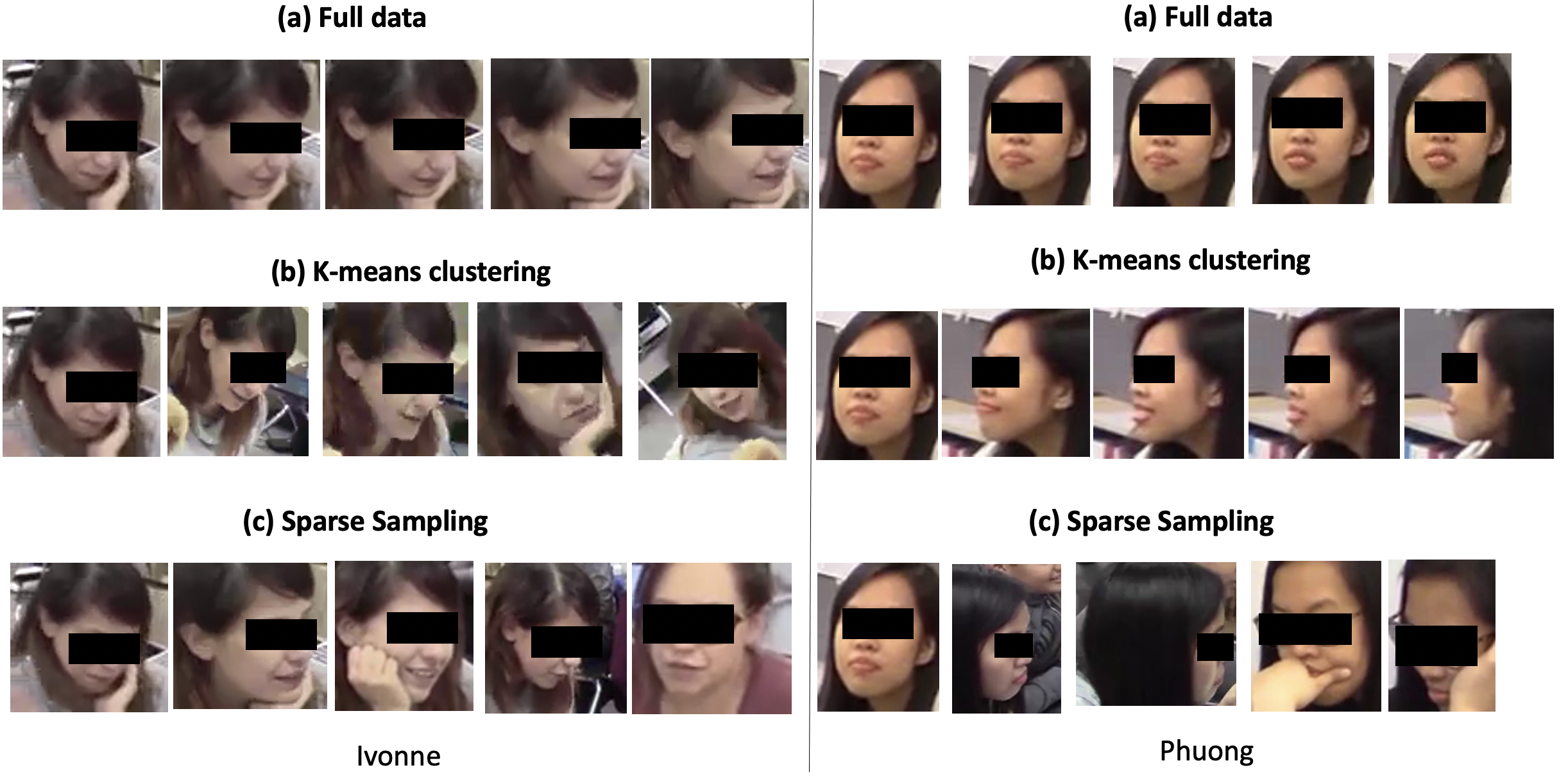}~\\[0.05 true in]
		\includegraphics[width=1\textwidth]{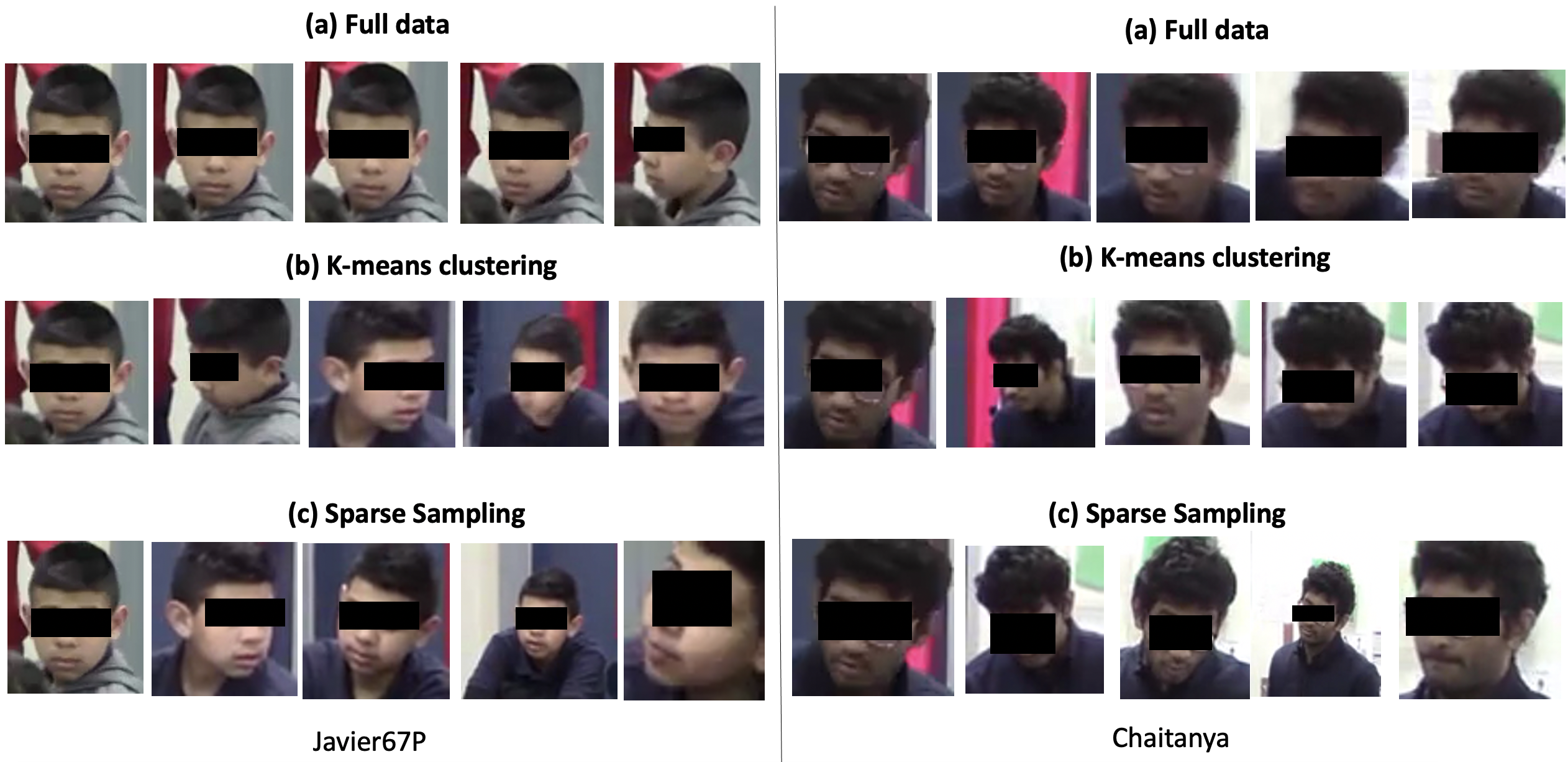}~\\
		\caption{Full data vs K-means Clustering vs Sparse Sampling samples.}
		\label{prototypes}
	\end{figure}

		The use of a single prototype per face did not work very well. The thesis used the entire training dataset for generating a collection of prototypes associated with image students to address the need to account for multiple poses. This baseline approach did not work very well as the number of faces for each participant was around 9,500; doing it this way would take very long to reach a number that can cover all the different poses from different sessions and videos. Therefore, this thesis introduced a way to compute the face prototypes using K-means Clustering. The thesis summarizes the K-means approach in Algorithm \ref{k-means}. For this approach, the algorithm was tested on Group C from Urban C1L1 (Video 1). This thesis uses K-means to cluster similar frames that appear when a student does not move very much. Hence, the thesis expects that the centroids that result from the K-means algorithms would represent a small number of diverse face poses for each participant. The algorithm computes the face prototypes using the training images closest to the cluster centroids to avoid unrealistic centroid images. After finding a prototype image closest to the mean, the algorithm aligns and resizes each image to $112\times 112$. Figure \ref{prototypes} provides examples of the first five multiples of ten samples (frame 0, 10, 20, 30, 40) in each dataset for Marisol122W, Phuong, Javier67P, and Chaitanya with and without using K-means Clustering. The example used K-means with 512 clusters applied to the training data. The thesis used multiples of tens as the consecutive frames for the entire dataset that is nearly identical. The frames with K-means are a lot more diverse than the ones in the entire dataset.
		
		The thesis uses a multi-objective optimization approach where the thesis jointly optimizes the number of clusters with the achieved accuracy to determine the optimal number of clusters.  For K-means Clustering, the thesis selects the combination that provides the best accuracy with the highest number of required clusters that do not yield a drop in performance. The Results chapter describes the optimization of the number of clusters. Initially, setting the number of clusters using multiples of five, i.e., 5, 10, ..., 100. Later, I considered a logarithmic search using base 2. The thesis uses a log-based search by varying the number of face prototypes from 2$^0$ to 2$^{11}$. This thesis also considers a finer logarithmic search by increasing the power by 0.25 instead of 1 to make the analysis more thorough. For example, the thesis does the analysis on  $2^{3}$, $2^{3.25}$, $2^{3.5}$, $2^{3.75}$ and $2^{4}$ instead of just $2^{3}$ and $2^{4}$.

	\noindent\textbf{Sparse Sampling:}\\
		The thesis considers a second baseline approach where Sparse Sampling is used from the training videos to define the face prototypes. The algorithm uses one sample image per second of video to achieve sparsity. Figure \ref{prototypes} provides examples of the first five samples for Marisol122W, Phuong, Javier67P, and Chaitanya with and without using Sparse Sampling. For the second approach, the thesis uses a video from Group D from Urban Cohort 2 Level 1 (C2L1, Video 4) and Group E from C3L1 (Video 5).
	
	\subsection{Video faces Recognition}
	The thesis presents the algorithm for video face recognition in Algorithm \ref{videoProcessing}. The input is an unlabeled video and the {\tt facePrototypes} that provides a list of images associated with each participant.  {\tt vidResult} contains the detected faces for each video frame. To address occlusion, the algorithm maintains the face recognition history in {\tt ActiveSet} and {\tt InactiveSet}.
	
	First, for the first two seconds of the videos, the algorithm detects all participants in each video frame using MTCNN \cite{MTCNN}. MTCNN computes five landmark points for each detected face: two for the eyes, one for the nose, and two for the mouth corners. The face detector uses a minimum area requirement to reject faces that belong to another group. Thus, the algorithm rejects more minor face detections that are either part of another group or are false positives because they appear smaller in the video. Second, each detected face is classified by selecting the participant that gives the minimum distance to their associated prototypes stored in {\tt facePrototypes}. 
	
	The algorithm uses the initial face recognition results to initialize {\tt ActiveSet} and {\tt InactiveSet}. {\tt ActiveSet} stores recognized faces that appear in more than half of the frames.  {\tt InactiveSet} stores the rest. For each face detection, the algorithm uses a dictionary to store: a pseudonym, location information, {\tt totAppearances} that stores the total number of frames where this face was detected, and {\tt totFrames\-Processed}, which represents the total number of frames processed since this face appeared for the first time. Hence, the algorithm uses the {\tt ActiveSet} to hold faces that appear consistently, whereas {\tt InactiveSet} contains faces that are still in doubt.   
	
	When a recognized face enters the {\tt ActiveSet}, it gets a maximum value of 10 for its corresponding {\tt contAppearances}. When a previously recognized face is missing, {\tt contAppearances} gets decremented by 1. When a face re-appears, {\tt contAppear\-ances} is incremented until it reaches 10 again. The algorithm also sets {\tt minApp\-earances} to 5 as the minimum requirement on the number of prior continuous appearances for addressing occlusion issues. Thus, for each face in the {\tt ActiveSet} that is not being detected in any frame, if  {\tt contAppearances} $\geq$ {\tt minAppearances}, the algorithm declares the face as occluded, the algorithm marks it as present, and updates {\tt vidResult}. Else, if {\tt contAppearances} $<$ {\tt minAppearances}, the algorithm declares the face as disappearing, and move it to the {\tt InactiveSet}. 
	
	The algorithm thus processes the rest of the {\tt video} based on the following four cases:
	
	(i) If a newly detected face corresponds to a minor movement of a prior detection, the algorithm keeps it in the {\tt ActiveSet}. This approach leads to a significant speedup in face recognition speed.
	
	(ii) If a newly detected face is in the {\tt InactiveSet}, the {\tt InactiveSet} is updated with the new detection, and looks at the ratio of {\tt totAppearances}/{\tt totFr- amesProcessed} to determine if it needs to move to the {\tt ActiveSet}. Otherwise, the face stays in {\tt InactiveSet}.
	
	(iii)  If a newly detected face does not belong to either set, then recognize it and move it to the {\tt ActiveSet}.
	
	(iv) If a face that belongs to the {\tt ActiveSet} no longer appears, the algorithm considers the case of occlusion or that of when a participant has left the frame. As described earlier, {\tt contAppearances} is checked to determine whether to declare the face occluded or not.
	
	The algorithm did some post-processing to not allow the same label to two different faces in the same frame. In this case, the algorithm assigns the label to the face that gives the minimum distance while the other(s) are declared Unknown. In addition, there usually does not have significant movement happening within a short amount of time in video processing. Thus, the algorithm assumes the participants do not move much and assign the same labels where the faces are in the previous frames in the {\tt ActiveSet}. The algorithm tested out on skipping no frames, 5, 10, 15, 20, 30, and 60 frames.
	
	\newpage
	\begin{algorithm}[!htp]
		\SetAlgoLined
		\textbf{Input:}\\
		$\quad$ {\tt video}: unlabeled video\\
		$\quad$ {\tt facePrototypes}: list of images associated with each student pseudonym.\\
		\textbf{Output:}\\
		$\quad$ {\tt vidResult}: student unique ID, face locations and landmarks.\\
		\textbf{Local Variables:}\\
		$\quad$  {\tt ActiveSet} and {\tt InactiveSet} store unique student
		identifiers, \\ 
		$\qquad$ face locations, {\tt totAppearances}, {\tt totFramesProcessed}, \\
		$\qquad$ {\tt contAppearances} \& distance.\\
		\textbf{while} \text{frame {\tt f} in initial part of video} \\
		$\quad$\textbf{Detect} faces in {\tt f} $\rhd\text{Detect and Recognize all faces initial duration}$\\
		$\quad$\textbf{Recognize} faces in {\tt f} using minimum distance to {\tt facePrototypes}\\
		$\quad$\textbf{Update} {\tt vidResult}\\
		\textbf{{\tt ActiveSet} = []; {\tt InactiveSet} = [];} $ \rhd\text{Initialization}$ \\
		{\tt ActiveSet}, {\tt InactiveSet} = \textbf{Initialization}(recognized faces)\\
		
		\textbf{for} \text{frame {\tt f} in rest of {\tt video}}\\
		$\quad$\textbf{Detect} faces in {\tt f}\\
		$\quad$\textbf{if} \text{minor movement in detected face}
		\textbf{then}  $\quad\rhd\text{Reuse face}$\\
		$\qquad$ \textbf{updateActiveSet}\text{(face, \tt{ActiveSet})}\\
		
		$\quad$\textbf{else if} \text{detected face found in {\tt InactiveSet}} \textbf{then} $\quad\rhd\text{Update face}$ \\
		$\qquad$ \textbf{updateInactiveSet}\text{(face, \tt{InactiveSet})} \\
		$\quad${\bf else} $\quad\rhd\text{Possible new face}$\\
		$\qquad${\bf Recognize} face in f using minimum distance to {\tt facePrototypes}\\
		$\qquad${\bf updateInactiveSet}\text{(face, \tt{InactiveSet})} \\
		
		$\quad${\bf checkForMissingFace}\text{(\tt{ActiveSet}, \tt{InactiveSet})} \\
		$\quad$ \textbf{for} \text{ all labels found in {\tt {f}}} $\quad
		\rhd\text{Consistent assignment check}$\\
		$\qquad$ \textbf{if} \text{ same label exists}\\
		$\qquad \quad \textbf{Set} \text{ label with larger distance to Unknown}$\\
		$\quad$ \textbf{Put} \text{a box around the eyes} $\quad\rhd\text{DeID faces}$ \\
		\caption{Video Faces Recognition.} 
		\label{videoProcessing}
	\end{algorithm}

\begin{algorithm}[H]
	\SetAlgoLined
	\textbf{Input:}\\
	$\quad$ {\tt faces} contains a list of faces's bounding box, label, landmarks, {\tt totAppearances}, and {\tt contAppearance}.\newline
	\textbf{Output:}\\
	$\quad$ {\tt ActiveSet}, {\tt InactiveSet}. \newline
	\textbf{while} {\tt face} in all faces \\
	$\quad${\bf if} {\tt totAppearances}({\tt face}) / {\tt totFramesProcessed}({\tt face}) $>=$ 50\%\\
	$\qquad${\bf Add} {\tt face} to {\tt ActiveSet} \\
	$\quad${\bf else}:\\
	$\qquad${\bf Add} {\tt face} to {\tt InactiveSet}\\
	\textbf{end}
	\label{initTrack}
	\caption{Initialization.}
\end{algorithm}

\begin{algorithm}[H]
	\SetAlgoLined
	\textbf{Input:}\\
	$\quad$  {\tt InactiveSet}, {\tt face} (location, label, landmarks, {\tt totAppearances}, {\tt contAppearance}) \newline
	\textbf{Output:}\\
	$\quad$ \text{updated \tt{InactiveSet}} \\
	{\bf Update} {\tt totAppearances}, {\tt contAppearance}, {\tt vidResult} for \tt{face}\\
	{\bf if} {\tt totAppearances}({\tt face})/{\tt totFramesProcessed}({\tt face})$>=$50\%\\
	$\quad${\bf Move} \tt{face} to {\tt ActiveSet} \\
	$\quad${\bf Remove} \tt{face} from {\tt InactiveSet} \\
	$\quad${\bf updateActiveSet}(face)\\
	\textbf{end}
	\label{inActiveSet}
	\caption{Update InactiveSet.}
\end{algorithm}

\begin{algorithm}[H]
	\SetAlgoLined
	\textbf{Input:}\\
	$\quad$  {\tt ActiveSet}, {\tt face} (location, label, landmarks, {\tt totAppearances}, {\tt contAppearance}) \newline
	\textbf{Output:}\\
	$\quad$ updated {\tt ActiveSet} \\
	\textbf{Reuse} face from  {\tt ActiveSet} \\
	\textbf {Update} {\tt totAppearances}, {\tt contAppearance}, {\tt vidResult} for \tt{face}\\
	\textbf{end}
	\label{activeSet}
	\caption{Update ActiveSet.}
\end{algorithm}

\begin{algorithm}[H]
	\SetAlgoLined
	\textbf{Input:}\\
	$\quad$ {\tt ActiveSet}, {\tt InactiveSet}.\newline
	\textbf{Output:}\\
	$\quad$ updated {\tt ActiveSet}, updated {\tt InactiveSet}. \newline
	\textbf{for} \text{face in  {\tt ActiveSet}}\\
	$\quad$\textbf{if} \text{ face not found in all detected faces} \textbf{ then} \\
	$\qquad$\textbf{if}  {\tt contAppearances} $>=$ {\tt minAppearances} \textbf{then}  $\qquad\rhd{\text{Occluded face}}$ \\
	$\quad$ $\qquad$\textbf{Add} \text{face to  {\tt ActiveSet}, {\tt vidResult}}\\
	$\quad$ $\quad$\textbf{if}  {\tt contAppearances} $<$ {\tt minAppearances} \textbf{then} $\quad\quad\rhd\text{Disappearing face}$ \\
	$\quad$ $\qquad$\textbf{Remove} face from  {\tt ActiveSet} \\
	$\quad$ $\qquad$\textbf{Add} face to  {\tt InactiveSet}\\
	\textbf{end}
	\label{missingFace}
	\caption{checkForMissingFace.}
\end{algorithm}

\subsection{Face recognition constraints: Rejecting out-of-group fa-ces and Assigning the same label to two different faces}

\textbf{Rejecting instances with same label assignment to two different faces:}\\
Sometimes, the algorithm constantly labeled some participants as others because of camera angles or facial features, creating confusion. This problem often resulted in having the same label applied to two different faces within the same video frame, which is impossible. Hence, when two faces got the same label, the most distant face was relabeled as unknown, whereas the minor distance face kept the correct label.\\~\\
\textbf{Results for rejecting out-of-group faces:} \\
The AOLME dataset defines an open-set face recognition problem since the videos include faces that do not appear in the training dataset. By rejecting faces that are not part of the collaborative groups, the accuracy increases while recognition time reduces. In addition to rejecting actual faces that do not belong to the current collaborative group, the method also rejected wrongly identified objects or body parts (e.g., ears, hands) as faces. The algorithm rejected small size boxes as the out-of-group faces are usually in the background and have much smaller face areas than the collaborative groups.

\section{Video Processing Frame-rate Optimizations }

In addition to Face prototypes with Sampling and video faces recognizer process, the thesis considered processing the video at reduced frame rates to reduce the number of face prototypes. The thesis considered processing the videos at the original frame rate and alternative frame rates. These alternative frame rates included skipping 5, 10, 15, 20, 30, and 60 frames after each processed frame. The goal was to speed up recognition times without sacrificing accuracy. This approach should work well in the cases where the participants do not move a lot. After testing each frame rate, the algorithm chooses the best one using the Pareto front. This optimization aims to maximize the number of skipped frames to save time while keeping the accuracy downwards.

\section{Optimal Data Augmentation}
With multiple faces that do not appear very frequently, or when a participant does not move too much, there exist many near-identical face prototypes. In addition, the ground truth process takes very long to do; thus, only small clips of length 10 seconds are picked, and thus, the problem of variety rises. This thesis tries to fix this problem by applying data augmentation onto the dataset to increase the face prototypes variety. With data augmentation, the dataset increased a factor of 10. This dataset then underwent the process of sparse sampling to get one frame per second, and lastly, it got applied to k-means clustering with log-based search to achieve the number of face prototypes from $2^4$ to 2$^{10}$. 

\subsection{Data Augmentation Setup}
Some students only show up for a couple of sessions, while some appear throughout multiple Cohorts and Levels; thus, the original dataset was not balanced. In addition, the algorithm often identified some students as different students. The thesis tried to fix this with data augmentation to simulate more poses per participant, hoping that it would boost accuracy. To further improve the accuracy, this thesis tried to combine all the approaches and adding data augmentation. The thesis combined the transformed augmented data with sparse sampling, k-means clustering, and frame rate skips (Fast Method). The process starts with the original data, which undergoes data augmentation to increase the dataset ten times. Next, sparse sampling was applied to get one frame per second. Finally, the algorithm applied k-means for face prototypes, and frame rate skipping was used from the above results to reduce recognition time further. 

\begin{figure}[]
	\centering
	\includegraphics[width=1\textwidth]{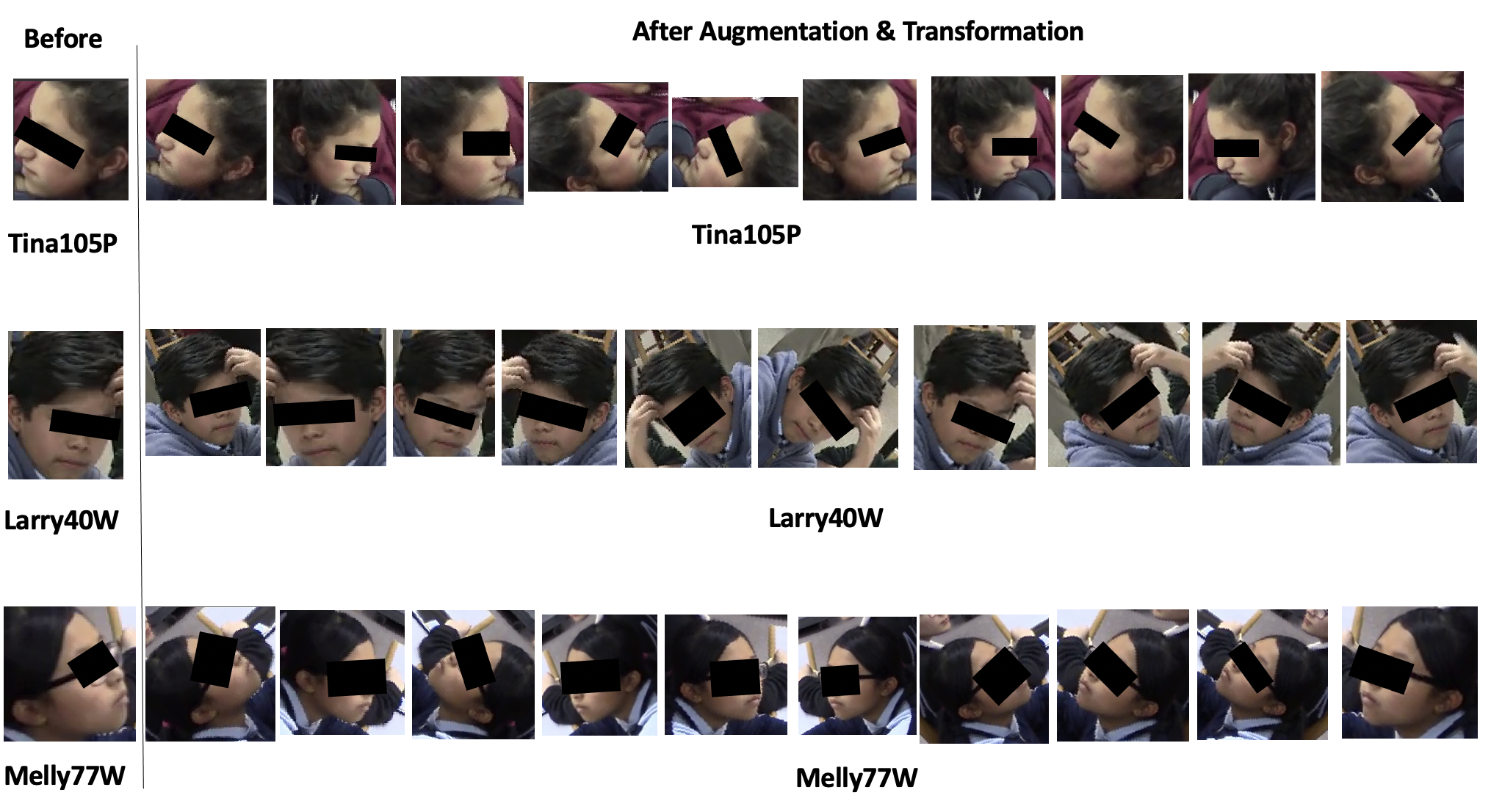}
	\caption[For data augmentation, the algorithm starts with one image on the left ]{For data augmentation, the algorithm starts with one image on the left and with transformation, the augmented data are on the right. The data size increases by a factor of ten.} 
	\label{augmented}
\end{figure}

The thesis used a combination of flipping, rotation, translation, scaling, and shearing in the same order to augment the AOLME dataset. The algorithm applied all of these transformations to an image frame.  Figure \ref{augmented} provides a demonstration of before and after augmentation is applied. With randomization, this method ensures an increase in the dataset variability. This approach trained on 11 participants from the 24 participants from the video faces recognition dataset. 

\textbf{Flip:}\\
This thesis implements a random horizontal flip where if a random token is 0.5 or less, the image is flipped and kept otherwise.

\textbf{Rotation:}\\
The algorithm chose a theta angle to be in between -45\textdegree and 45\textdegree. The thesis tests out multiple values, and for faces, these provide the most likely results. For faces, choosing too big of a rotation angle might end up with worse results.

\textbf{Scaling:} \\
For scaling, to not distort an image too much, this thesis decides to stay with a range between 0.8 and 1.2.

\textbf{Translation:}\\
This thesis chooses a random value for translating x and y values between -10 and 10.

\textbf{Shearing:}\\
The thesis chose the shear range between -0.1 to 0.1. to avoid disfiguring the faces.

\chapter{Results}

\section{Image Face Recognition Dataset}
This section describes the development of PCA and SVM as a baseline method \cite{PCA}.

\subsection{Baseline Ground Truth Process}
For ground truth generation, cropped out images were chosen manually from an Urban and a Rural school. As the dataset is video-based, there are many different poses a student can have. Thus, for each student, several different poses were chosen. The images then got resized to 28x28 before going through the training process.

\subsection{Baseline Image Dataset}
\begin{figure}[!t]
	\centering
	\includegraphics[width=1\textwidth]{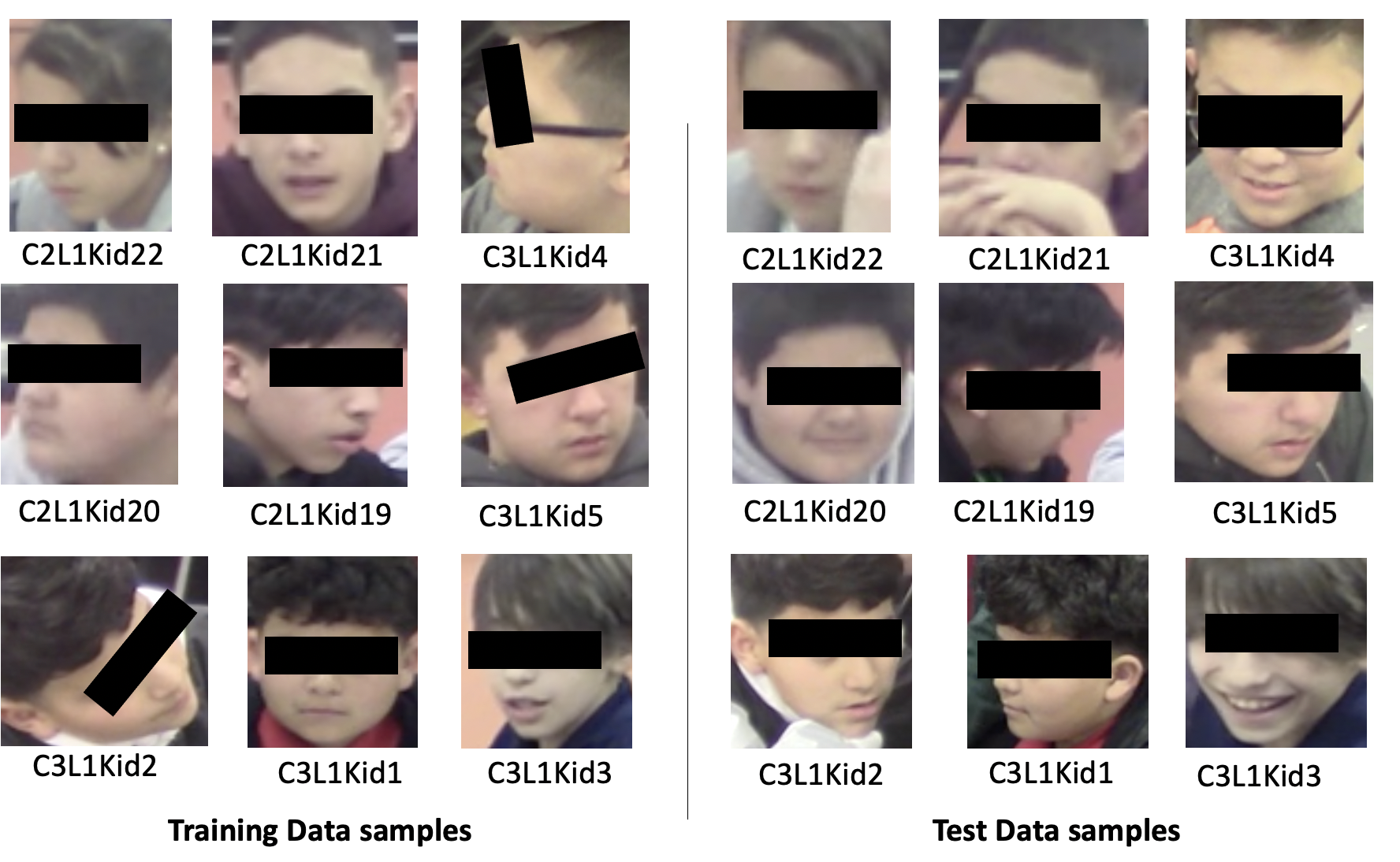}
	\caption{Baseline image dataset for face recognition from cropped face images} 
	\label{frameData}
\end{figure}

The training and testing dataset consisted of 900 manually cropped images from nine students’ faces, in which each face has 100 images from multiple sessions during the program. The thesis trained and tested on an SVM classifier model using the code provided by scikit-learn \cite{PCA} and implemented eight more classifiers to make a comparison between these different models. Out of the 784 pixels from the original cropped dataset, PCA computed 150 principal components that capture 95\% of the variance. For optimization, the approach applied GridSearchCV using combinations of different parameters to pick the best parameter values for optimal results. \\
The training and testing dataset images came from the same sessions, but they got split in the beginning to come from different video segments. The ratio chosen was 75-25\% for training and testing, respectively. The videos came from Group E of the Rural school from Cohort 2 Level 1, and Group A of the Rural school from Cohort 3 Level 1. \\
The naming convention for this dataset is different from the video face recognition method as there had not been a uniform way of naming the students. Figure \ref{frameData} shows an illustration of the training and testing samples.

\begin{figure}[!t]
	\centering
	\includegraphics[width=1\textwidth]{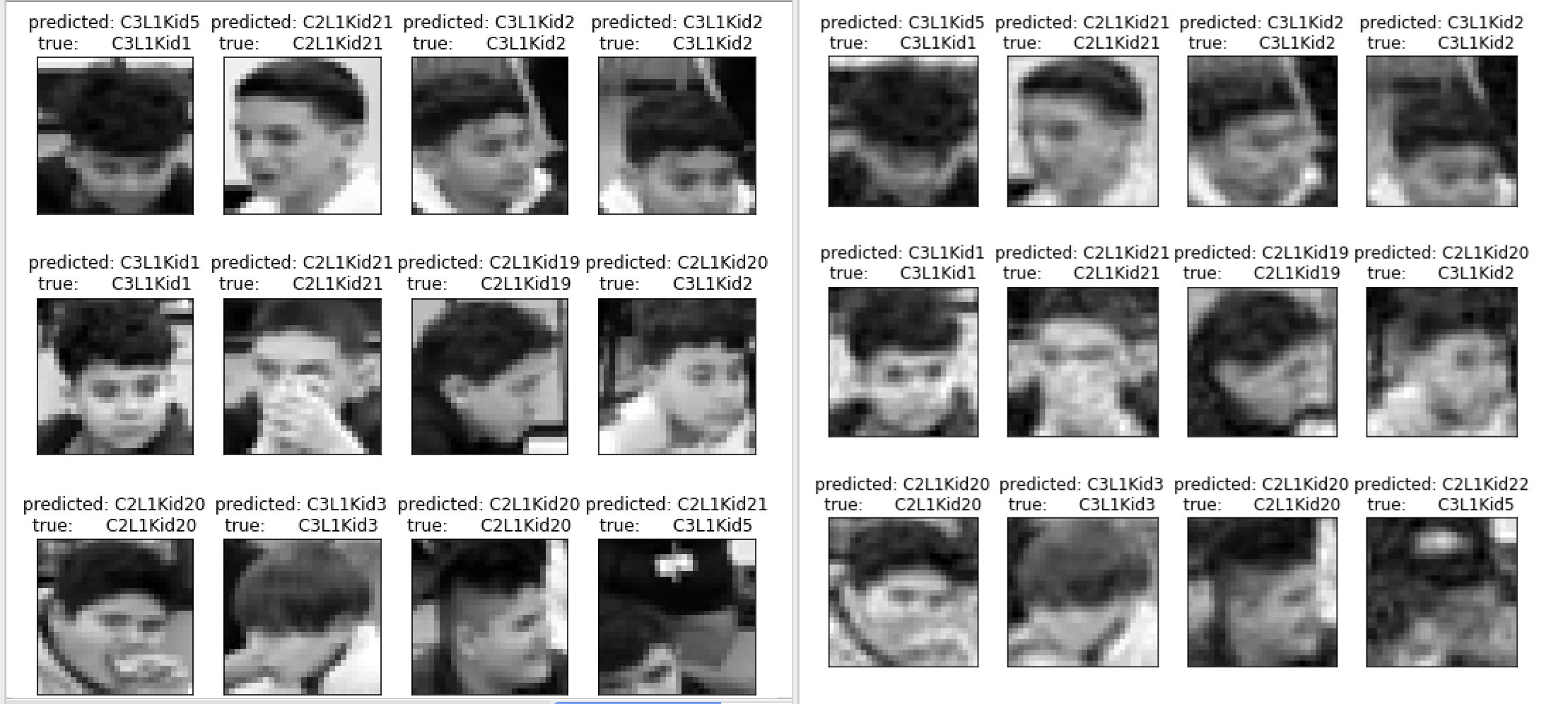}
	\caption[The first four columns are the Original Images ]{The first four columns are the Original Images and the last four columns are the PCA compressed representations.} 
	\label{fig:EigenTransform}
\end{figure}

Table \ref{PCA} summarizes the results of 9 different classifiers in descending order. SVM did the best in recognizing these students with an average of 74\%  accuracy, followed by MLP and Gaussian Process of 71\%. QDA did not do very well when it only got 22\% accuracy. Figure \ref{fig:EigenTransform} showed original data on the left and reduced dimensional data on the right. PCA reduced the dimension complexity to almost six times, but it still captured the essential features where the outline of each face is still clearly seen.

\begin{table}[]
	\caption{\label{PCA} Single-frame face recognition results.}
	
	\begin{tabular}{c|c}
		\hline
		\textbf{Classifier}                             & \textbf{Recognition Result} \\ \hline
		SVM (Support Vector Machine)           & 73\%               \\
		MLP (Multi-Layer Perceptron)           & 71\%               \\
		Gaussian Process                       & 71\%               \\
		Gaussian NB (Gaussian Naive Bayes)     & 70\%               \\
		Random Forest                          & 67\%               \\
		KNN (k-Nearest Neighbors)              & 52\%               \\
		Decision Tree                          & 50\%               \\
		Ada Boost                              & 42\%               \\
		QDA (Quadratic Discriminant Analyasis) & 22\%               \\ \hline
	\end{tabular}
\end{table}

\section{Video Face Recognition Results} 

\subsection{Face recognition constraints: Rejecting out-of-group fa-ces and Assigning the same label to two different faces}

\begin{figure}[]
	\centering
	\includegraphics[width=1\textwidth]{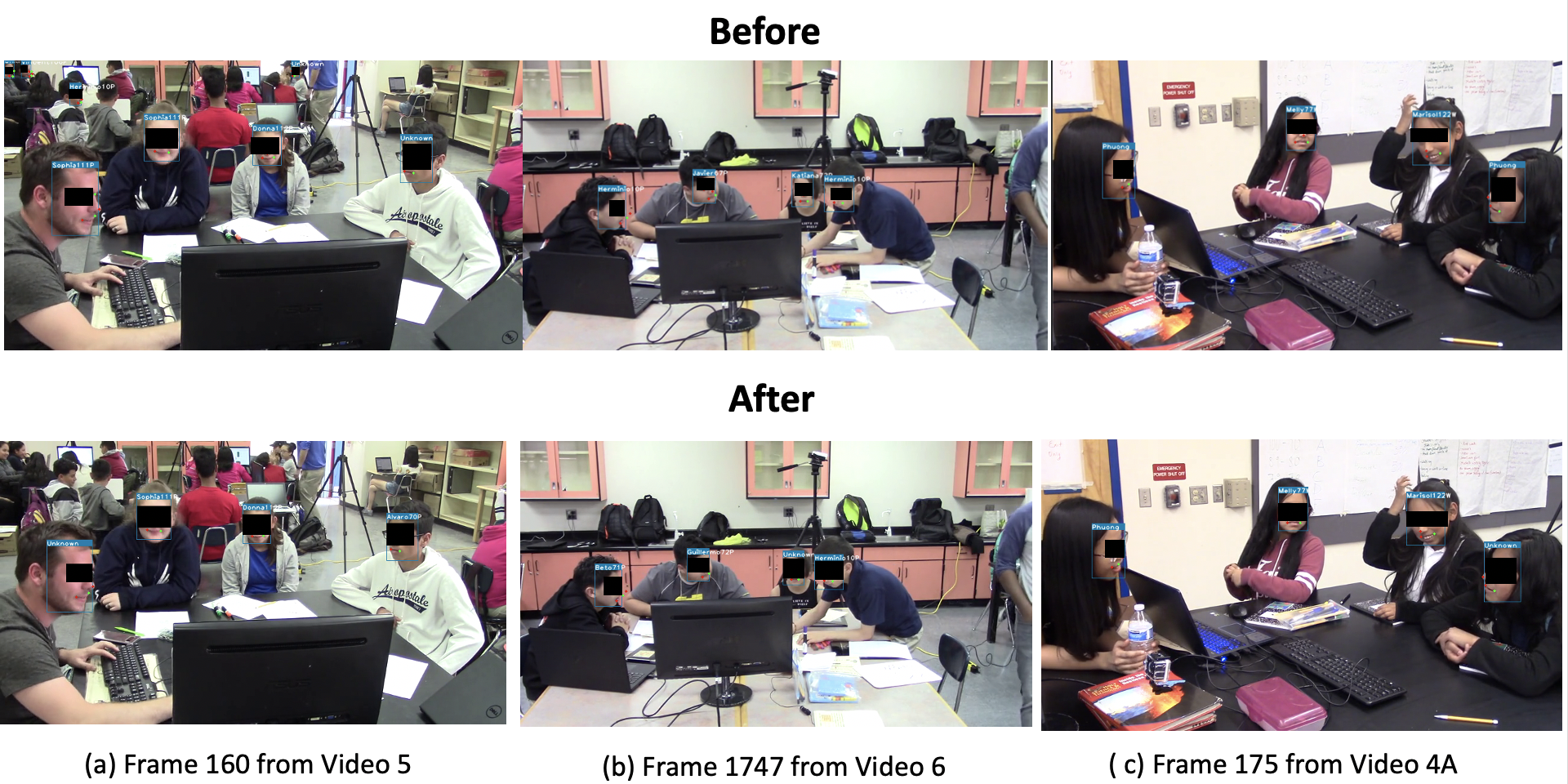}
	\caption[Improvement of face recognition results ]{Improvement of face recognition results by eliminating the use of the same label for two different people in the same frame.} 
	\label{fig:sameLables}
\end{figure}

\begin{figure}[b]
	\centering
	\includegraphics[width=1\textwidth]{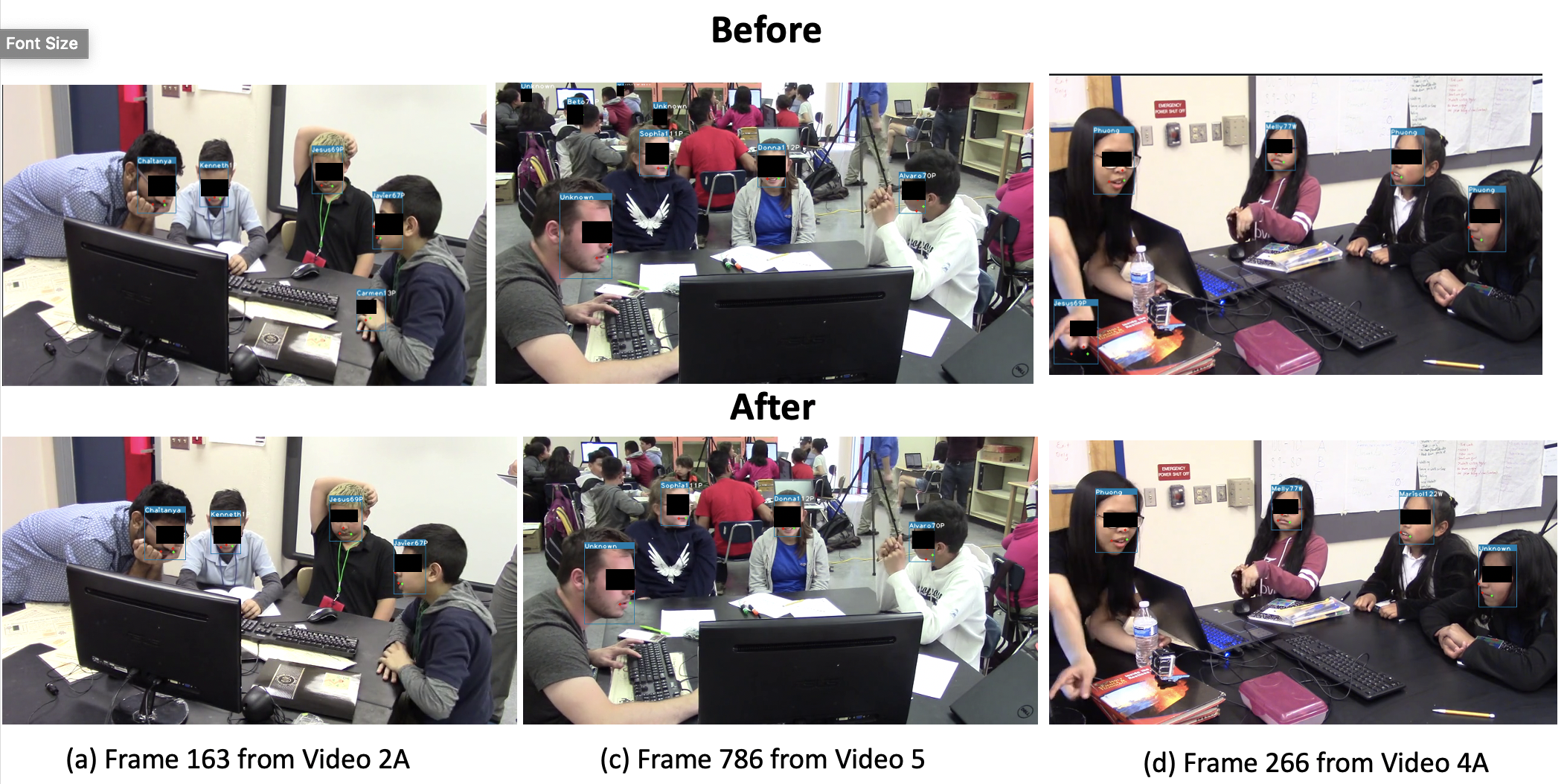}
	\caption[Face recognition improvement by rejecting out-of-group faces.]{Face recognition improvement by rejecting out-of-group faces. In the middle-column, faces that do not belong to group closest to the camera are rejected. In the first and third columns, background objects labeled as faces are rejected.}
	\label{fig:bkgrdFaces}
\end{figure}

\textbf{Results for rejecting instances where the same label is assigned to two different faces:}\\
Figure \ref{fig:sameLables} shows an example of before and after the implementation of rejecting the same labels. The top row images demonstrate the problem of wrongly labeling the facilitator (far left) as Sophia111P (blue sweater), labeled both students on the far sides as Herminio10P when only the far-right student is labeled correctly, or labeled Cristie123W (far right) with the same label as the facilitator (far left) (see a, b, and c, respectively). The bottom row demonstrates the success of the approach with the two wrongly labeled faces got IDed as 'Unknown' (see a and c) while correctly identified the participant as 'Beto71P' (see b).\\~\\
\textbf{Results for rejecting out-of-group faces:} \\
Figure \ref{fig:bkgrdFaces} represents the results of correctly rejecting non-face objects (see a and d) and correctly rejecting out-of-group faces (see b and c). By rejecting out-of-group faces, the recognition time is also significantly reduced as computing distances to face prototypes is extremely time-consuming.

\subsection{Face Prototypes Results}

\subsubsection{Face Prototypes with K-means results}
\begin{figure}[]
	\centering
	\includegraphics[width=1\textwidth]{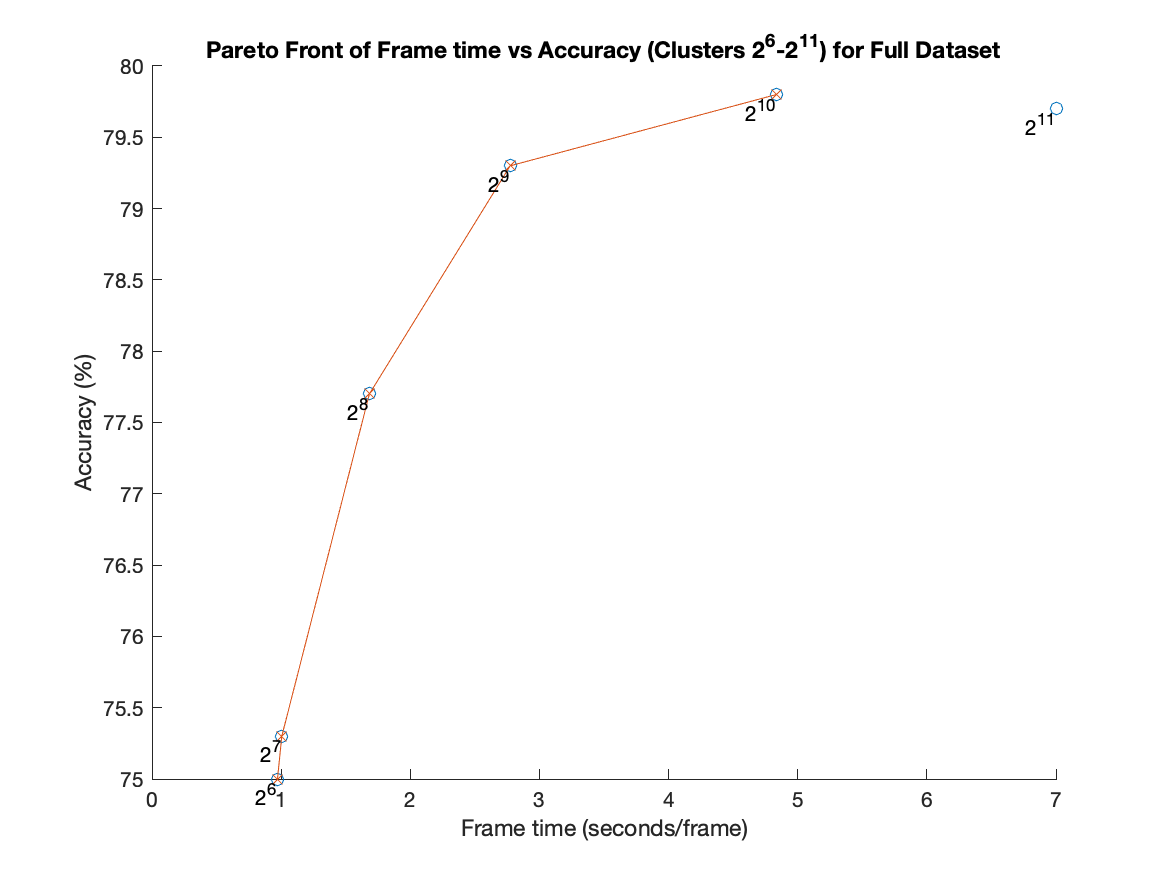}
	\caption{Pareto front for Face Prototypes using K-means.} 
	\label{fig:ParetoFront}
\end{figure}

The results show that using too large clusters did not yield significant improvements in face detection accuracy. 
This process ran on a personal Macbook Pro that ran Mac-OS with 2.3GHz, 4-core, Intel i5 processors. The thesis represents the results for a representative video from group C, level 1, in Figure \ref{fig:ParetoFront}. The accuracy peaks at 79.8\% for 1024 face prototypes, with a recognition rate of 4.8 seconds per frame. 

\subsubsection{Face Prototypes with Sampling results}
For Sparse Sampling, the videos came from Group D from Rural Cohort 2 Level 1 (C2L1, Videos 2A, 2B \& 2C), Group D from Urban Cohort 3 Level 1 (C3L1, Video 3A \& 3B), Group E from Rural Cohort 3 Level 1 (C3L1, Video 4), Group D from Rural Cohort 3 Level 1 (C3L1, Video 6C) and Group C from Urban Cohort 3 Level 1 (C3L1 Video 7). Tables \ref{AccRes} and \ref{AccRes2} show the results where it is clear that the thesis's approach did a little better than the baseline approach, except for Video 7. This result might occur because the algorithm went through an initialization process, and thus, if no face was detected, the tracking does not work very well. In addition, consistent tracking allowed for a tolerance of five missing frames, and if the detector failed to find the face over a long time (complex occlusion), the algorithm would not perform well. However, the algorithm performed better than the baseline method on the rest of the testing video clips. The difference in accuracy ranged from as low as $2\%$ to as large as $26\%$. In addition, the baseline method failed to process long videos (Video 3B) where it did not converge. Out of 24 participants in these eight videos, the thesis achieved higher or the same accuracy for 16 participants.
Overall, the thesis achieved an average of  $71.8\%$ compared to  $62.3\%$ for the baseline method if the thesis did not take into account video 3B (did not converge); otherwise, the baseline method averages at 54.6\% if it is considered 0\% accuracy for the baseline method at video 3B since it did not run.

In addition, Figure \ref{TimeRes} represents the recognition time taken when running Face Prototypes with K-means vs. the baseline approach. Using the entire dataset as face prototypes instead of Sparse Sampling, the time taken using the baseline approach was a lot slower. The baseline method required an average of 9.3 seconds/frame, whereas our proposed method required 0.8 seconds/frame. On average, the proposed method was 11.1$\times$ faster. Video 2B  shows the best speedup factor of 52.7$\times$. The speedups are due to the reduced number of Face prototypes with Sampling and the fact that we do not rerun the minimum distance classifier if there is little movement in the detected faces. For example, for videos 2B and 3A, InsightFace took a very long time (more than ten seconds/frame) because it compared each participant with (almost) ten thousand images. For video 4, in addition to comparisons to about ten thousand images for the leading group, InsightFace also had to compare against faces from the background groups.
In comparison, this thesis's approach rejected the need to recognize background groups by applying a minimum face size constraint. The baseline method was tested on the last two videos, videos 6C and 7, with Sparse Sampling. Thus, recognition time got reduced significantly compared to other videos.

\begin{table}[!t]
	\caption[Accuracy for Facial Recognition using the thesis's approaches (Ours) ]{\label{AccRes}
		Accuracy for Facial Recognition using the thesis's approaches (Ours) for processing the videos at the original frame-rate vs using the baseline method (Insightface). Face Prototypes with K-means was used in Video 1 and the rest applied Face Prototypes with Sampling (I of II).
		Each video represents a different group session segment. Repeated ID represents different short clips cut from the same video at different time.}
	\begin{center}
		\begin{tabular}{p{0.24\textwidth}p{0.13\textwidth}p{0.2\textwidth}p{0.13\textwidth}p{0.15\textwidth}}
			\toprule
			\textbf{Video}  & \textbf{Duration}                  & \textbf{Person Label}           & \textbf{Ours} & \textbf{Insightface} \\ \toprule
			\multirow{4}{*}{\begin{tabular}[c]{@{}l@{}}\textbf{1}\\ (Face prototypes \\with K-means)\end{tabular}} & \multirow{4}{*}{10 seconds}&\textit{\textbf{Antone39W}}& \textbf{36.5\%}    & \textbf{36.5\%}      \\
			& & \textit{\textbf{Jaime41W}}   &   \textbf{86.7\%}   & 84.2\%       \\
			& & \textit{\textbf{Larry40W}}  &    \textbf{99.3\%}                          & 98.3\%         \\
			& & \textit{\textbf{Ernesto38W}}   &    \textbf{96.5\%}                          & 95.3\%
			\\ \cmidrule{3-5}
			& & \textbf{Average}   & \textbf{79.8\%}                          & 78.6\%      \\ \midrule
			
			\multirow{4}{*}{\begin{tabular}[c]{@{}l@{}}\textbf{2A}\\ (Face prototypes \\with Sampling)\end{tabular}}  & \multirow{4}{*}{10 seconds}& \textit{\textbf{Chaitanya}}          &\textbf{95.3\%}    & 80.3\%       \\
			&& \textit{\textbf{Kenneth1P}}     & \textbf{91\%}                          &83.1\%                    \\
			&& \textit{\textbf{Jesus69P}}   & \textbf{100\%}                         &\textbf{100\%}                         \\
			&& \textit{\textbf{Javier67P}}    & \textbf{100\%}                          &69.1\%                    
			\\ \cmidrule{3-5}
			&& \textbf{Average}   & \textbf{96.5\%}                          & 83.1\%           \\ \midrule
			
			\multirow{4}{*}{\begin{tabular}[c]{@{}l@{}}\textbf{2B}\\ (Face prototypes \\with Sampling)\end{tabular}}  & \multirow{4}{*}{60 seconds} & \textit{\textbf{Chaitanya}}          &\textbf{80.0\%}    & 56.1\%       \\
			&& \textit{\textbf{Kenneth1P}}     & \textbf{98.3\%}                          &61.5\%                    \\
			&& \textit{\textbf{Jesus69P}}   & \textbf{99.3\%}                         &\textbf{99.3\%}                         \\
			&& \textit{\textbf{Javier67P}}    & \textbf{80.6\%}                          &39.0\%                    
			\\ \cmidrule{3-5}
			&& \textbf{Average}   & \textbf{89.5\%}                          & 63.2\%           \\ \midrule
			
			\multirow{4}{*}{\begin{tabular}[c]{@{}l@{}}\textbf{3A}\\ (Face prototypes \\with Sampling)\end{tabular}}  & \multirow{4}{*}{10 seconds} & \textit{\textbf{Melly77W}}          &\textbf{96.0\%}    & 59.7\%       \\
			&& \textit{\textbf{Marisol112W}}     & \textbf{84.0\%}                          &60.5\%                    \\
			&& \textit{\textbf{Cristie123W}}   & 8.67\%                       &\textbf{27.3\%}                       \\
			&& \textit{\textbf{Phuong}}    & \textbf{77.4\%}                          &21.4\%                    
			\\ \cmidrule{3-5}
			&& \textbf{Average}   & \textbf{66.5\%}                          & 42.2\%            \\ \midrule
			
			\multirow{4}{*}{\begin{tabular}[c]{@{}l@{}}\textbf{3B}\\ (Face prototypes \\with Sampling)\end{tabular}}  & \multirow{4}{*}{10 minutes} & \textit{\textbf{Melly77W}}          &\textbf{64.9\%}    & N/A       \\
			&& \textit{\textbf{Marisol112W}}     & \textbf{0.2\%}                          &N/A                    \\
			&& \textit{\textbf{Cristie123W}}   & \textbf{96.5\%}                       &N/A                  \\
			&& \textit{\textbf{Phuong}}    & \textbf{86.3\%}                          &N/A                  
			\\ \cmidrule{3-5}
			&& \textbf{Average}   & \textbf{62\%}                          & N/A            \\ 
			\bottomrule
		\end{tabular}
	\end{center}
\end{table}

\begin{table}[!t]
	\caption[Accuracy for Facial Recognition using the thesis's approaches (Ours) ]{\label{AccRes2}
		Accuracy for Facial Recognition using the thesis's approaches (Ours) for processing the videos at the original frame-rate vs using the baseline method (Insightface). Face Prototypes with K-means was used in Video 1 and the rest applied Face Prototypes with Sampling  (II of II).
		Each video represents a different group session segment. Repeated ID represents different short clips cut from the same video at different time.}
	\begin{center}
		\begin{tabular}{p{0.24\textwidth}p{0.13\textwidth}p{0.21\textwidth}p{0.13\textwidth}p{0.15\textwidth}}
			\toprule
			\textbf{Video}  & \textbf{Duration}                  & \textbf{Person Label}           & \textbf{Ours} & \textbf{Insightface} \\ \toprule
			
			\multirow{3}{*}{\begin{tabular}[c]{@{}l@{}}\textbf{4}\\ (Face prototypes \\with Sampling)\end{tabular}}  & \multirow{3}{*}{60 seconds} & \textit{\textbf{Alvaro70P}}          &\textbf{96.4\%}    & 60.8\%       \\
			&& \textit{\textbf{Donna112P}}     & \textbf{100\%}                          &99.8\%                    \\
			&& \textit{\textbf{Sophia111P}}   & 99.5\%                       &\textbf{99.9\%}                       \\
			\cmidrule{3-5}
			&& \textbf{Average}   & \textbf{98.6\%}                          & 86.8\%            \\ \midrule

			\multirow{5}{*}{\begin{tabular}[c]{@{}l@{}}\textbf{6C}\\ (Face prototypes \\with Sampling)\end{tabular}}  & \multirow{5}{*}{60 seconds} & \textit{\textbf{Ivonne}}  & 91.5\%        &\textbf{94.6\%}          \\
			
			&& \textit{\textbf{Juanita107P}}         &\textbf{61.7\%}    & 42.4\%       \\
			&&\textit{\textbf{Katiana73P}}                            &\textbf{1.1\%}        & 0.02\%              \\
			&& \textit{\textbf{Maya108P}}   & 0.3\%                     &\textbf{20.7\%}                       \\
			&& \textit{\textbf{Marcia109P}}                    &\textbf{55.6\%}    & 18.5\%                    \\
			\cmidrule{3-5}
			&& \textbf{Average}                      &     \textbf{42.0\%} & 35.3\%            \\ \midrule
			
			\multirow{6}{*}{\begin{tabular}[c]{@{}l@{}}\textbf{7}\\ (Face prototypes \\with Sampling)\end{tabular}}  & \multirow{4}{*}{60 seconds} & \textit{\textbf{Phuong}}         &\textbf{89.2\%}    & 67.7\%       \\
			&& \textit{\textbf{Juanita107P}}    & \textbf{98.1\%}    &89.6\%                    \\
			&& \textit{\textbf{Josephina104P}}   & 8.6\%                       &\textbf{50.1\%}                       \\
			&& \textit{\textbf{Tina105P}} & 0.4\%                       &\textbf{10.1\%}                       \\
			&& \textit{\textbf{Vincent106P}} & 0.5\%                       &\textbf{23.6\%}                       \\
			&& \textit{\textbf{Jacob103P}} & 40.8\%                       &\textbf{42.3\%}                       \\
			\cmidrule{3-5}
			&& \textbf{Average}   & 39.6\%                          & \textbf{47.2}\%            \\ \midrule

			&&\textbf{Overall Average}                             & \textbf{71.8\%}         & \textbf{62.3}\%    \\
			
			\bottomrule
		\end{tabular}
	\end{center}
\end{table}

\begin{table}[]
	\caption[Recognition times using the thesis's approaches (Ours) ]{\label{TimeRes}
		Recognition times using the thesis's approaches (Ours) for processing the videos at the original frame-rate vs using the baseline method (Insightface). Face Prototypes with K-means was used in Video 1 and the rest applied Face Prototypes with Sampling. Each video ID represents a different group session segment. Repeated ID represents different short clips cut from the same video at different time.}
	\begin{center}
		\begin{tabular}{p{0.08\textwidth}p{0.125\textwidth}p{0.13\textwidth}p{0.18\textwidth}p{0.18\textwidth}p{0.118\textwidth}}\toprule
			{\begin{tabular}[c]{@{}l@{}}\textbf{Video}\\ \textbf{}\end{tabular}}    & {\begin{tabular}[c]{@{}l@{}}\textbf{Duration}\\ \textbf{}\end{tabular}}   &{\begin{tabular}[c]{@{}l@{}}\textbf{GT Faces}\\ \textbf{}\end{tabular}}  & {\begin{tabular}[c]{@{}l@{}}\textbf{Insightface}\\ \textbf{(sec/frame)}\end{tabular}}   & {\begin{tabular}[c]{@{}l@{}}\textbf{Ours}\\ \textbf{(sec/frame)}\end{tabular}} & {\begin{tabular}[c]{@{}l@{}}\textbf{Speedup}\\ \textbf{factor}\end{tabular}}  \\ \toprule
			\textbf{1}          & 10  & 4  & 9.91      & 2.8    & 3.5x \\
			\textbf{2A}          & 10 & 4 & 9.96    & 0.8  & 12.5x  \\
			\textbf{2B}          & 60  &4 & 15.8    & 0.3  & 52.7x   \\
			\textbf{3A}          & 10 &4      & 10.1      &  0.9  & 11.2x    \\
			\textbf{4}          & 60   &3   & 15.2         & 0.3 & 50x \\
			\textbf{6C}  & 60   & 4   &    0.61      & 0.3 &  2.0x\\
			\textbf{7}  & 60   & 6   &     0.82    & 0.4 & 2.1x
			\\ \cmidrule{4-6}
			& & \textbf{Average}   & 8.9                         & \textbf{0.8}   &   \textbf{11.1x}
			\\\bottomrule
		\end{tabular}
	\end{center}
\end{table}

\subsection{Video processing frame-rate optimizations}
The algorithm tested on different frame skipping values, and results shown in Table \ref{testFrameRate} represent different average accuracy with different rates. The algorithm tested on videos from Group D from Rural Cohort 2 Level 1 (C2L1), Group E from Rural Cohort 2 Level 1 (C2L1), Group E from Rural Cohort 3 Level 1 (C3L1), and Group D from Urban Cohort 3 Level 1 (C3L1).

\begin{table}[]
	\caption[Average Accuracy and Recognition Time for Facial Recognition ]{\label{FrameRates}
		Average Accuracy and Recognition Time for Facial Recognition using different Frame Rates. 
		Each video represents a different group session segment. Repeated ID represents different short clips cut from the same video at different time (I of II).}
	\begin{center}
		\begin{tabular}{p{0.23\textwidth}p{0.13\textwidth}p{0.16\textwidth}p{0.16\textwidth}p{0.165\textwidth}}
			\toprule
			\textbf{Video}  & \textbf{Duration}                  & \textbf{Frame Rates(FR)}           & \textbf{Accuracy}  & \textbf{Recognition Time (sec)} \\ \toprule

			\multirow{7}{*}{\begin{tabular}[c]{@{}l@{}}\textbf{2B}\\ (Face prototypes  \\ with sampling) \end{tabular}}  & \multirow{7}{*}{60 seconds} & \textit{\textbf{No Skip}}         & 89.5\%    & 955      \\			&& \textit{\textbf{5 frames}}    & 90.5\%  & 366                   \\
			&& \textit{\textbf{10 frames}}   & \textbf{88.8\%}  & \textbf{309 }                 \\
			&& \textit{\textbf{15 frames}}   & 83.0\%  & 335                 \\
			&& \textit{\textbf{20 frames}}   & 86.9\%  & 345                     \\
			&& \textit{\textbf{30 frames}}   & 88.9\%  & 325                  \\
			&& \textit{\textbf{60 frames}}   & 65 \%   & 310                  \\
			\cmidrule{3-5}
			&& \textbf{Best:}   & \textbf{Skipping 10} & \textbf{ frames}          \\ \midrule
			
			\multirow{7}{*}{\begin{tabular}[c]{@{}l@{}}\textbf{2C}\\ (Face prototypes  \\ with sampling)  \end{tabular}}  & \multirow{7}{*}{70 seconds} & \textit{\textbf{No Skip}}         & 37.5\%    & 1059       \\
			&& \textit{\textbf{5 frames}}    & 38.4\%  & 385                   \\
			&& \textit{\textbf{10 frames}}   & \textbf{40.0\%}  & \textbf{370 }                 \\
			&& \textit{\textbf{15 frames}}   & 33.5\%  & 335                  \\
			&& \textit{\textbf{20 frames}}   & 33.9\%  & 311                     \\
			&& \textit{\textbf{30 frames}}   & 35.1\%  & 298                  \\
			&& \textit{\textbf{60 frames}}   & 40.0 \% & 297                   \\
			\cmidrule{3-5}
			&& \textbf{Best:}   & \textbf{Skipping 10} & \textbf{ frames}            \\ \midrule

			\multirow{7}{*}{\begin{tabular}[c]{@{}l@{}}\textbf{3C}\\(Face prototypes  \\ with sampling)  \end{tabular}}  & \multirow{7}{*}{10 minutes} & \textit{\textbf{No Skip}}         & 34.1\%    & 3447      \\
			&& \textit{\textbf{5 frames}}    & 33.8\%  & 1595                  \\
			&& \textit{\textbf{10 frames}}   & 25.8\%  & 1151  \\
			&& \textit{\textbf{15 frames}}   & \textbf{36.9\%}  & \textbf{125}                 \\
			&& \textit{\textbf{20 frames}}   & 32.2\%  & 106                    \\
			&& \textit{\textbf{30 frames}}   & 34.7\%  & 95                 \\
			&& \textit{\textbf{60 frames}}   & 31 \% & 89    \\             
			\cmidrule{3-5}
			&& \textbf{Best:}   & \textbf{Skipping 15}                          & \textbf{ frames}          \\ \midrule

			\multirow{7}{*}{\begin{tabular}[c]{@{}l@{}}\textbf{4}\\(Face prototypes  \\ with sampling)   \end{tabular}}  & \multirow{7}{*}{60 seconds} & \textit{\textbf{No Skip}}         & 98.6\%    & 530      \\
			&& \textit{\textbf{5 frames}}    & 99.1\%  & 303                   \\
			&& \textit{\textbf{10 frames}}   & \textbf{99.5\%}  & \textbf{290 }                \\
			&& \textit{\textbf{15 frames}}   & 98.5\%  & 245                  \\
			&& \textit{\textbf{20 frames}}   & 99.3\%  & 281                     \\
			&& \textit{\textbf{30 frames}}   & 96.8\%  & 276                  \\
			&& \textit{\textbf{60 frames}}   & 97.7\%  & 274                   \\
			\cmidrule{3-5}
			&& \textbf{Best:}   & \textbf{Skipping 10} &\textbf{  frames}           \\

			\bottomrule
		\end{tabular}
	\end{center}
\end{table}

\begin{figure}[]
	\centering
	~\includegraphics[width=0.47\textwidth]{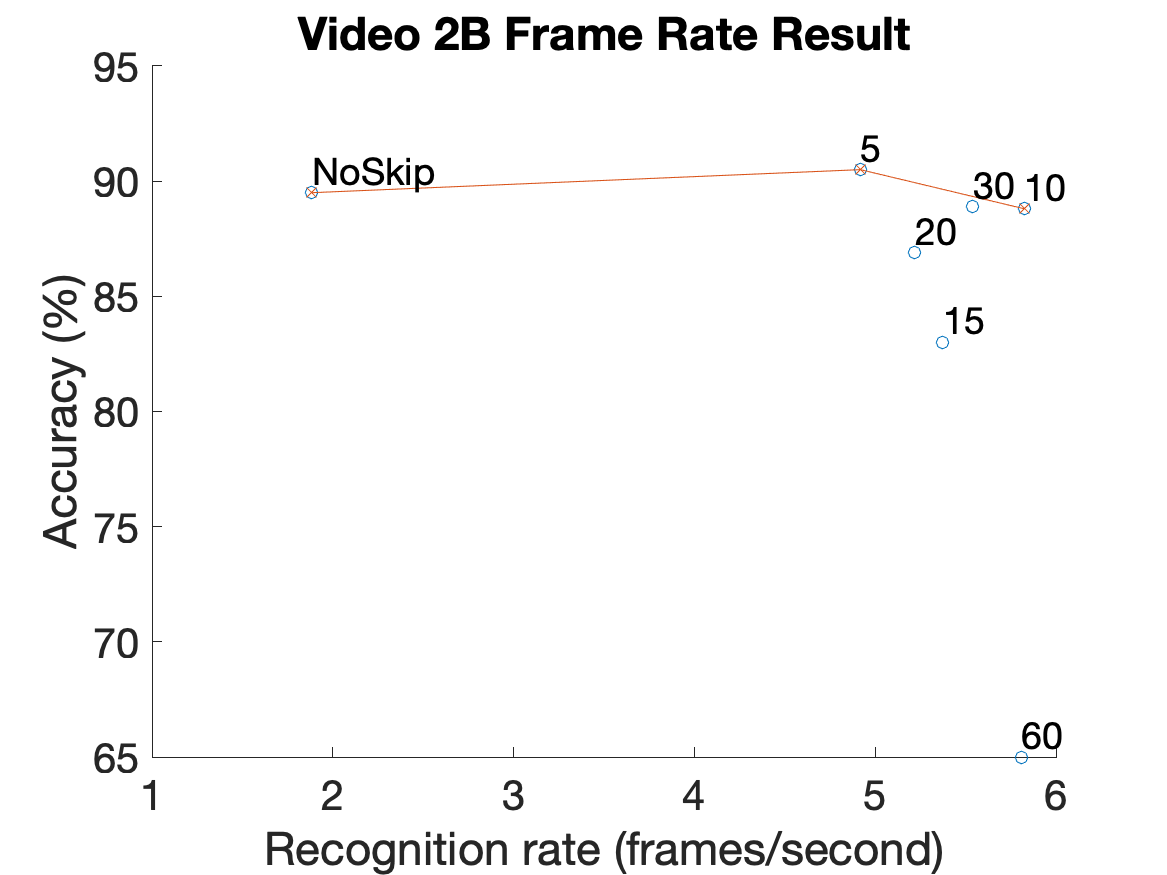}~
	~\includegraphics[width=0.47\textwidth]{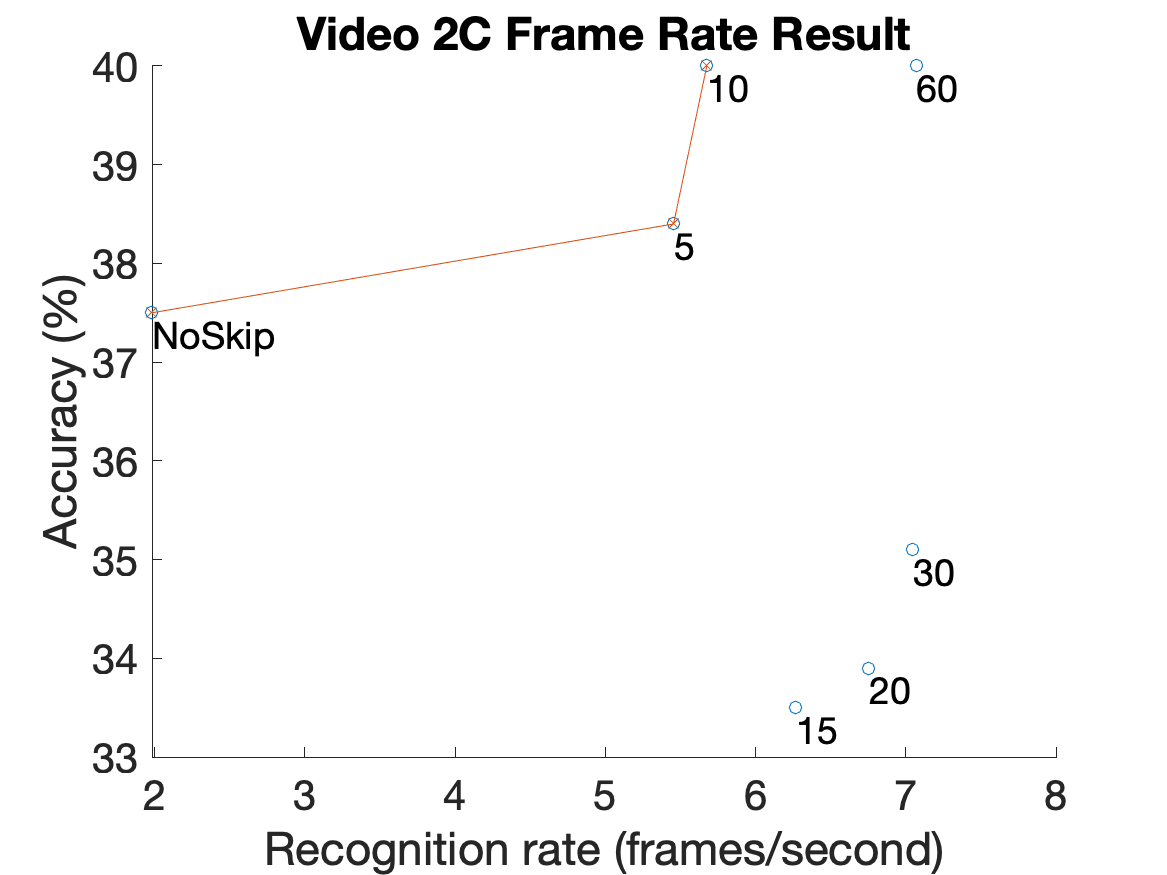}~\\[0.05 true in]
	~\includegraphics[width=0.47\textwidth]{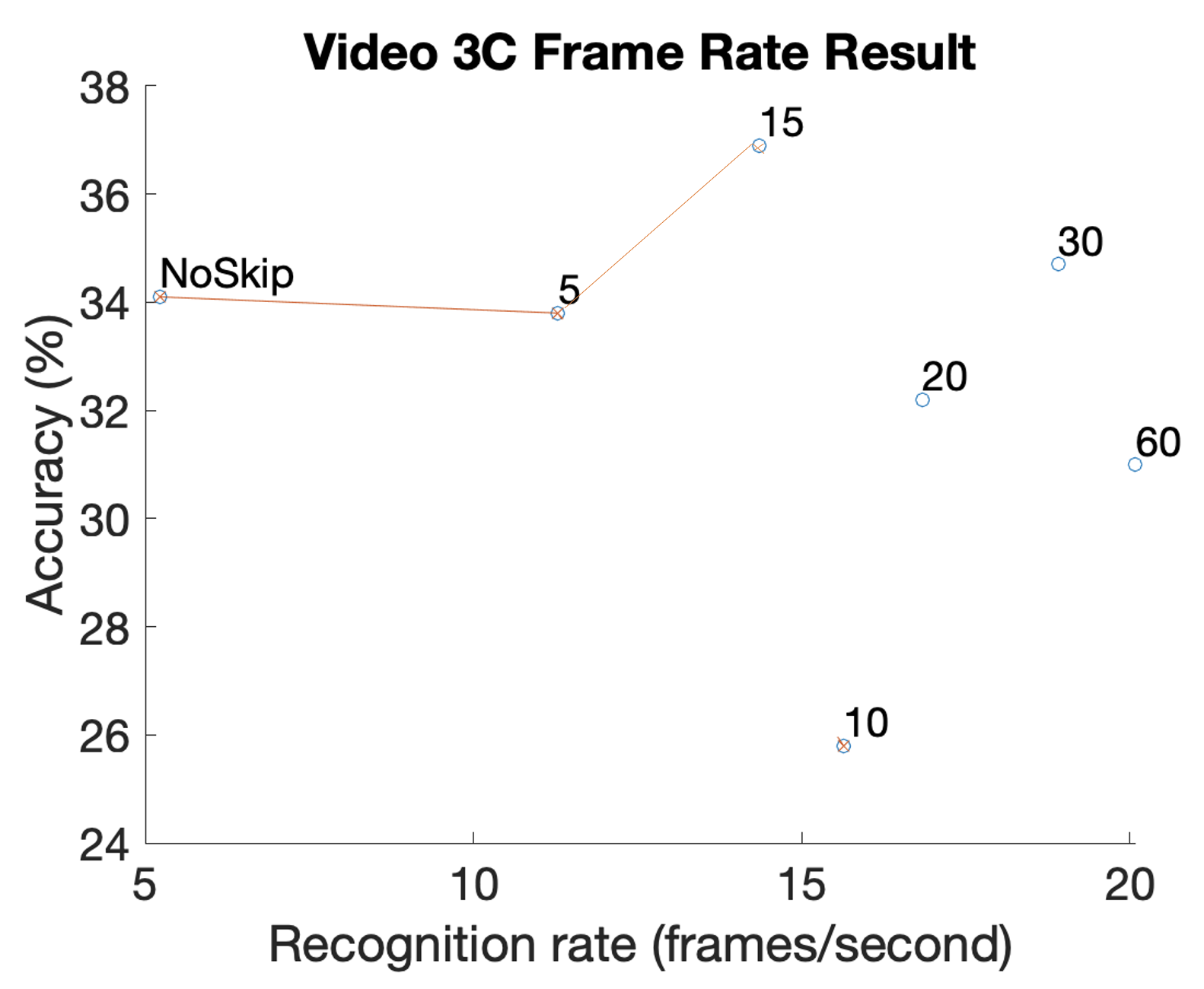}~
	~\includegraphics[width=0.47\textwidth]{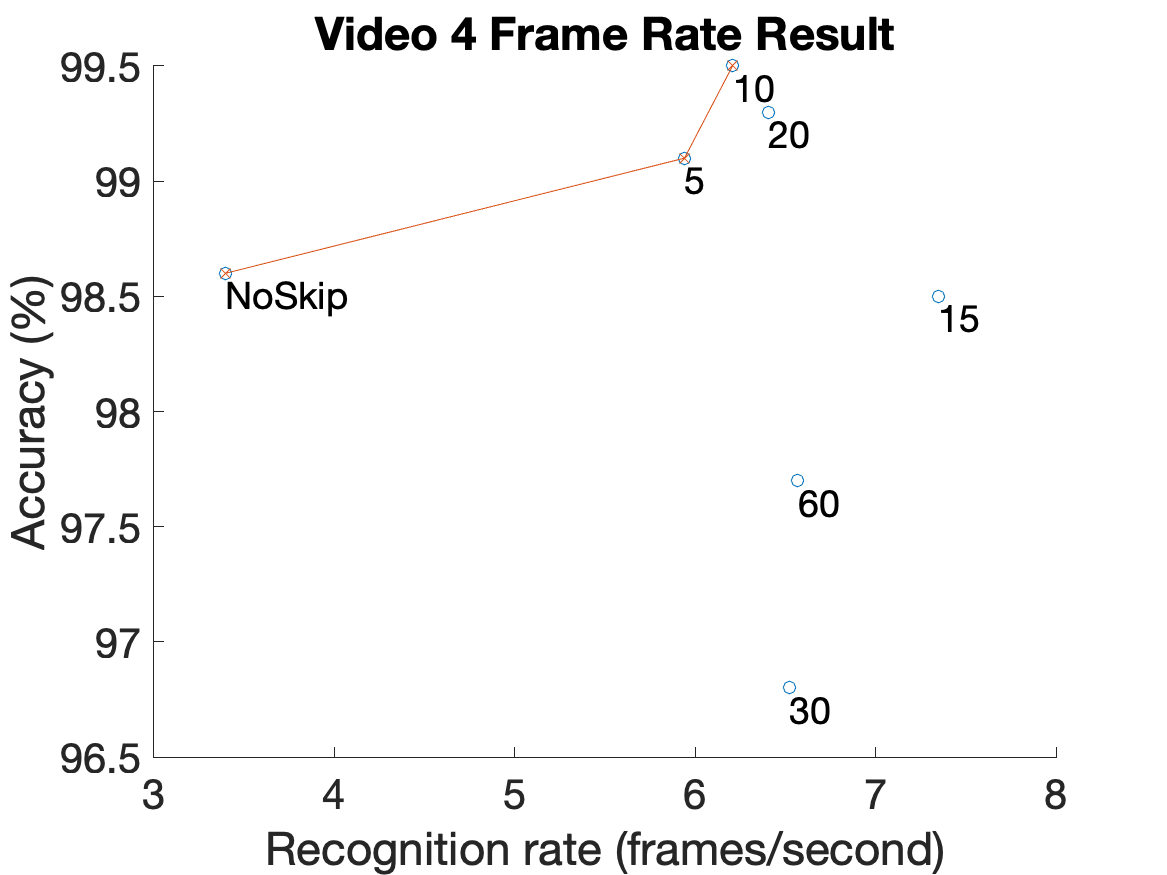}~\\[0.05 true in]
	\caption{Varied Frame Rates Results.} 
	\label{frameSkip}
\end{figure}

This thesis uses multi-objective optimization to determine which frame skipping value performs the best. Figure \ref{frameSkip} represents the plots of four videos from Table \ref{FrameRates}, which shows that at frame rates 10 (75\%) and 15 (25\%), the algorithm performs the best. Then, the thesis uses a frame rate of 10 to test videos 5A and 3A. The result is shown in Table \ref{testFrameRate}. Both videos show a much faster recognition time of 2.11x and 2.26x compared to this thesis's method without skipping, respectively, whereas compared to the baseline method, they did 11.3x and 54x times better. 

\begin{table}[]
	\caption[Average Accuracy and Recognition Time for Facial Recognition ]{\label{testFrameRate}
		Average Accuracy and Recognition Time for Facial Recognition at best frame rate with Face Prototypes with Sampling method. Baseline method does not use video processing for tracking. The table summarizes the results of determining the optimal number of frames to skip without sacrificing recognition accuracy. Each video represents a different group session segment.}
	\begin{center}
		\begin{tabular}{p{0.22\textwidth}p{0.13\textwidth}p{0.15\textwidth}p{0.15\textwidth}p{0.20\textwidth}}
			\toprule
			\textbf{Video}  & \textbf{Duration}                  & \textbf{Frame Rates(FR)}           & \textbf{Accuracy}  & \textbf{Recognition Time} \\ \toprule

			\multirow{3}{*}{\begin{tabular}[c]{@{}l@{}}\textbf{3A} \\ (Face Prototypes \\ with Sampling) \end{tabular}}  & \multirow{2}{*}{10 seconds} &
			\textit{\textbf{Baseline}}   & 66.5\%  & 3030 seconds                 \\ &&
			\textit{\textbf{No Skip}}         & 65\%    & 566 seconds     \\                \\
			&& \textit{\textbf{10 frames}}   & 64.3\%  & 268 seconds                 \\
			
			\cmidrule{3-5}
			&& \textbf{Overall:}   & \textbf{-2.2\%} & \textbf{11.3x speedup}           \\ \midrule
			
			\multirow{2}{*}{\begin{tabular}[c]{@{}l@{}}\textbf{5A}\\ (Face Prototypes \\ with Sampling) \end{tabular}}  & \multirow{2}{*}{60 seconds} &
			\textit{\textbf{Baseline}}   & 25.2\%  & 42751 seconds       \\ && \textit{\textbf{No Skip}}         & 65.2\%    & 1791 seconds     \\                \\
			&& \textit{\textbf{10 frames}}   & 60.1\%  & 791 seconds                 \\
			
			\cmidrule{3-5}
			&& \textbf{Overall:}   & \textbf{+35\%} & \textbf{54x speedup}           \\
			\bottomrule
		\end{tabular}
	\end{center}
\end{table}

\subsection{Improvement of Face Recognition Results Using Data Augmentation} 

\begin{figure}[]
	\centering
	~\includegraphics[width=0.47\textwidth]{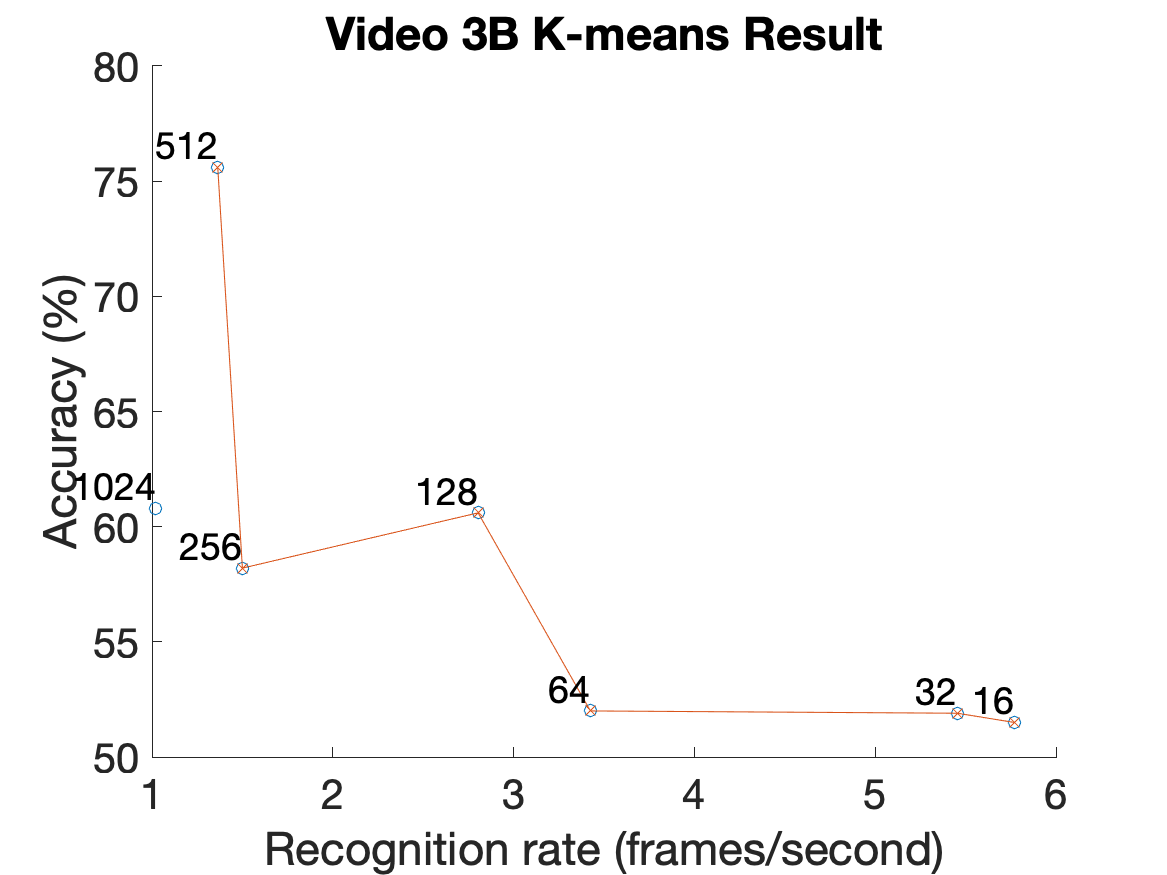}~
	~\includegraphics[width=0.47\textwidth]{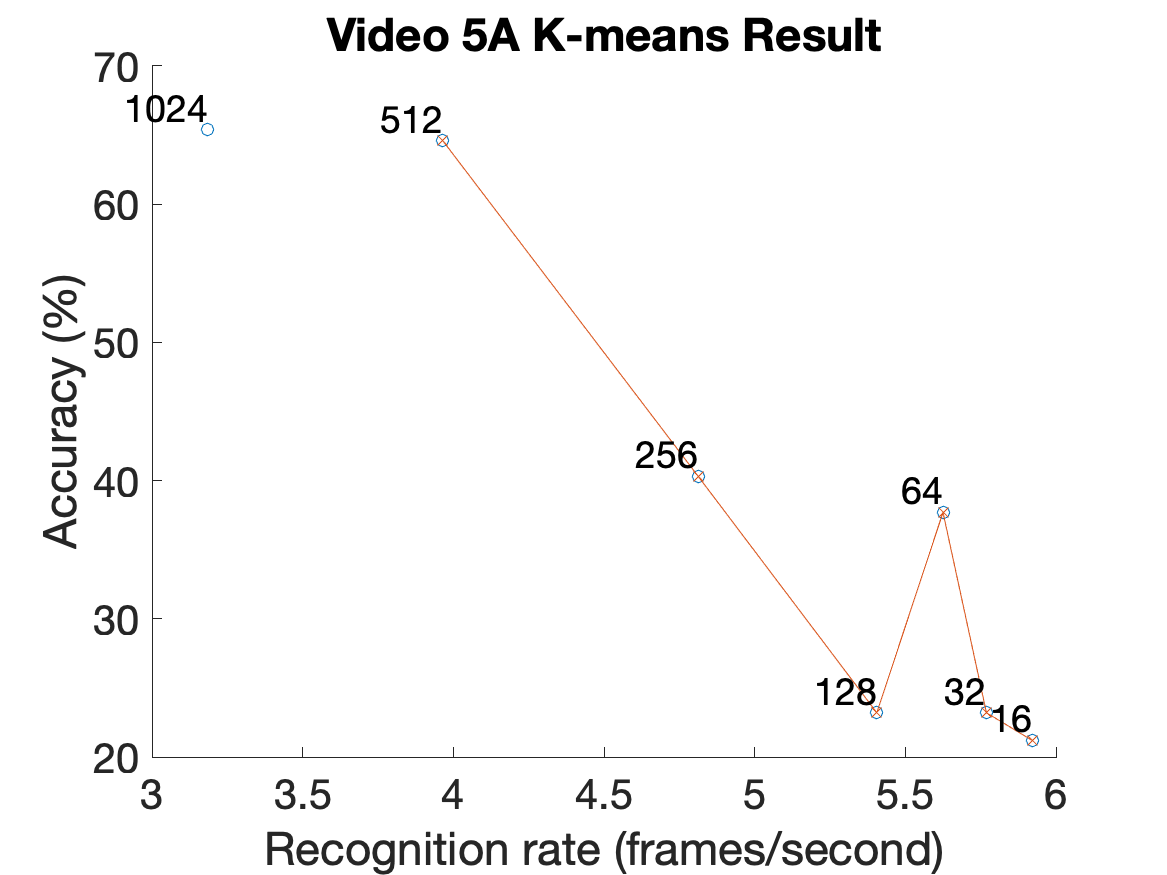}~\\[0.05 true in]
	~\includegraphics[width=0.5\textwidth]{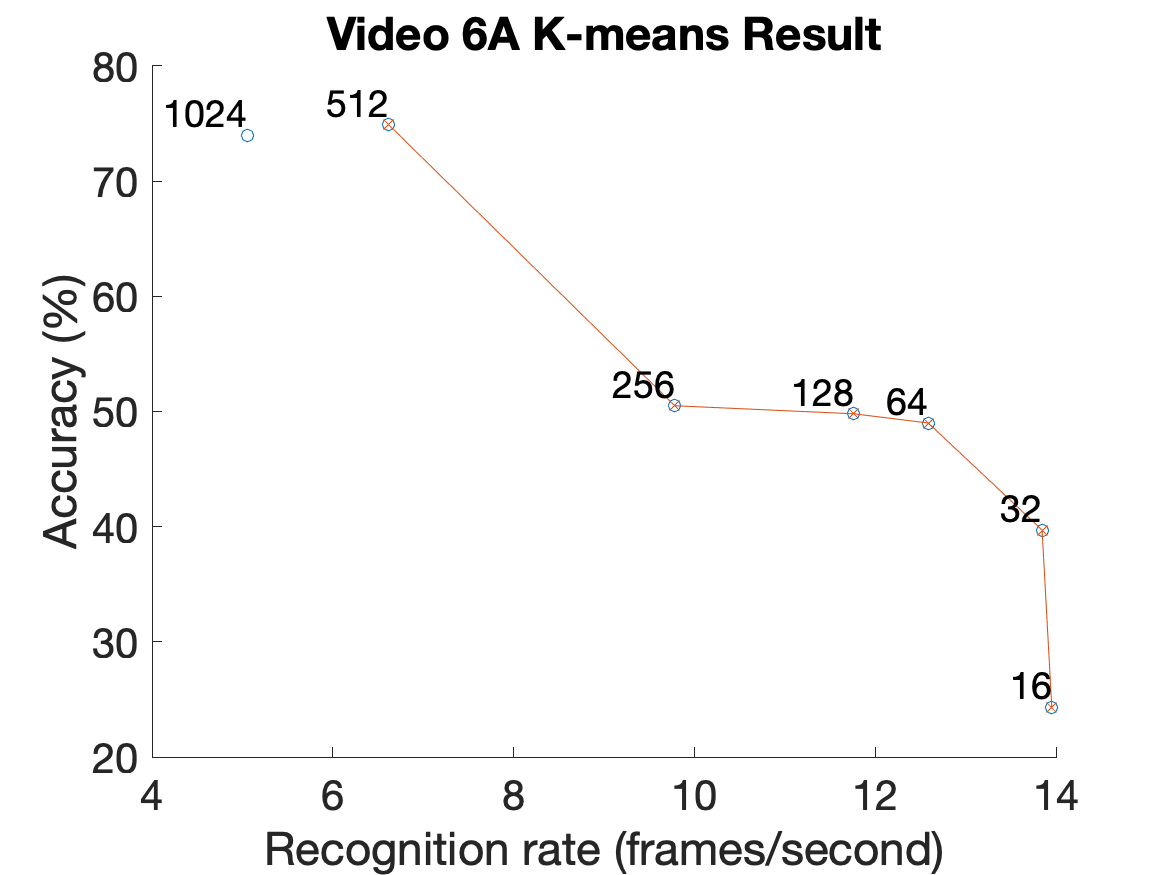}~\\[0.05 true in]
	\caption{K-means Clustering on the augmented data.} 
	\label{augmentedCluster}
\end{figure}

From the previous section, the best number of frames to skip is 10. Thus, the thesis incorporated that to run on three video clips from Group D from Rural Cohort 2 Level 1 (60 seconds), Group D from Rural Cohort 3 Level 1 (60 seconds), and Group D from Urban Cohort 3 Level 1 (10 seconds) to find best clusters. Table \ref{AugKmeans} provides the accuracy along with time results of these groups when applied k-means. Figure \ref{augmentedCluster} shows the results with multi-objective optimization with Pareto front, which shows that cluster 512 is the best.

\begin{table}[]
	\caption[Fast method optimization for the optimal number of clusters that ]{\label{AugKmeans}
		Fast method optimization for the optimal number of clusters that do not sacrifice a significant amount of accuracy.
		Each video represents a different group session segment. Repeated ID represents different short clips cut from the same video at different time.}
	\begin{center}
		\begin{tabular}{p{0.18\textwidth}p{0.13\textwidth}p{0.21\textwidth}p{0.13\textwidth}p{0.18\textwidth}}
			\toprule
			\textbf{Video}  & \textbf{Duration}                  & \textbf{Clusters}           & \textbf{Accuracy} & \textbf{Recognition Time} \\ \toprule
			\multirow{7}{*}{\begin{tabular}[c]{@{}l@{}}\textbf{6A}\\ (Fast method) \end{tabular}}  & \multirow{7}{*}{60 seconds} 
			& \textit{\textbf{16}}   & 24.3\%  & 129 seconds           \\
			&& \textit{\textbf{32}}   & 39.7\%  & 130 seconds                 \\
			&& \textit{\textbf{64}}   & 49\%  & 143 seconds                  \\
			&& \textit{\textbf{128}}  & 49.8\%  & 153 seconds                 \\
			&& \textit{\textbf{256}}   & 50.5\%  & 184 seconds                 \\
			&& \textit{\textbf{512}}   & \textbf{74.9\%}  & \textbf{272 seconds}           \\
			&& \textit{\textbf{1024}}   & 74\%  & 356 seconds                 \\
			
			\cmidrule{3-5}
			&& \textbf{Optimal:}   & \textbf{512}            \\ \midrule
			
			\multirow{7}{*}{\begin{tabular}[c]{@{}l@{}}\textbf{5A}\\ (Fast method)\end{tabular}}  & \multirow{7}{*}{60 seconds}  
			& \textit{\textbf{16}}   & 21.2\%  & 304 seconds           \\
			&& \textit{\textbf{32}}   & 23.2\%  & 312 seconds                 \\
			&& \textit{\textbf{64}}   & 37.7\%  & 320 seconds                  \\
			&& \textit{\textbf{128}}  & 23.2\%  & 333 seconds                 \\
			&& \textit{\textbf{256}}   & 40.3\% & 374 seconds                \\
			&& \textit{\textbf{512}}   & \textbf{64.6\%} & \textbf{454 seconds}           \\
			&& \textit{\textbf{1024}}   & 65.4\%  & 565 seconds                 \\

			\cmidrule{3-5}
			&& \textbf{Optimal:}   & \textbf{512}                            \\ \midrule
			
			\multirow{7}{*}{\begin{tabular}[c]{@{}l@{}}\textbf{3B}\\ (Fast method)\end{tabular}}  & \multirow{7}{*}{10 seconds} 
			& \textit{\textbf{16}}   & 51.5\%  & 52 seconds           \\
			&& \textit{\textbf{32}}   & 51.9\%  & 55 seconds                 \\
			&& \textit{\textbf{64}}   & 52\%  & 87.6 seconds                  \\
			&& \textit{\textbf{128}}  & 60.6\%  & 107 seconds                 \\
			&& \textit{\textbf{256}}   & 58.2\%  & 111 seconds                 \\
			&& \textit{\textbf{512}}   & \textbf{75.6\%}  & \textbf{251 seconds}           \\
			&& \textit{\textbf{1024}}   & 60.8\%  & 294 seconds                 \\
			
			\cmidrule{3-5}
			&& \textbf{Optimal:}   & \textbf{512}          \\ 
			
			\bottomrule
		\end{tabular}
	\end{center}
\end{table}

Next, the combination of data augmentation, sparse sampling, k-means clustering of 512, and frame rate of 10 (Fast method) was tested on the same three groups, but with different clips at different times. Tables \ref{augAcc} and \ref{augTime} showed the performance in accuracy and time of the three test clips, respectively. 

The time taken when running baseline methods for these three videos are shorter than the one in Table \ref{TimeRes} because these three ran on the Sparse Sampling dataset with fps = 30, which is 30 times less than the original data size. For video 3C, as the baseline method not converging, the thesis compared only to the thesis's approach without using k-means and frame rate optimization for with and without augmentation datasets. Thus, Table \ref{augTime} is the speedup summary on Fast, Augmented, and Face Prototypes with Sampling methods. Videos 5B and 6B with 1.5 and 1 minute long, respectively, showed an additional decrease in time with a speedup of 12x and 5.8x compared to Augmented and Face Prototypes with Sampling without consistent recognition using the Active sets. In addition, as the baseline method does not converge on the 10-minute test video, the thesis compared this using consistent recognition instead of just the baseline method. The speedup is 1.9x and 3.8x compared to the Face Prototypes with Sampling and Augmented methods, respectively. 

For accuracy, the baseline method for video 3C, where each frame is processed individually, did not run. Thus, the algorithm ran on the consistent tracking method. The accuracy for video 3C with Augmented is better than that of the Fast method. This accuracy reduction might be because of the lack of face poses (more than 512 from k-means results) or because of the intense movement of the participants (frame rate skipping of 10 is too much). However, the Fast method showed astonishing results compared to the two methods in the last two test videos. The accuracy improved by 49.4\% and 31.7\% for videos 5B and 6C with Augmented. Similarly, for Face Prototypes with Sampling, the Fast method showed much more accuracy with 20.4\% and 29.6\% for videos 5B and 6C. Overall, the accuracy improved by 17\% and 24.6\% when running on the Fast method compared to Augmented and Face Prototypes with Sampling.

\begin{table}[!t]
	\caption[Recognition times for facial recognition with Fast method ]{\label{augTime}
		Recognition times for facial recognition with Fast method compared to Augmented and Face Prototypes with Sampling methods. In this comparison, the fast method uses 512 clusters and skips 10 frames. Each video ID represents a different group session segment. Repeated ID represents different short clips cut from the same video at different time.}
	\begin{center}
		\begin{tabular}{p{0.08\textwidth}p{0.16\textwidth}p{0.19\textwidth}p{0.19\textwidth}p{0.24\textwidth}}\toprule
			{\begin{tabular}[c]{@{}l@{}}\textbf{Video}\\ \textbf{}\end{tabular}}    & {\begin{tabular}[c]{@{}l@{}}\textbf{Duration}\\ \textbf{}\end{tabular}}  & {\begin{tabular}[c]{@{}l@{}}\textbf{Fast Method}\\ \textbf{(sec/frame)}\end{tabular}}   & {\begin{tabular}[c]{@{}l@{}}\textbf{Augmented}\\ \textbf{(sec/frame)}\end{tabular}}   &
			{\begin{tabular}[c]{@{}l@{}}\textbf{{\begin{tabular}[c]{@{}l@{}}\textbf{Face Prototypes }\\ \textbf{ with Sampling}\end{tabular}}}\\
					\textbf{(sec/frame)}\end{tabular}}  \\ \toprule
			
			\textbf{3C}  & 600   &   0.06    &    0.28 & 0.19 \\
			\cmidrule{3-5} 
			&  \textbf{Speed-Up} & & \textbf{4.7x} & \textbf{3.2x} \\~\\
			\textbf{5B}          & 90& 0.11    & 1.27  & 0.64  \\
			\cmidrule{3-5}
			&  \textbf{Speed-Up}& & \textbf{12x} & \textbf{5.8x} \\~\\
			\textbf{6B}          & 60  &  0.24  & 0.91 & 0.46   \\
			\cmidrule{3-5}
			&\textbf{Speed-Up}&  &  \textbf{3.8x} & \textbf{1.9x}
			\\
			\midrule
			&{\begin{tabular}[c]{@{}l@{}}\textbf{Average}\\ \textbf{Speedup}\end{tabular}}      &           & \textbf{6.3x}         & \textbf{3.3x}    \\
			\bottomrule
		\end{tabular}
	\end{center}
\end{table}

\begin{table}[]
	\caption[Accuracy for facial recognition with Fast method]{\label{augAcc}
		Accuracy for facial recognition with Fast method compared to Augmented and Face Prototypes with Sampling methods. In this comparison, the fast method uses 512 clusters and skips 10 frames. 
		Each video represents a different group session segment. Repeated ID represents different short clips cut from the same video at different time.}
	\begin{center}
		\begin{tabular}{p{0.08\textwidth}p{0.068\textwidth}p{0.194\textwidth}p{0.109\textwidth}p{0.157\textwidth}p{0.225\textwidth}}
			\toprule
			\textbf{Video}  & \textbf{Span} &\textbf{Person Label}           & {\begin{tabular}[c]{@{}l@{}}\textbf{Fast} \\\textbf{Method} \end{tabular}} & \textbf{Augmented} &
			{\begin{tabular}[c]{@{}l@{}}\textbf{Face Prototypes} \\\textbf{ with Sampling} \end{tabular}} \\ \toprule

			\multirow{4}{*}{\begin{tabular}[c]{@{}l@{}}\textbf{3C}\\ \end{tabular}}  & \multirow{4}{*}{\begin{tabular}[c]{@{}l@{}}\textbf{10} \\\textbf{min} \end{tabular}} & \textit{\textbf{Melly77W}}          & 29\%    & \textbf{99.4\%}   & 99.4\%    \\
			&& \textit{\textbf{Marisol112W}}     & 93.3\%           &\textbf{96.8\%}    &5.3\%     \\
			&& \textit{\textbf{Cristie123W}}   &98.8\% &  \textbf{99.7\%} &          0.1\%             \\
			&& \textit{\textbf{Phuong}}    & 10.5\%                     &\textbf{56.1\%}      &31.7\%               
			\\ \cmidrule{3-6}
			&& \textbf{Average}   & 57.9\%   & \textbf{88\%}      & 34.1\%        \\ \midrule
			
			\multirow{4}{*}{\begin{tabular}[c]{@{}l@{}}\textbf{5B}\\ \end{tabular}}  & \multirow{4}{*}{\begin{tabular}[c]{@{}l@{}}\textbf{60} \\\textbf{sec} \end{tabular}} & \textit{\textbf{Herminio10P}}         &\textbf{79.5\%}    & 3.9\%  & 55.0\%       \\
			&& \textit{\textbf{Beto71P}}    & \textbf{96.2\%}                          & 31.3\% & 52.6\%                    \\
			&& \textit{\textbf{Guillermo72P}}   & \textbf{93.6}\%                       & 36.7\% & 33.8\%                     \\
			&& \textit{\textbf{Katiana73P}} & 0.1\%                      & 0.1\% & \textbf{46.7\%}                  \\
			\cmidrule{3-6}
			&& \textbf{Average}   & \textbf{67.4\%}                          & 18\%     & 47\%        \\ \midrule

			\multirow{4}{*}{\begin{tabular}[c]{@{}l@{}}\textbf{6B}\\ \end{tabular}}  & \multirow{4}{*}{\begin{tabular}[c]{@{}l@{}}\textbf{60} \\\textbf{sec} \end{tabular}} & \textit{\textbf{Ivonne}}         &\textbf{96.8\%}    & 72.3\%     & 94.7\%    \\
			&& \textit{\textbf{Katiana73P}}    & 0\%                         &\textbf{0.2}\%     & \textbf{0.2\%}                 \\
			&& \textit{\textbf{Maya108P}}   & \textbf{99.3}\%                       &8.2\%       & 21.1\%                  \\
			&& \textit{\textbf{Marcia109P}} & \textbf{58.2}\%                       &46.9\%     & 20.0\%                  \\
			\cmidrule{3-6}
			&& \textbf{Average}   & \textbf{63.6\%}                          & 31.9\%     & 34\%         \\ \midrule

			&&{\begin{tabular}[c]{@{}l@{}}\textbf{Overall}\\ \textbf{Average}\end{tabular}}                 & \textbf{63\%}         & 46.0\% & 38.4\%    \\
			
			\bottomrule
		\end{tabular}
	\end{center}
\end{table}

\subsection{Results Summary}
The thesis ran on two different sets of test videos, as summarized in Table \ref{summary}. The first set of test videos was run only on the baseline method and Face Prototypes with Sampling, which achieved an 11x speedup in time and a 9.5\% accuracy improvement. The second set ran on the Fast Method, Augmented, and Face Prototypes with Sampling. The baseline method did not converge on the second set of test videos. The lack of convergence by the baseline method may be because the second test set contained longer videos (10 minutes). The Fast method achieved a 3.3x speedup compared to Face Prototypes with Sampling and a 6.3x speedup compared to the Augmented method. The fast method improved accuracy by 17\% against the Augmented method and 24.6\% against the Face prototypes with the Sampling method.

\begin{table}[]
	\caption[Result summary for face recognition.]{\label{summary}
		Result summary for face recognition. Each row was run on a different set of test videos.}
	\begin{tabular}{cccccc}
		\hline
		\textbf{}  & \textbf{Fast} & \textbf{Augmented} & \textbf{\begin{tabular}[c]{@{}c@{}}Face Prototypes\\ with Sampling\end{tabular}} & \textbf{Baseline} & \textbf{Table} \\ \hline
		\multirow{2}{*}{\textbf{Accuracy}} & N/A           & N/A                & 71.8\%                                                                           & 62.3\%            & 5.2, 5.3       \\
		& 63\%          & 46\%               & 38.4\%                                                                           & N/A               & 5.9            \\ \hline
		\multirow{2}{*}{\textbf{\begin{tabular}[c]{@{}c@{}}Time\\ (sec/frame)\end{tabular}}}    & N/A           & N/A                & 0.8                                                                              & 8.9               & 5.4            \\
		& 0.13          & 0.82               & 0.43                                                                             & N/A               & 5.8            \\ \hline
	\end{tabular}
\end{table}

\begin{figure*}[!t]
	\centering
	(a)~\includegraphics[width=0.25\textwidth]{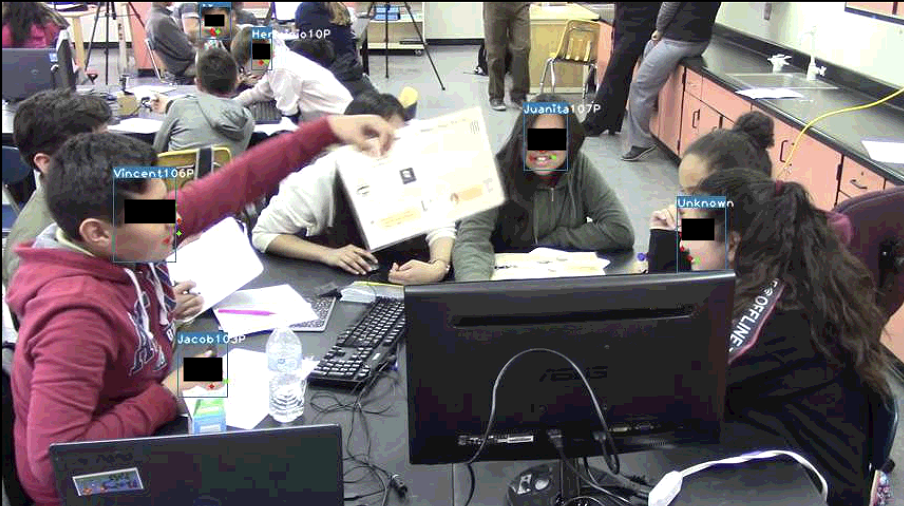}~
	(b)~\includegraphics[width=0.25\textwidth]{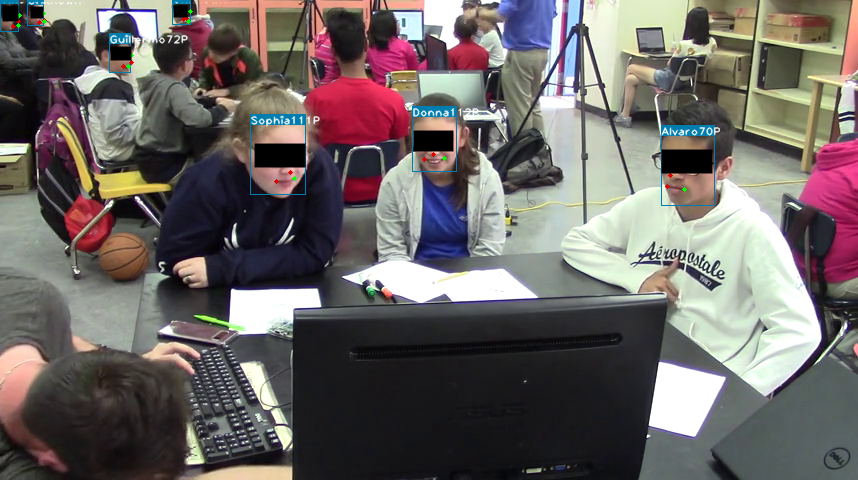}~
	(c)~\includegraphics[width=0.25\textwidth]{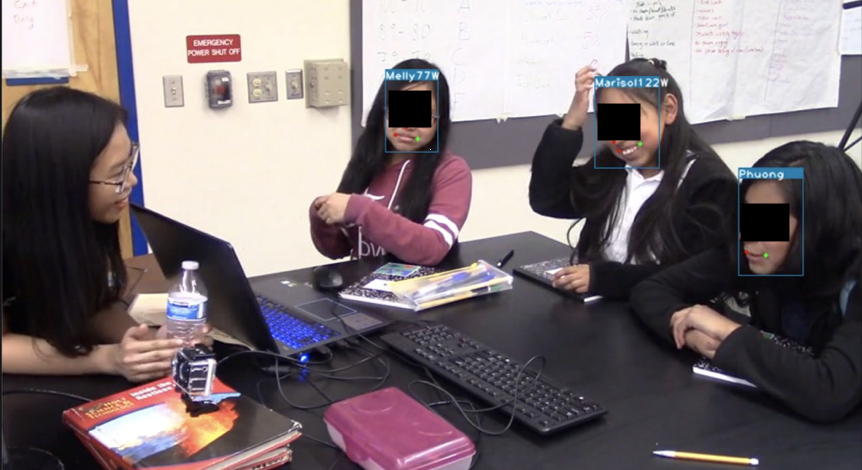}~\\[0.05 true in]
	(d)~\includegraphics[width=0.25\textwidth]{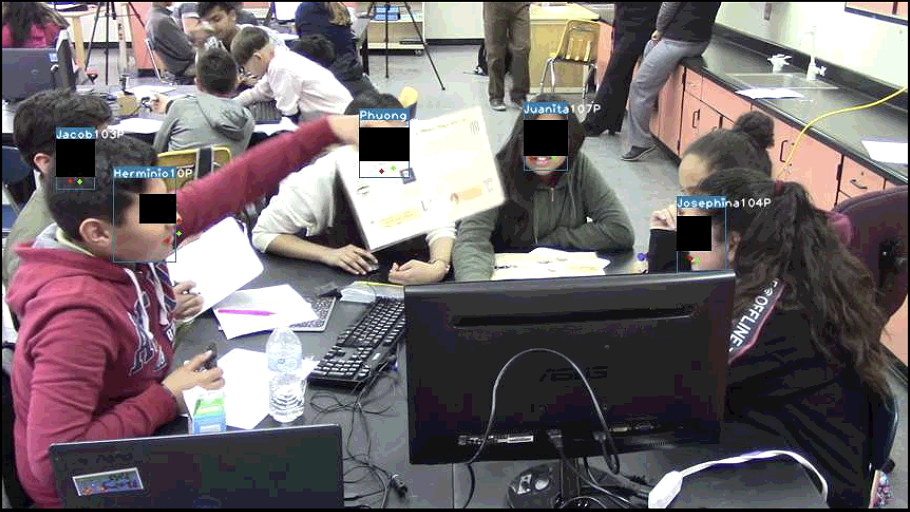}~
	(e)~\includegraphics[width=0.25\textwidth]{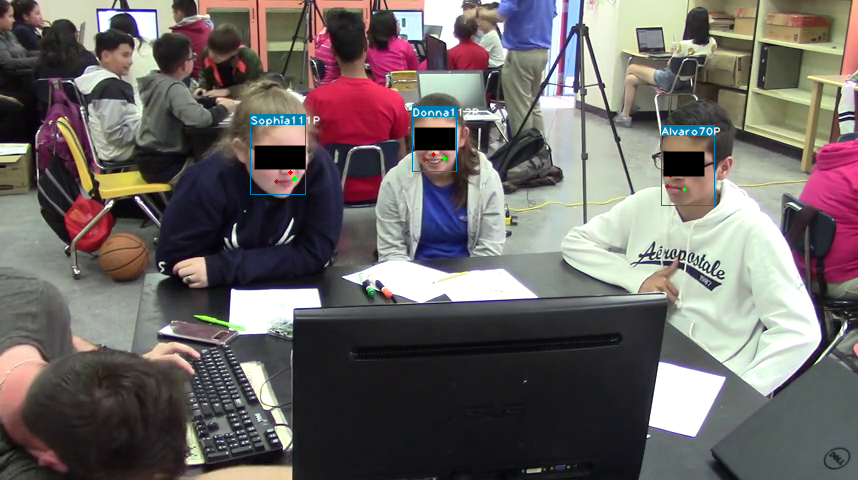}~
	(f)~\includegraphics[width=0.25\textwidth]{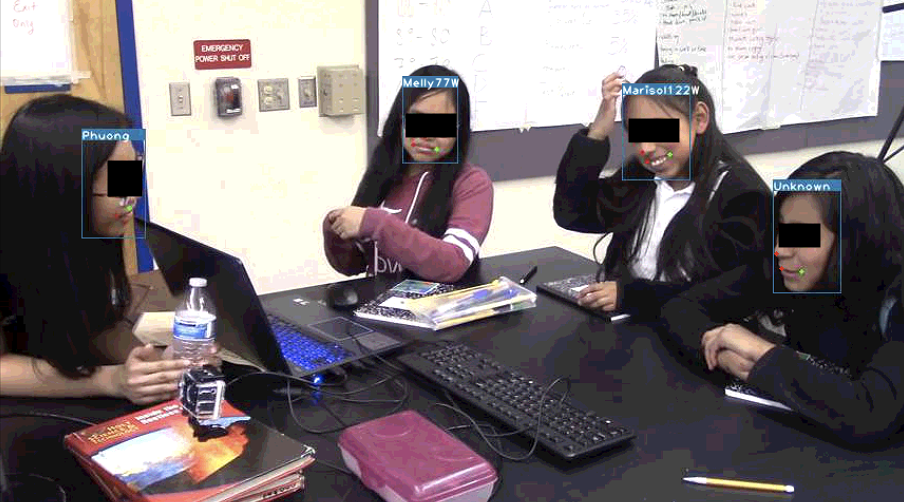}~\\[0.05 true in]
	(g)~\includegraphics[width=0.25\textwidth]{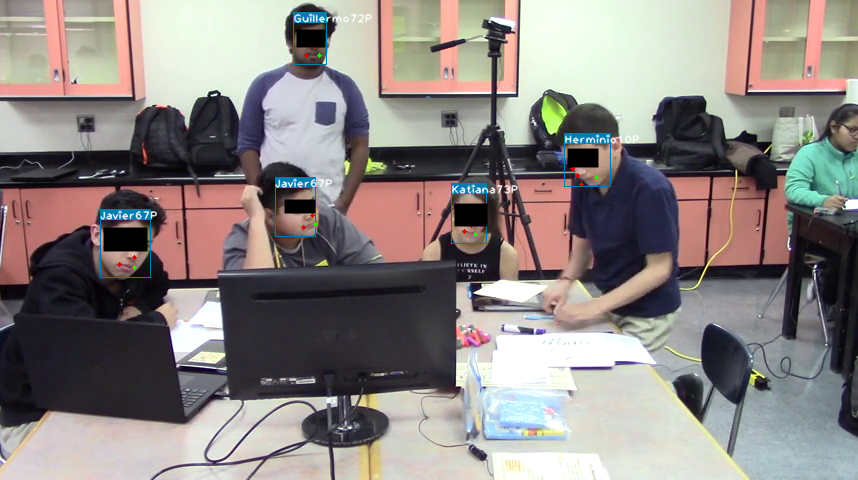}~
	(h)~\includegraphics[width=0.25\textwidth]{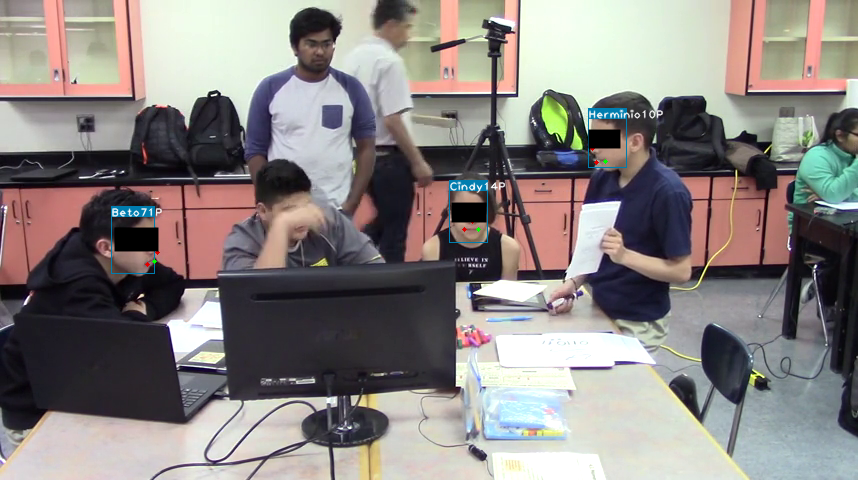}~
	(i)~\includegraphics[width=0.25\textwidth]{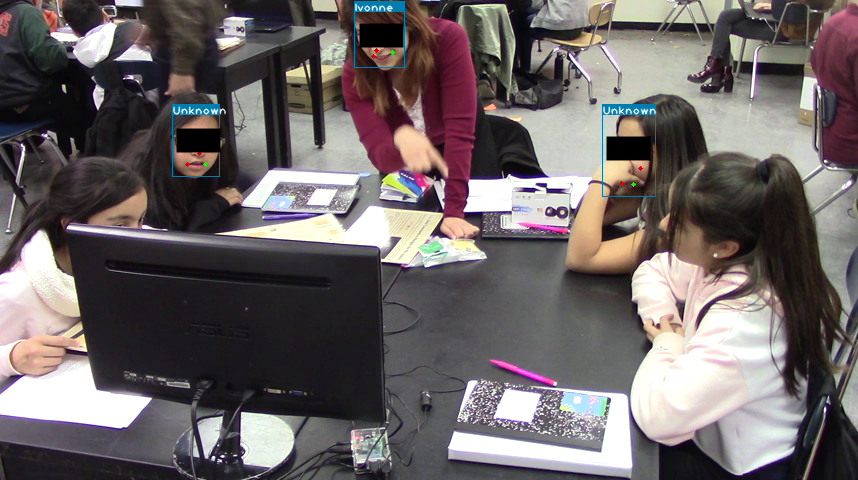}~\\[0.05 true in]
	(j)~\includegraphics[width=0.25\textwidth]{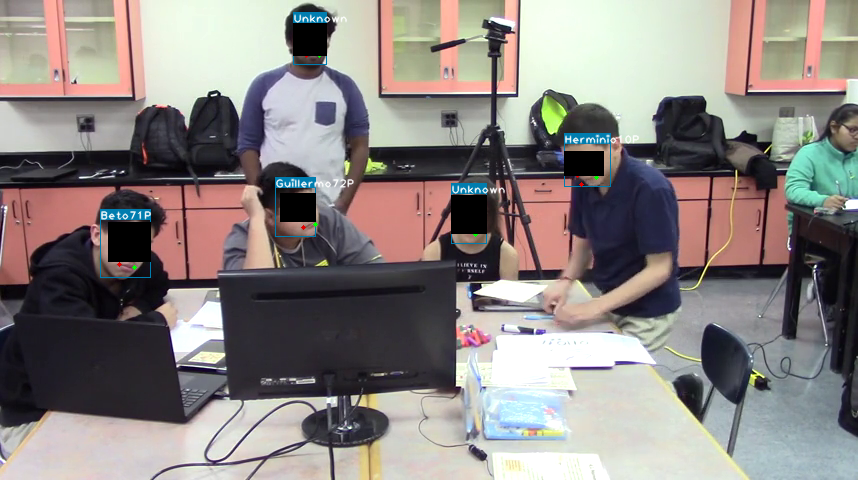}~
	(k)~\includegraphics[width=0.25\textwidth]{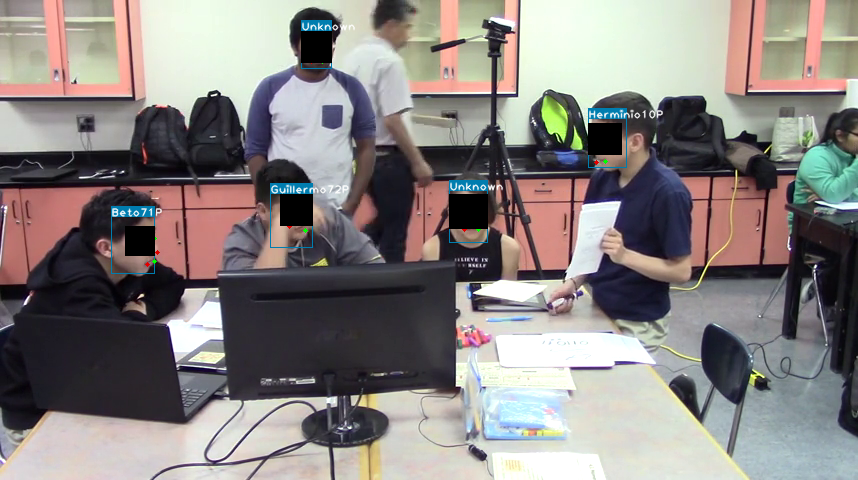}~
	(l)~\includegraphics[width=0.25\textwidth]{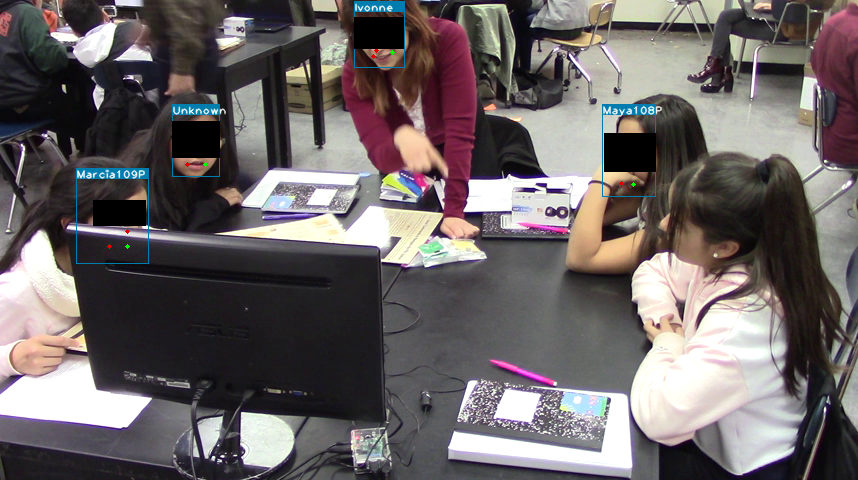}~\\[0.05 true in]
	(m)~\includegraphics[width=0.25\textwidth]{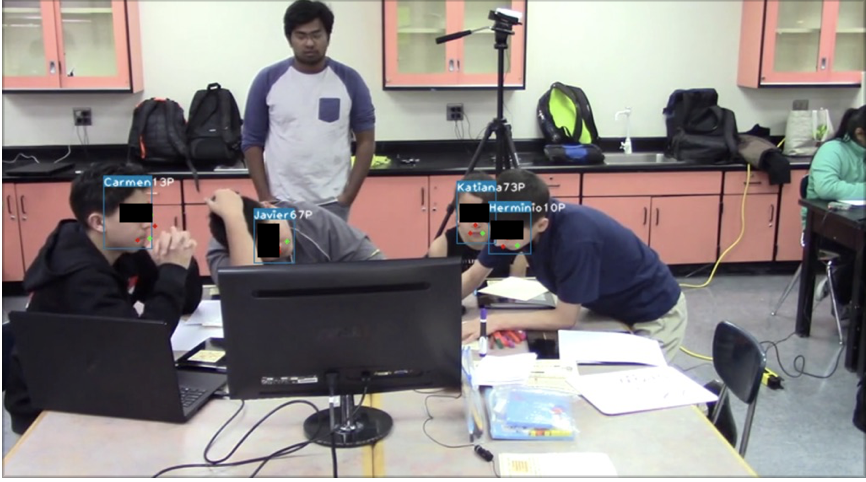}~
	(n)~\includegraphics[width=0.25\textwidth]{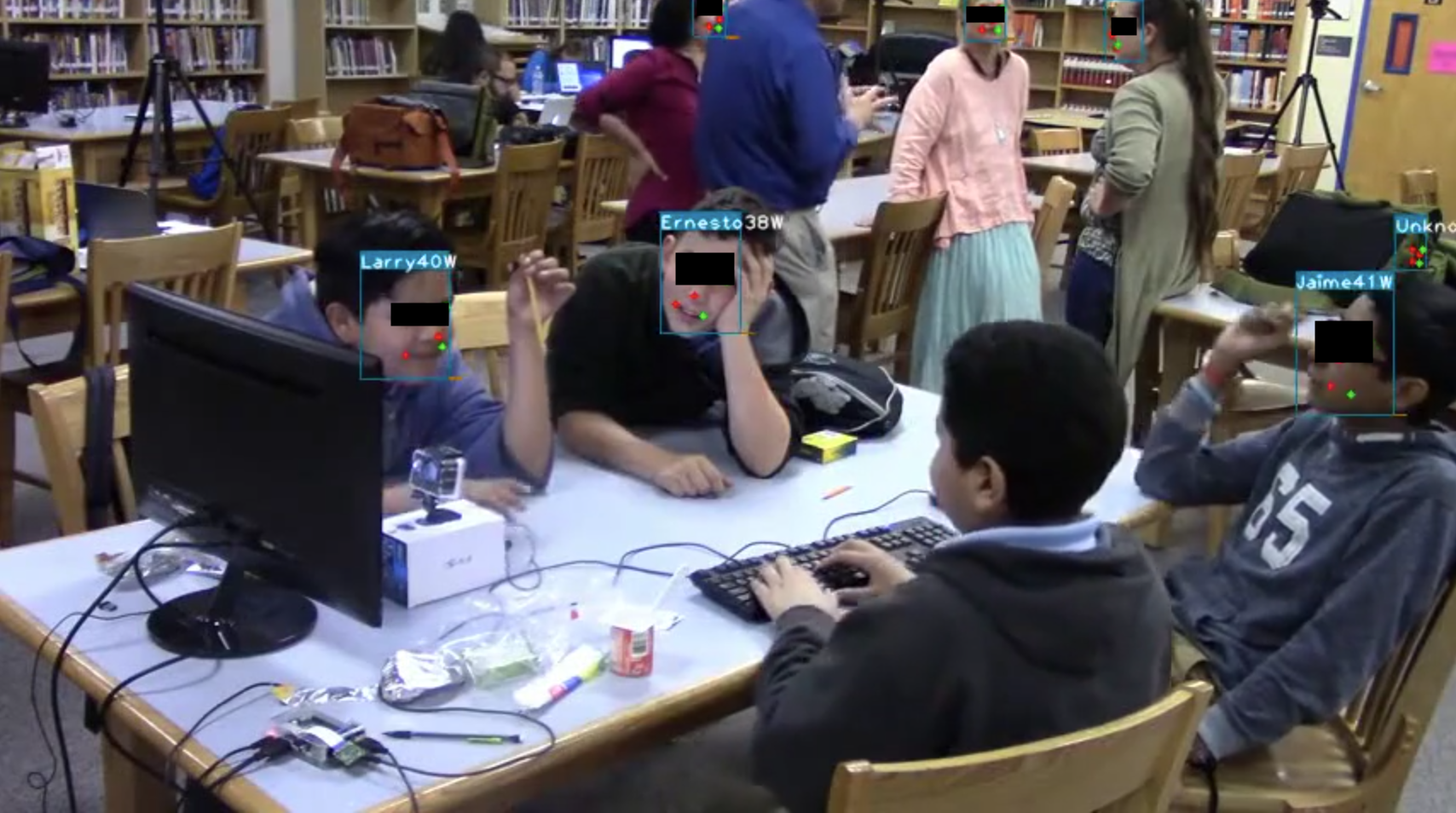}~
	(o)~\includegraphics[width=0.25\textwidth]{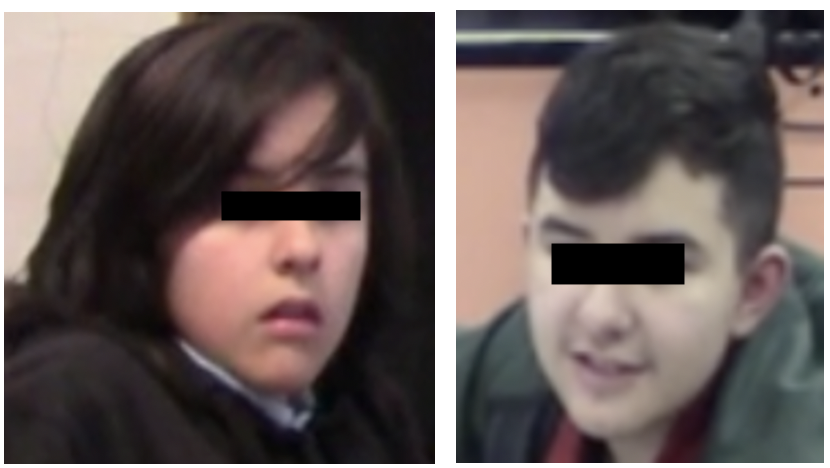}~\\[0.05 true in]
	\caption[Video face recognition results for three collaborative groups.]{
		Video face recognition results for three collaborative groups.
		The first row shows results from the use of InsightFace (baseline).
		The second row shows the results using the sampling method.
		In (k) and (l), we show successful detections despite 
		occlusions.
		Results from the use of K-means Clustering are shown in (n).
		Then, we show dramatic changes in appearance in (o).
	} 
	\label{fig:Interaction_sample}
\end{figure*}

The comparison between the baseline and our approach is shown in Figure \ref{fig:Interaction_sample}. The baseline method could not detect Phuong with a completely covered face by a sheet of paper in the baseline method, whereas our algorithm recognized Phuong correctly using consistent video tracking (see a and d for Video 7 frame example). The second example is taken from Video 4 (see b and e). The baseline method correctly identified all three people. However, the baseline method also detected and incorrectly claimed recognition of background participants that we did not train. Our proposed methods used projection and small-area elimination to reject these false-positives. 

A third example is from Video 3 (see c and f). The baseline method only recognized Melly77W (pink sweater) and wrongly recognized Cristie123W (lower right) as Phuong, wearing glasses on the far left. Our method used historical information to address the partial occlusion issue and correctly recognized Phuong, who is in the far left of Figure \ref{fig:Interaction_sample}(f). Furthermore, the algorithm rejected the wrong assignment of Phuong because it does not allow the assignment of the same identifier to two different faces.
Instead, the wrong assignment was re-assigned to Unknown. Figure \ref{fig:Interaction_sample}(m) shows a fourth example of the thesis's method. We can see that the algorithm works in occlusion cases. Herminio10P (dark blue polo, right) and Guillermo72P (blue T-shirt) were correctly recognized even though their faces were partial. 

The algorithm also succeeded in identifying occluded faces, whereas the baseline failed to do. Guillermo72P (second from the left) had his arm covering his face, but we detected him correctly. In addition, Marcia109P had her face covered by the monitor, but the algorithm caught it as well (see k and l in Figure \ref{fig:Interaction_sample}).
We also present challenges in Figure \ref{fig:Interaction_sample}.  Antone39W did not get recognized because he had his back facing the camera where as Kirk28P was not recognized due to significant changes in appearance through time (see n and o).

\chapter{Conclusion and Future Work}
\section{Conclusion} 
The thesis presented a new method for video face recognition that is significantly faster and more accurate than the baseline method. This thesis has introduced (i) clustering methods to identify image clusters for recognizing faces from different poses, (ii) robust tracking with multi-frame processing for occlusions, (iii) multi-objective optimization and frame skipping to reduce recognition time, (iv) data augmentation to increase the size of the training dataset, especially for students that did not have enough training samples, and (v) DeID faces method in digital videos for protecting the identity of the participants.

Compared to the baseline method, the final optimized method resulted in speedy recognition times with significant improvements in face recognition accuracy. The proposed method achieved an accuracy of 71.8\% compared to 62.3\% for the baseline system while running 11.6 times faster than the baseline with Face prototype with Sampling only. The Fast method achieved a 3.3x speedup compared to Face Prototypes with Sampling and a 6.3x speedup compared to the Augmented method. The Fast method improved accuracy by 17\% compared to the Augmented method and 24.6\% compared against Face Prototypes with Sampling.

\section{Future Work}
In future work, the aim is to extend the thesis's approach to all 150 participants in about 1,000 hours of videos and process the participants in AOLME83. 

The thesis assigned the face with a higher distance to Unknown for the same label assigned to two different faces. We want to improve on this to retrieve the correct face. After rejecting, a proposed method is to try face recognition again on the second closest face.

\begin{figure}[]
	\centering
	\includegraphics[width=1\textwidth]{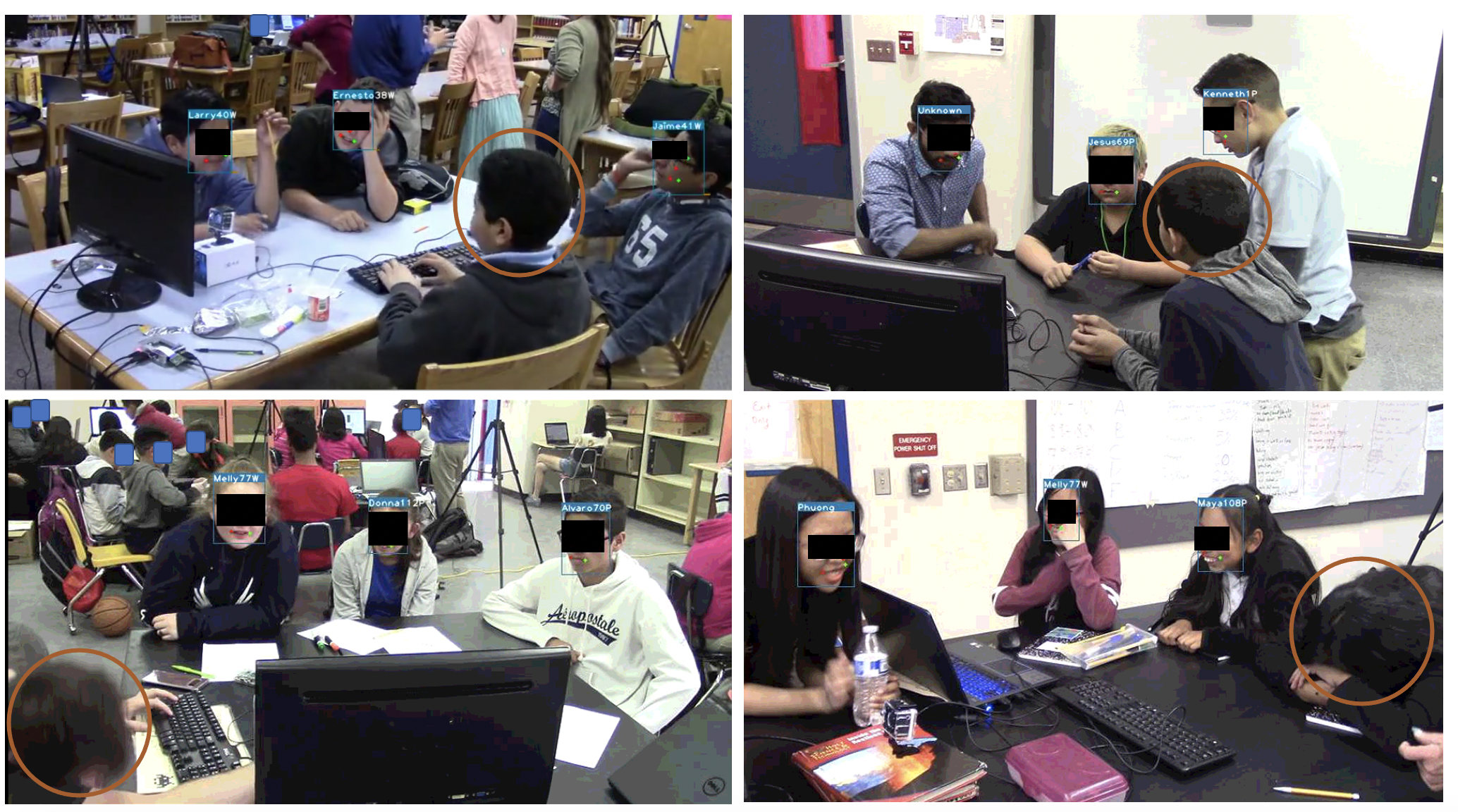}
	\caption[Person recognition problem for extending the fast video face recognition ]{Person recognition problem for extending the fast video face recognition method proposed in this thesis. The unrecognized faces are marked as unknown. The undetected persons are marked using orange circles. The DeID method draws the black-filled rectangles to protect the identities of the recognized students. The blue-filled rectangles were manually drawn to protect the identities of the out-of-group students.}
	\label{backOfHead}
\end{figure}

\begin{figure}[]
	\centering
	\includegraphics[width=1\textwidth]{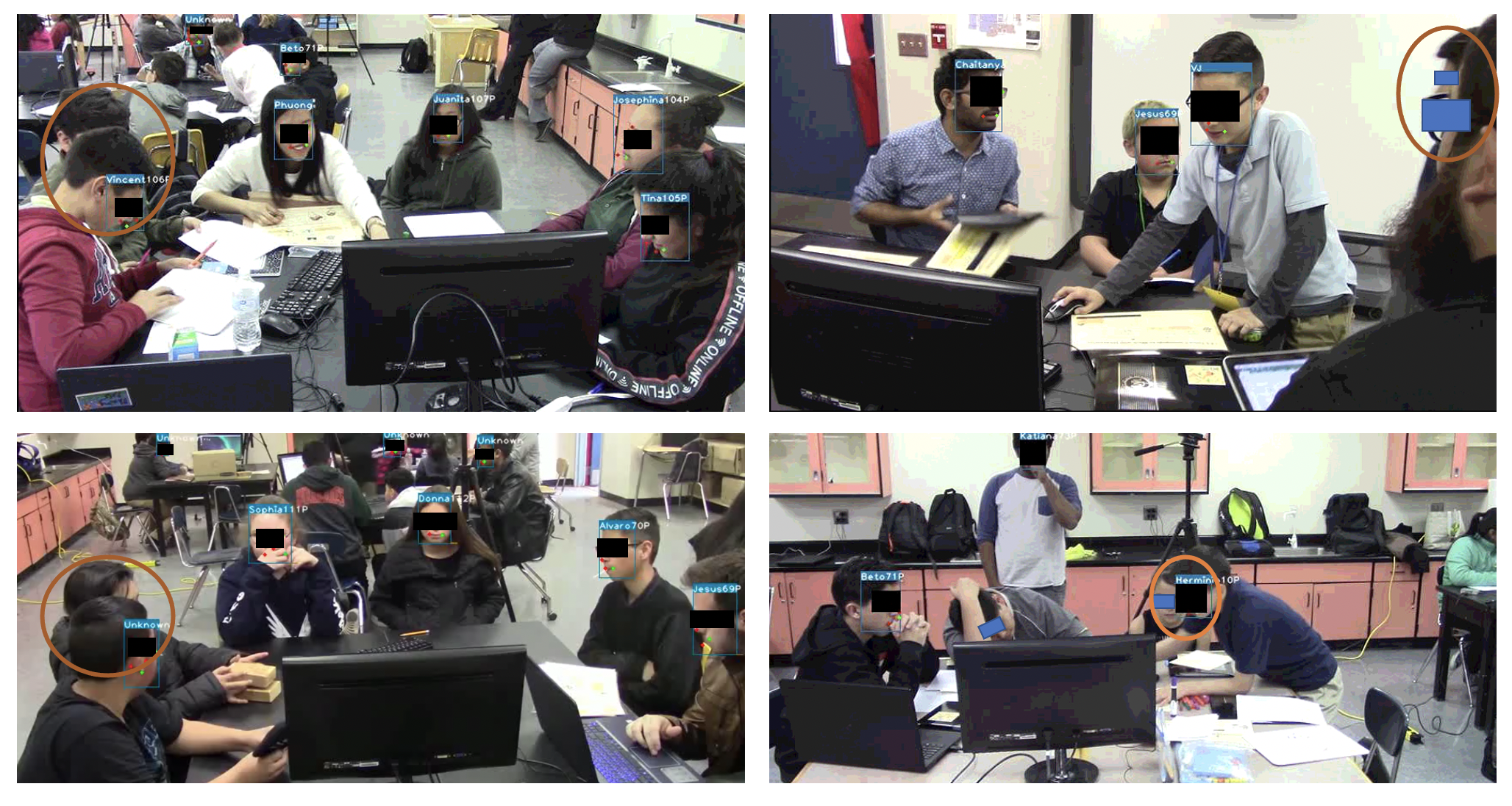}
	\caption[Face occlusion examples where a participant's face is occluded by another participant.]{Face occlusion examples where a participant's face is occluded by another participant. The fast video recognition algorithm draws the black-filled boxes to protect the identities of the participants. The blue-filled boxes were manually drawn to protect the identities of the out-of-group participants. The orange circles are drawn over the occluded faces that need to be considered in future work. }
	\label{covered}
\end{figure}

We also want to improve on the back-of-the-head detection and recognition as shown in Figure \ref{backOfHead} in cases of mostly covered faces of participants. Thus, future work would implement methods that can improve back-of-the-head recognition rates to combine with the current methods to achieve better results. 
Lastly, we want to further improve on cases when multiple students are covering each other, as shown in Figure \ref{covered}. The thesis would like to improve using overlapping areas based on our method's tracked bounding boxes.

\addcontentsline{toc}{chapter}{Appendices}
\appendix

\chapter{Participation Maps}
\begin{figure}[]
	\centering
	\includegraphics[width=1\textwidth]{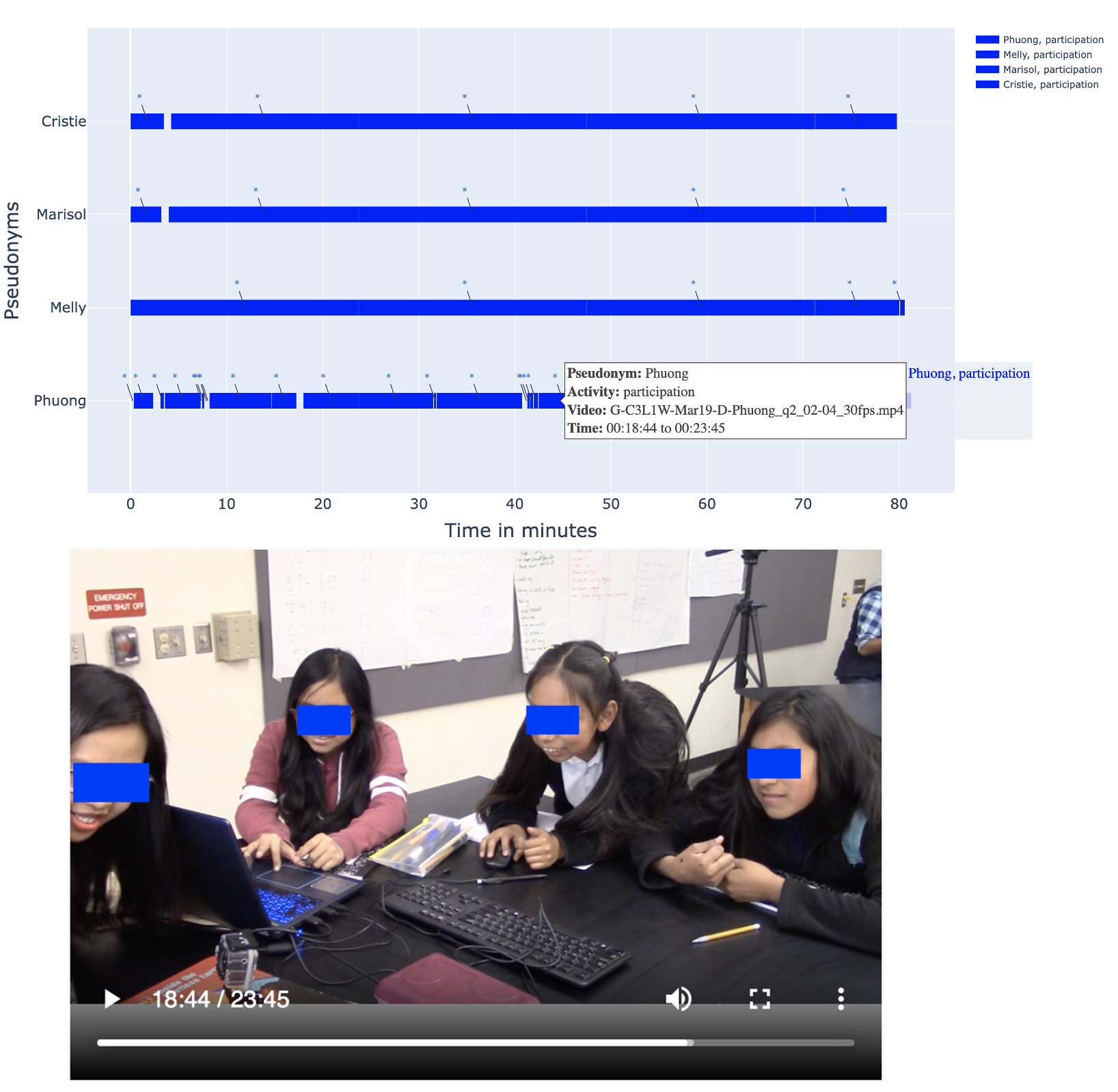}
	\caption[Participation Map Example of Group D, C3L1 on March 19, 2019.]{Participation Map Example of Group D, C3L1 on March 19, 2019.}
	\label{partMap1}
\end{figure}

\begin{figure}[]
	\centering
	\includegraphics[width=1\textwidth]{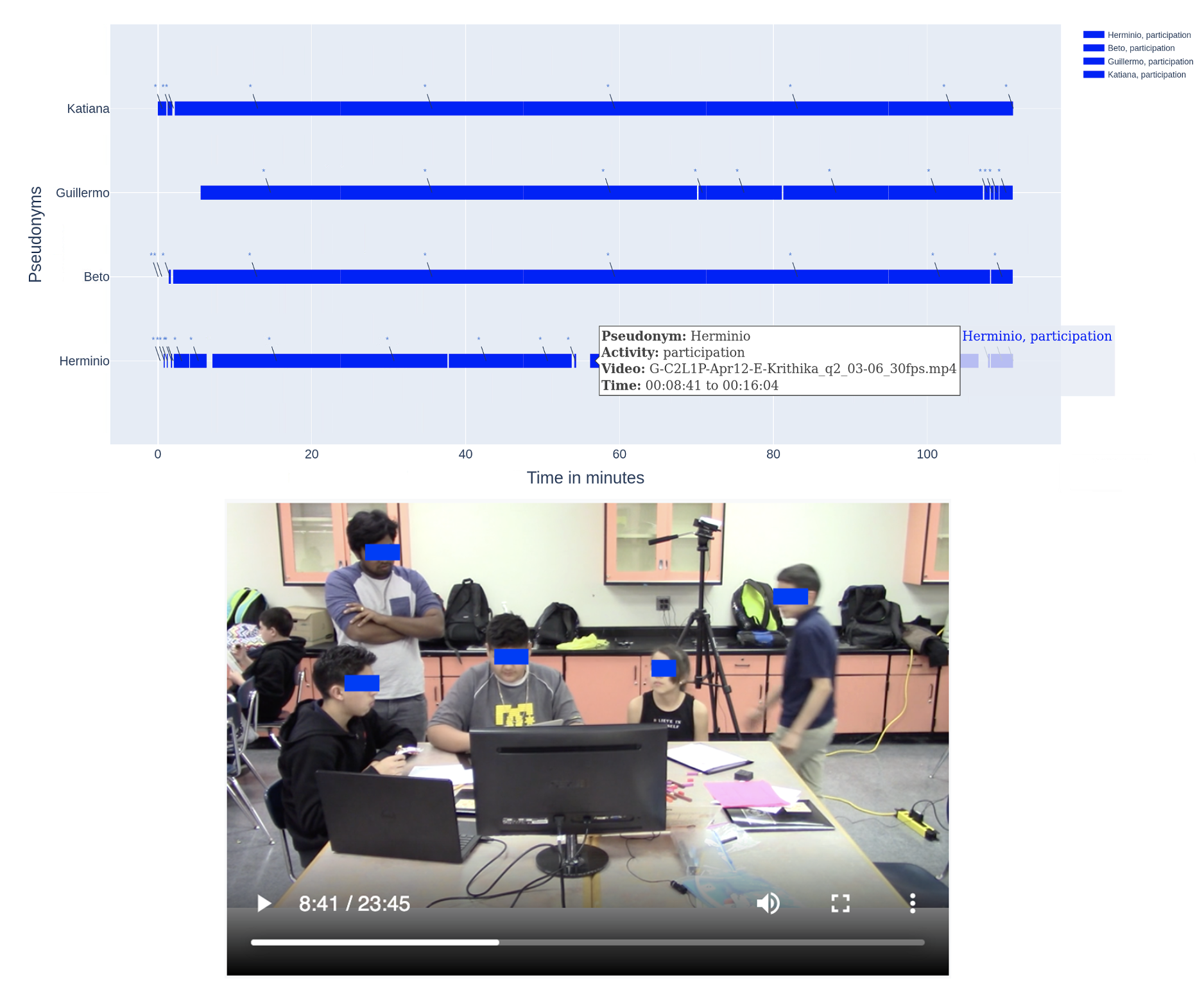}
	\caption[Participation Map Example of Group E, C2L1 on April 12, 2018]{Participation Map Example of  Group E, C2L1 on April 12, 2018.}
	\label{partMap2}
\end{figure}

The appendix presents the participation maps of the AOLME groups as one of the goals of this thesis is to assist educational researchers with fast access to the students' participation and how often they appear in a session. More specifically, the participation maps are the representation of how long a student from a specific collaborative group stays within the camera range. Figure \ref{partMap1} and \ref{partMap2} show two examples of Group D from Cohort 3 Level 1 on March 19, 2019 and Group E from Cohort 2 Level 1 on  April 12, 2018. This participation map allows researchers to see at what specific time did the student leave and come back. The blue blocks show that the students are present at the current time, whereas the gaps in between mean that the student has left or they are not visible within the camera range. In addition to visualizing the time range, this map is embedded as a hyperlink. Thus, it will also let the user hover over the time range, and if the user clicks on the asterisks on top of each student, it will redirect to the AOLME website that contains that specific video at that time. 

\chapter{DeID faces}
The appendix presents a method for DeID the participants after the fast video recognition algorithm has recognized them. The goal here is to protect the privacy of the student participants. More specifically, the DeID method protects the use of the data in bio-metric or surveillance applications. For each face found, a box is put around the eyes to avoid using the images in bio-metric applications. The method uses eye detection of the left and right eyes to define a bounding box that covers the eyes and the surrounding regions. Figure \ref{DeID} shows the outputs of the detected faces with their eyes covered. The top row shows an example of using external blurring tools. The bottom row shows this thesis's method that automatically draws the bounding boxes around the faces as they get detected. Algorithm \ref{DeID} provides the process of achieving this.

\begin{algorithm}[H]
	\SetAlgoLined
	\textbf{Input:}\\
	$\quad$ video frame.\newline
	\textbf{Output:}\\
	$\quad$ video frame with a boxes covering participants' eyes .\newline
	\textbf{for} \text{each detected participant in frame} \\
	\quad$\textbf{Get} \text{ coordinates, features, and landmarks from face detection method}$ \\
	\quad$\textbf{Find} \text{ left eye and right eye coordinates from landmarks}$\\
	\quad$\textbf{Put} \text{ a black-filled rectangle centered on both eyes.}$\\
	\textbf{end}
	\label{DeID}
	\caption{Face DeID Algorithm.}
\end{algorithm}

\begin{figure}[!t]
	\centering
	\includegraphics[width=1\textwidth]{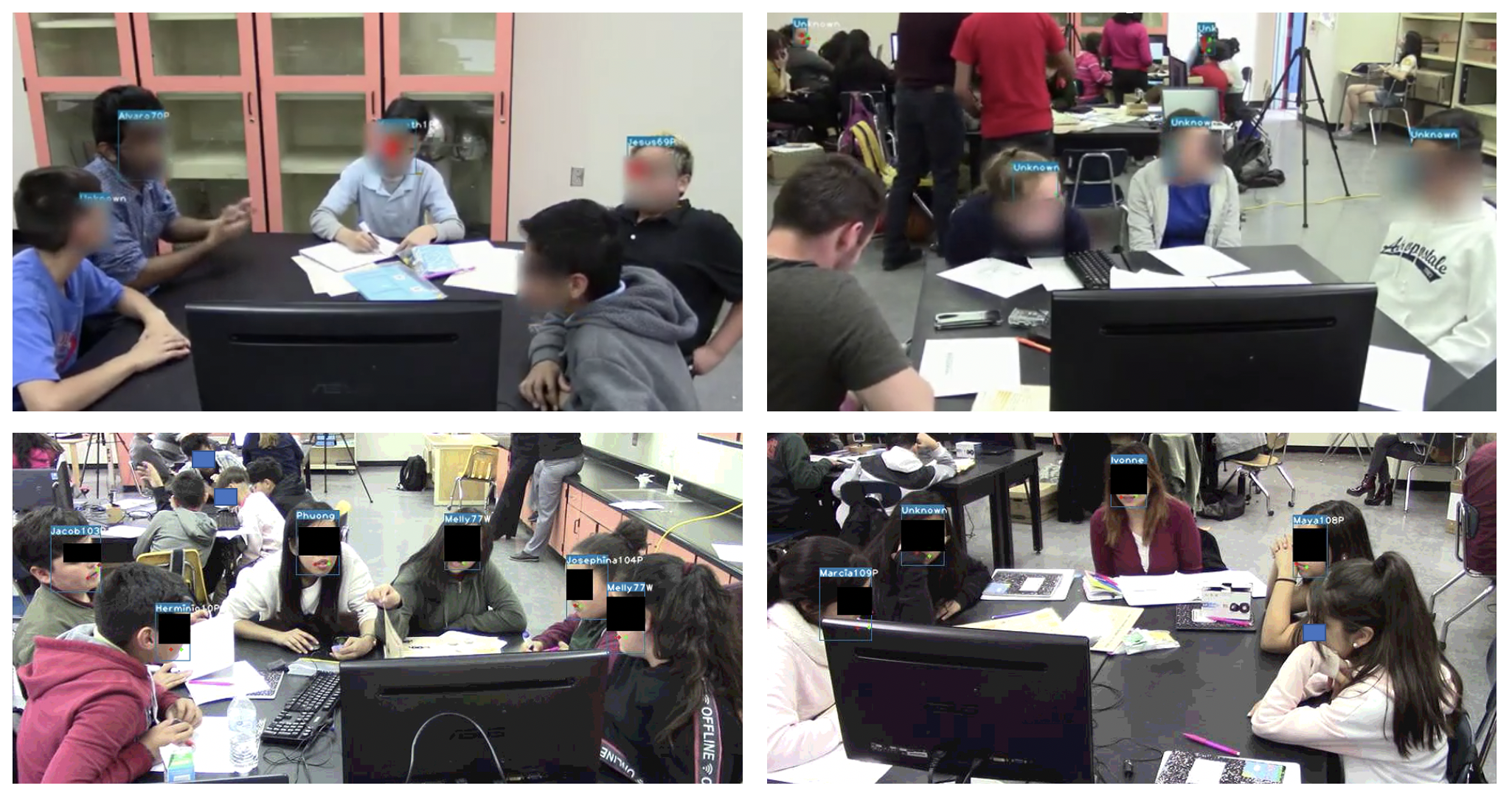}
	\caption[Face images before and after applying DeID.]{Face images before and after applying DeID. The blue bounding boxes are manually drawn to protect the participants' identities as DeID does not work if a face is not detected.} 
	\label{DeID}
\end{figure}

\chapter{AOLME Dataset}
\begin{figure*}[!t]
	\centering
	\includegraphics[width=.9\textwidth]{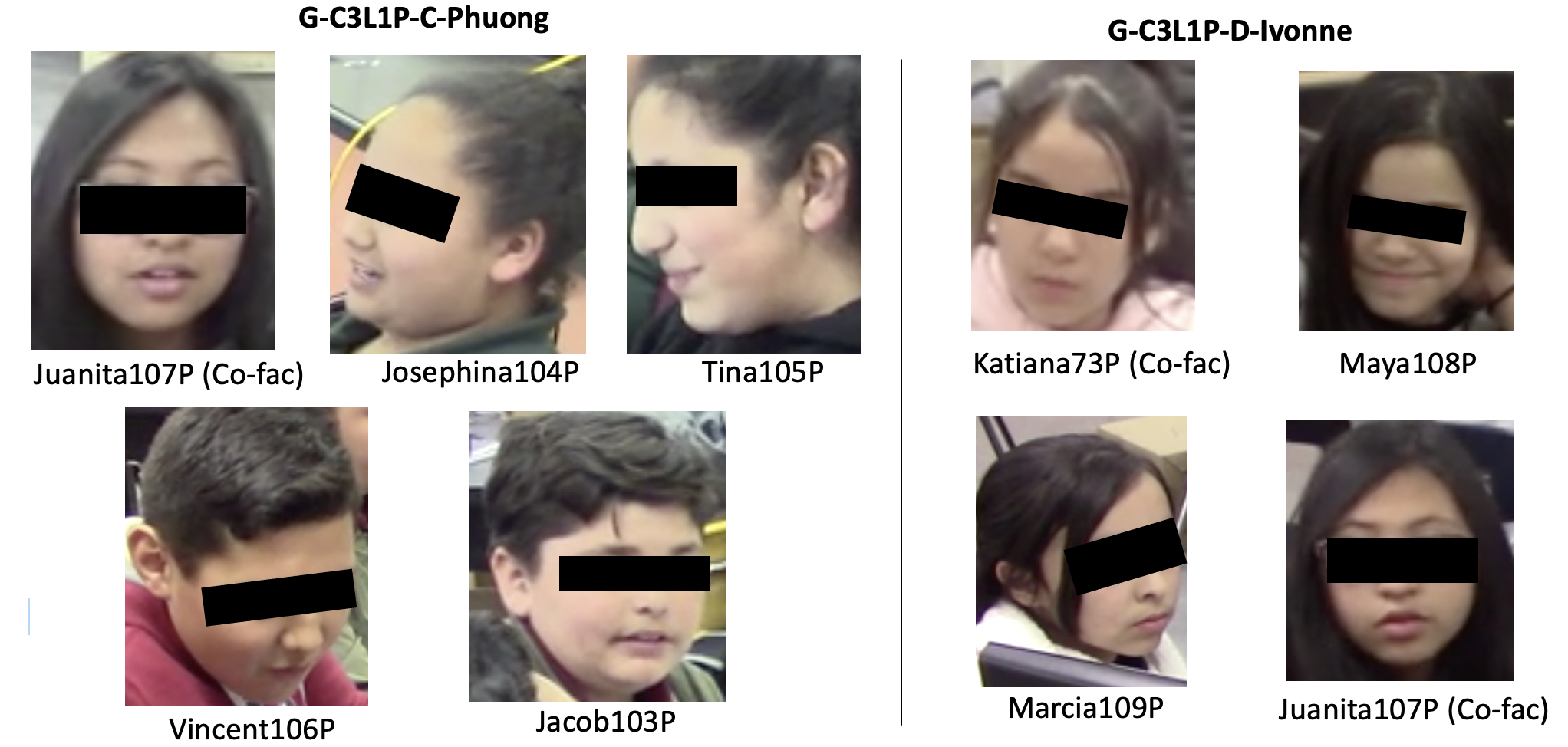}~\\[0.05 true in]
	\includegraphics[width=.9\textwidth]{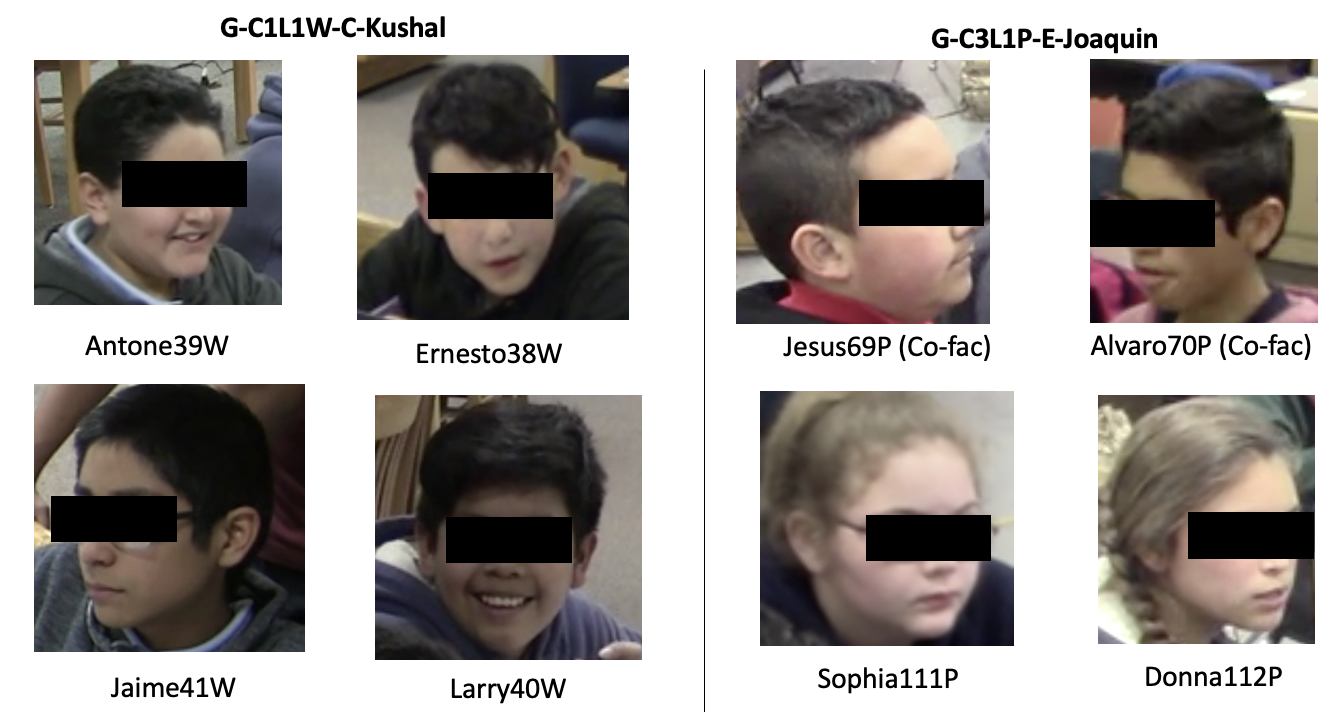}~\\[0.05 true in]
	\includegraphics[width=.85\textwidth]{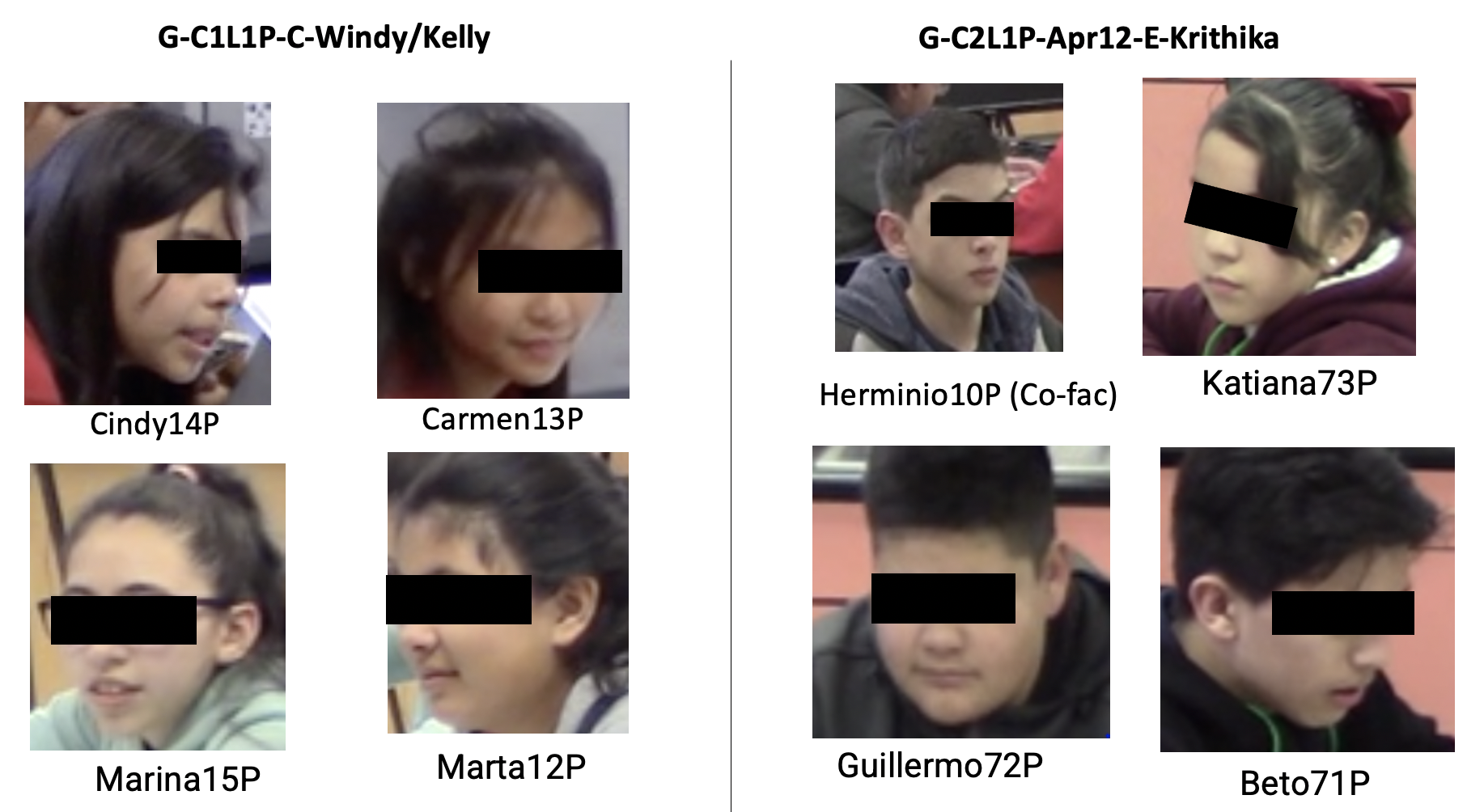}~\\[0.05 true in]
	\caption{
		AOLME32 dataset (I of II).} 
	\label{fig:trainedDatabase}
\end{figure*}

\begin{figure*}[!t]
	\centering
	\includegraphics[width=1\textwidth]{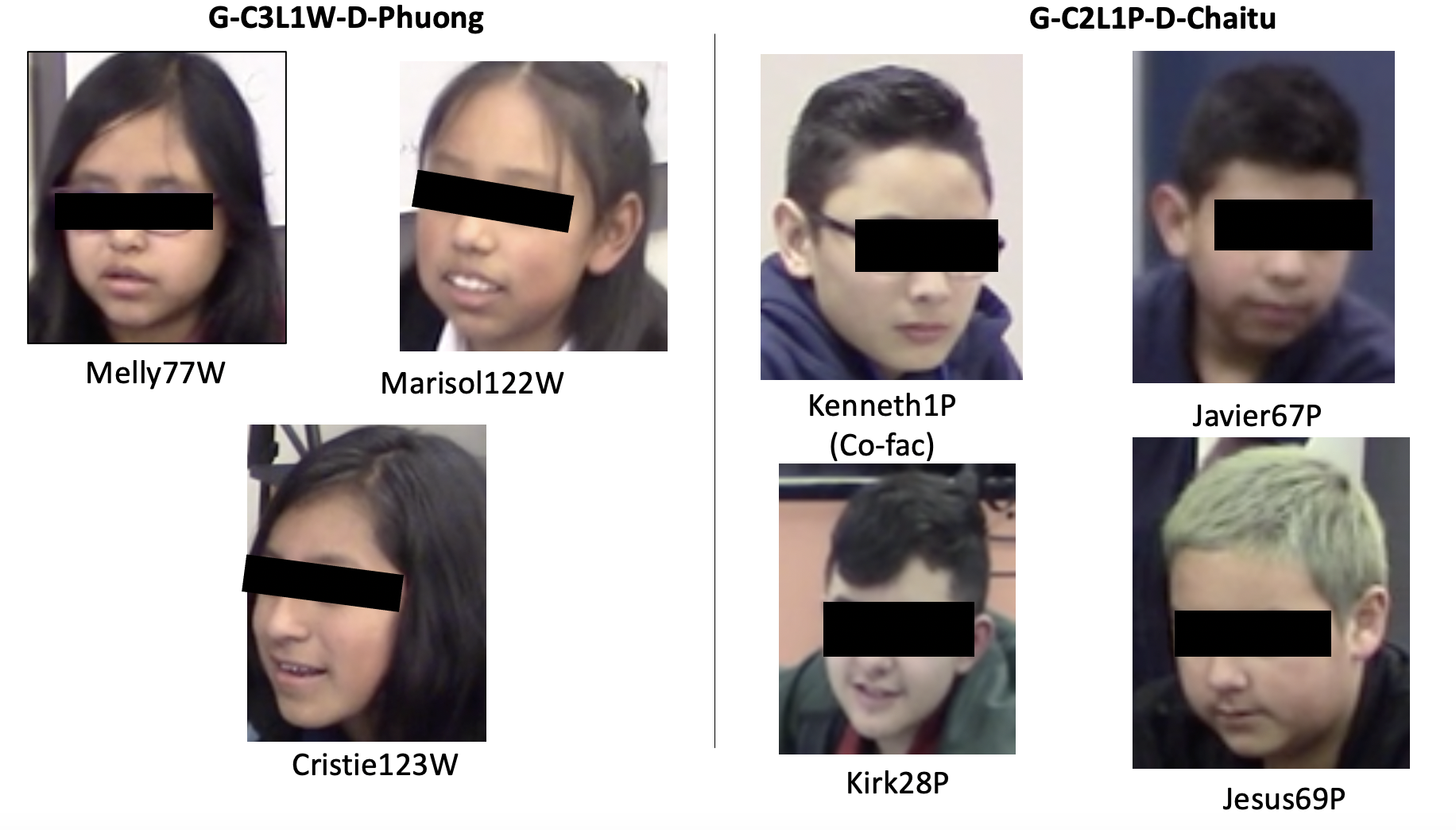}~\\[0.05 true in]
	\includegraphics[width=1\textwidth]{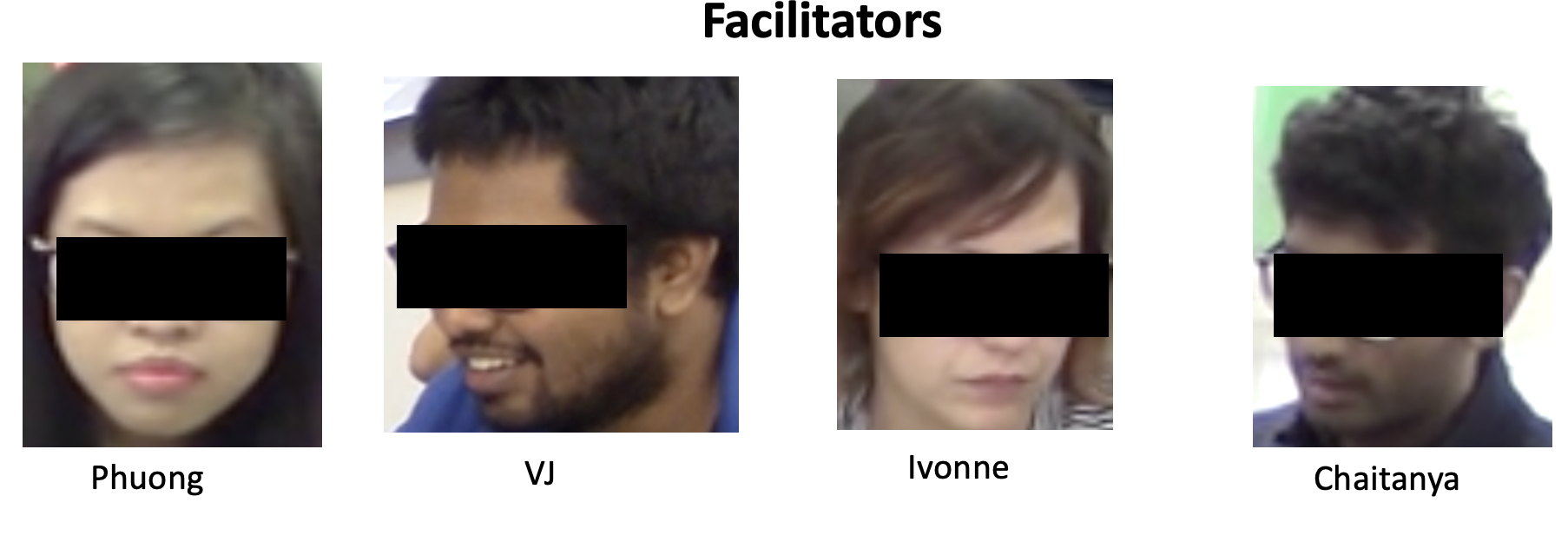}~\\[0.05 true in]
	\caption{AOLME32 dataset (II of II).}
	\label{fig:trainedCont}
\end{figure*}

\begin{figure*}[!t]
	\centering
	\includegraphics[width=.9\textwidth]{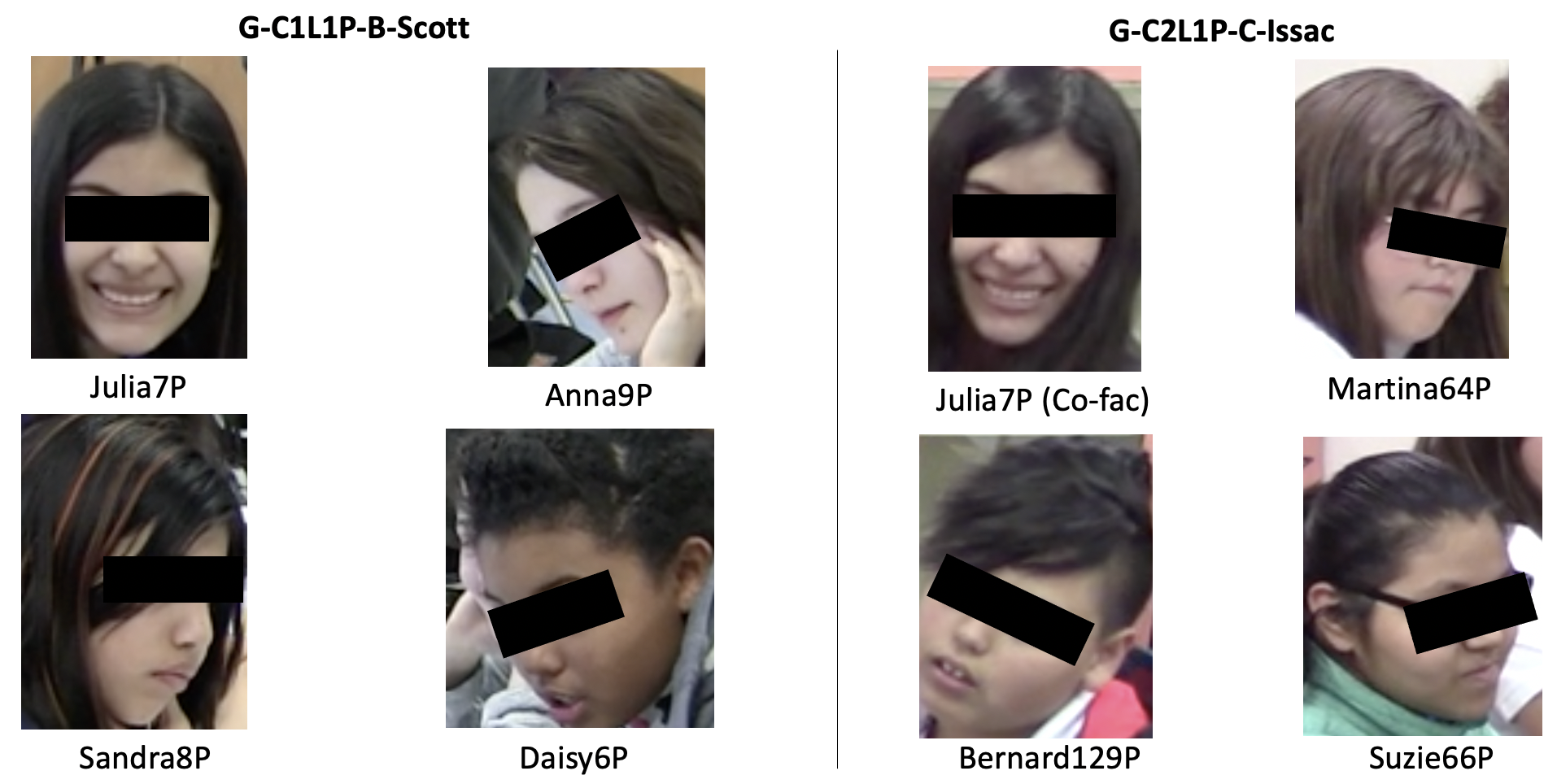}~\\[0.05 true in]
	\includegraphics[width=1\textwidth]{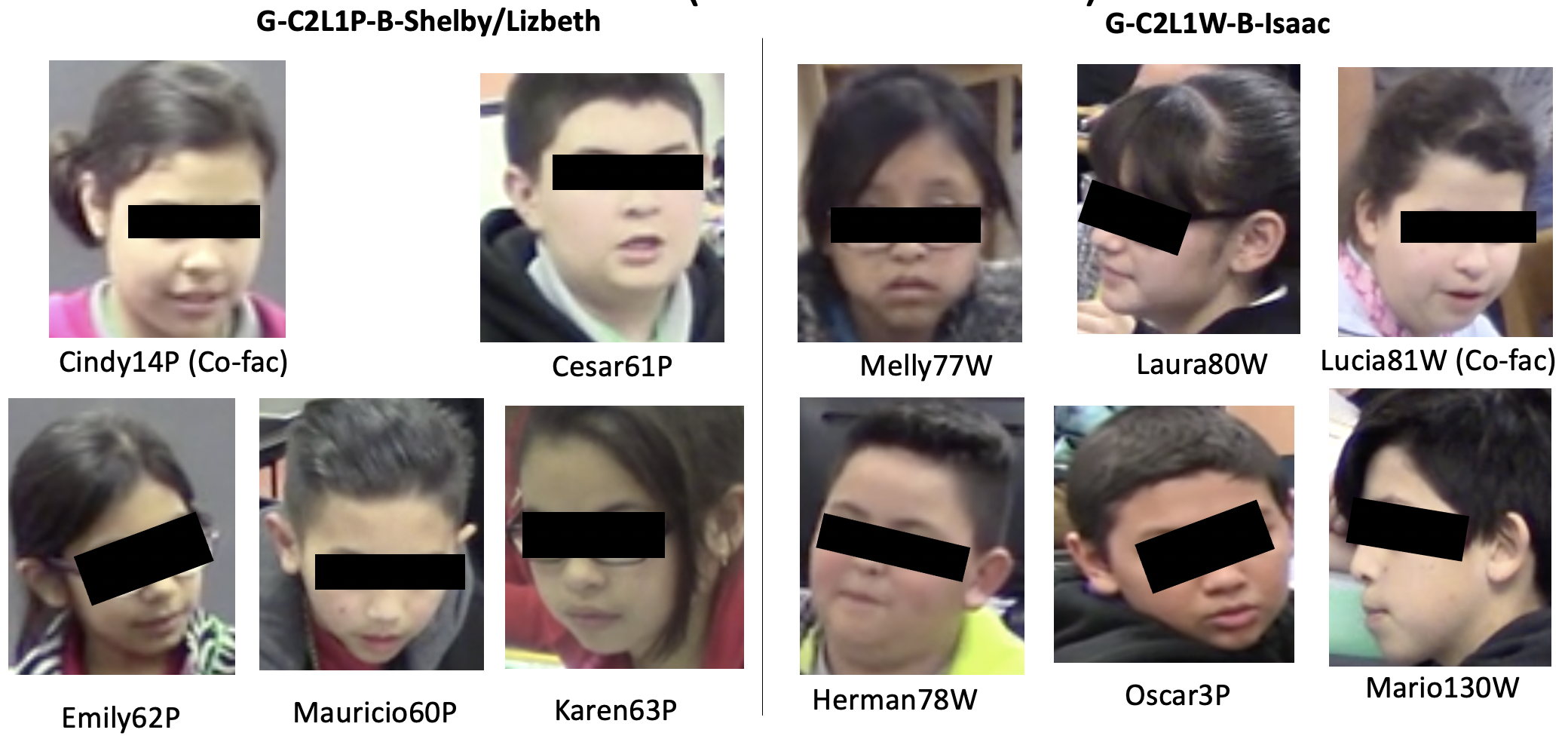}~\\[0.05 true in]
	\includegraphics[width=.9\textwidth]{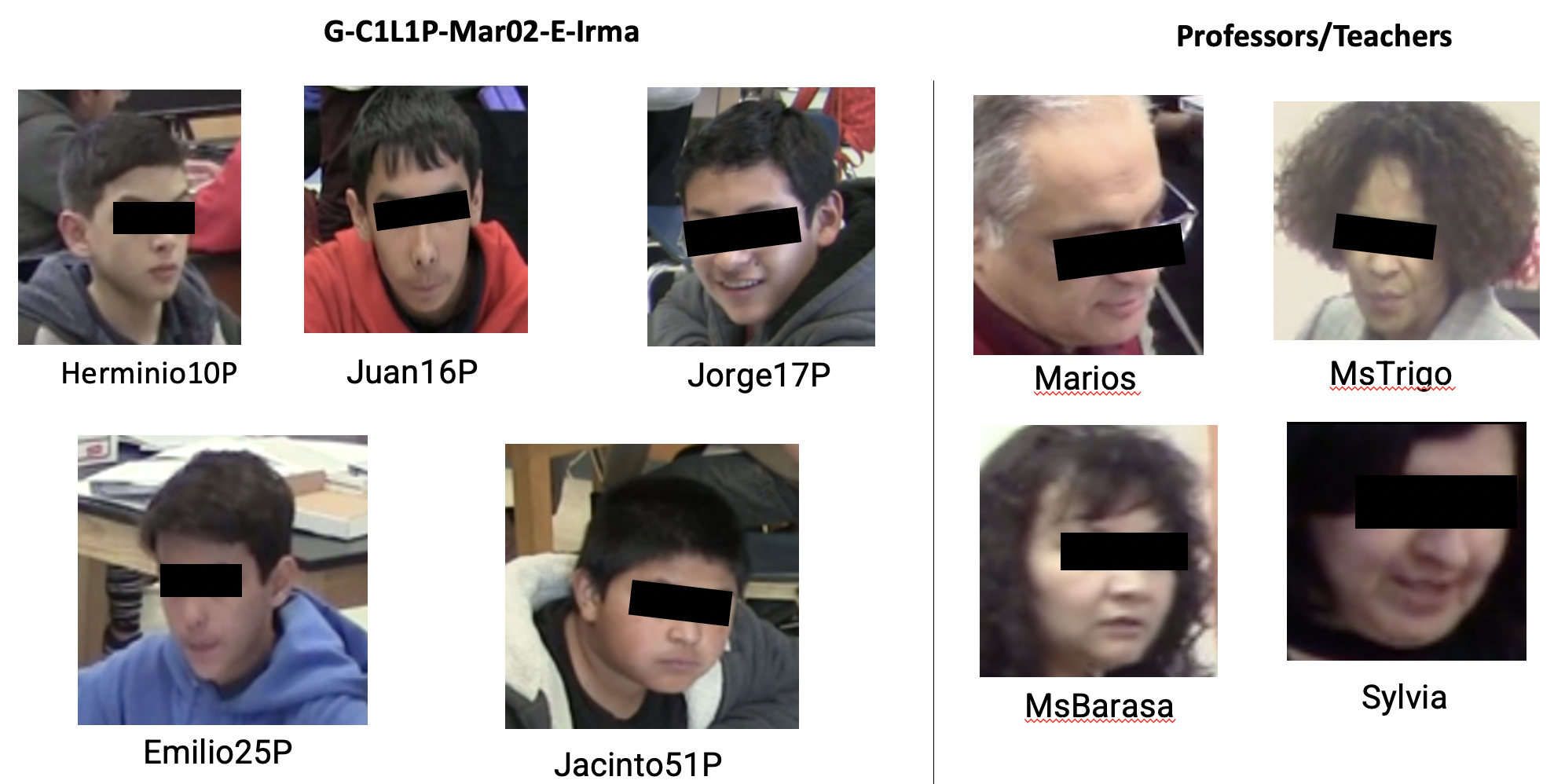}~\\[0.05 true in]
	\caption{\label{wait}AOLME41 dataset (I of II).}
\end{figure*}

\begin{figure*}[!t]
	\centering
	\includegraphics[width=1\textwidth]{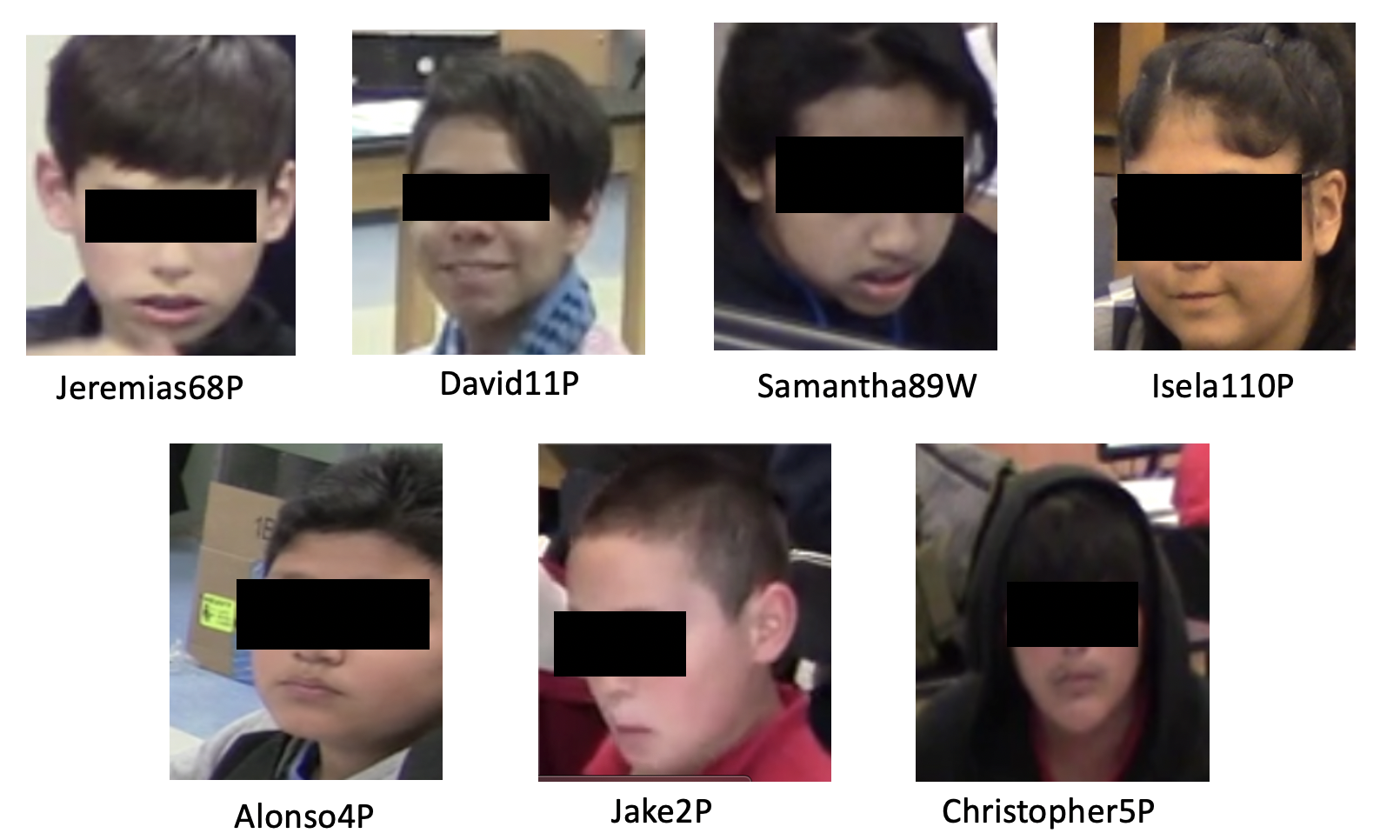}~\\[0.05 true in]
	\includegraphics[width=1\textwidth]{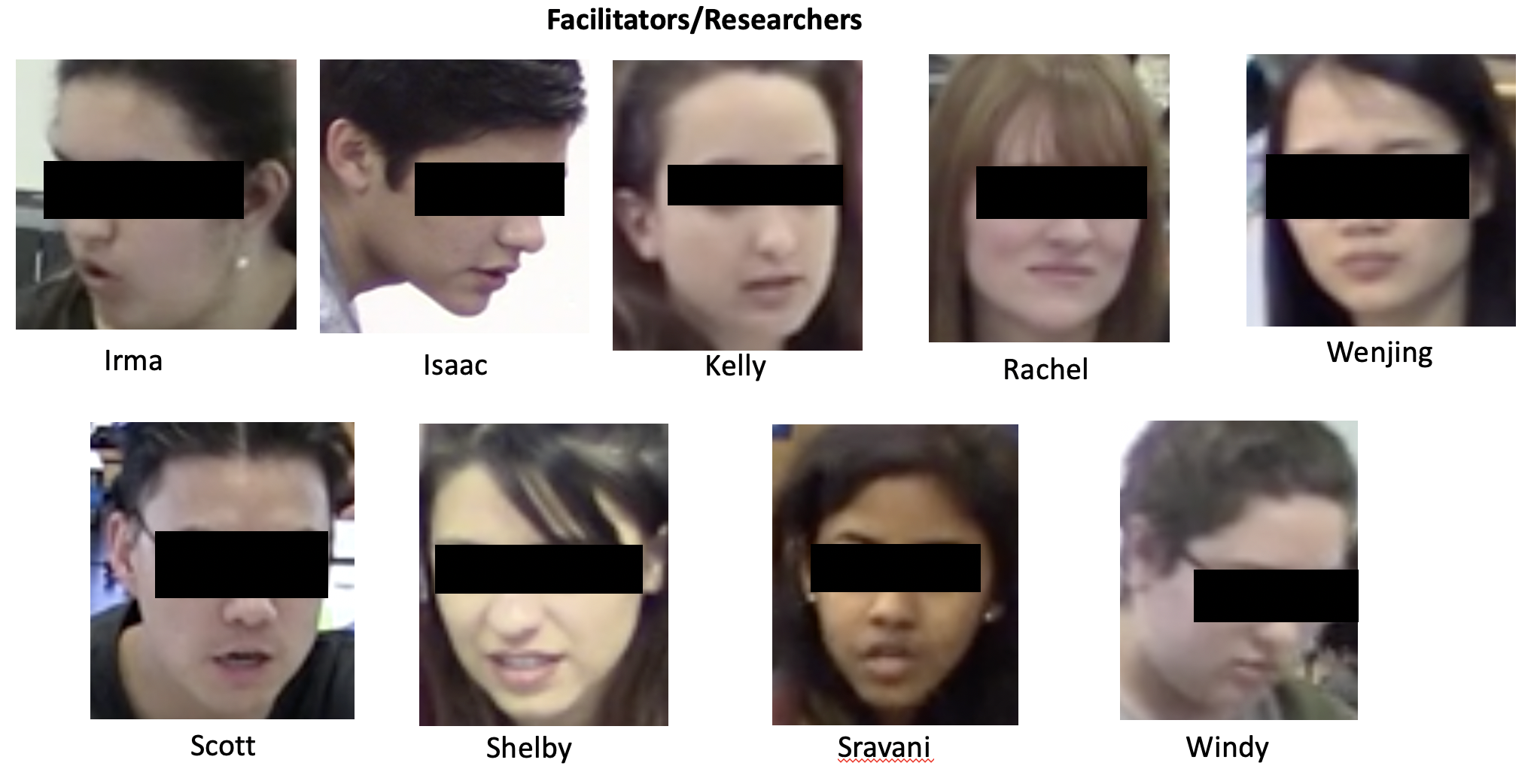}~\\[0.05 true in]
	\caption{\label{waitCont}AOLME41 dataset (II of II).}
\end{figure*}

\begin{figure*}[!t]
	\centering
	\includegraphics[width=1\textwidth]{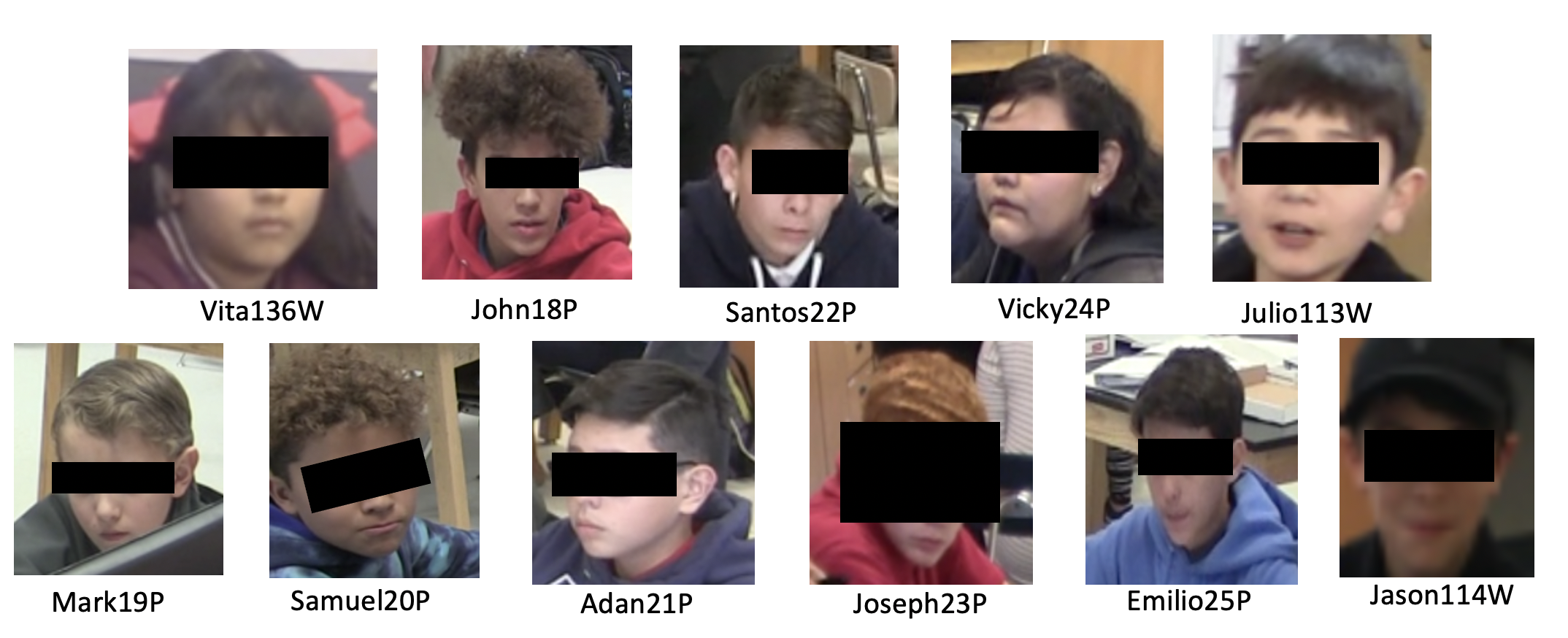}~\\[0.05 true in]
	\includegraphics[width=1\textwidth]{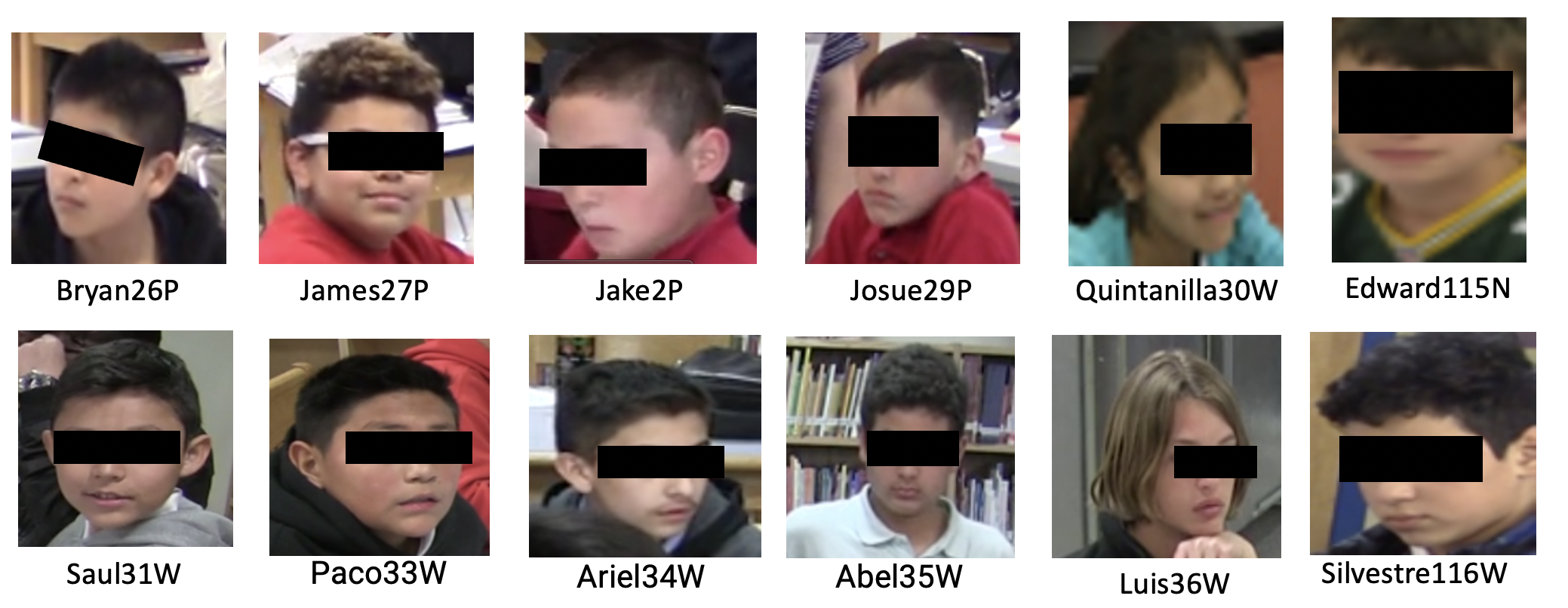}~\\[0.05 true in]
	\includegraphics[width=1\textwidth]{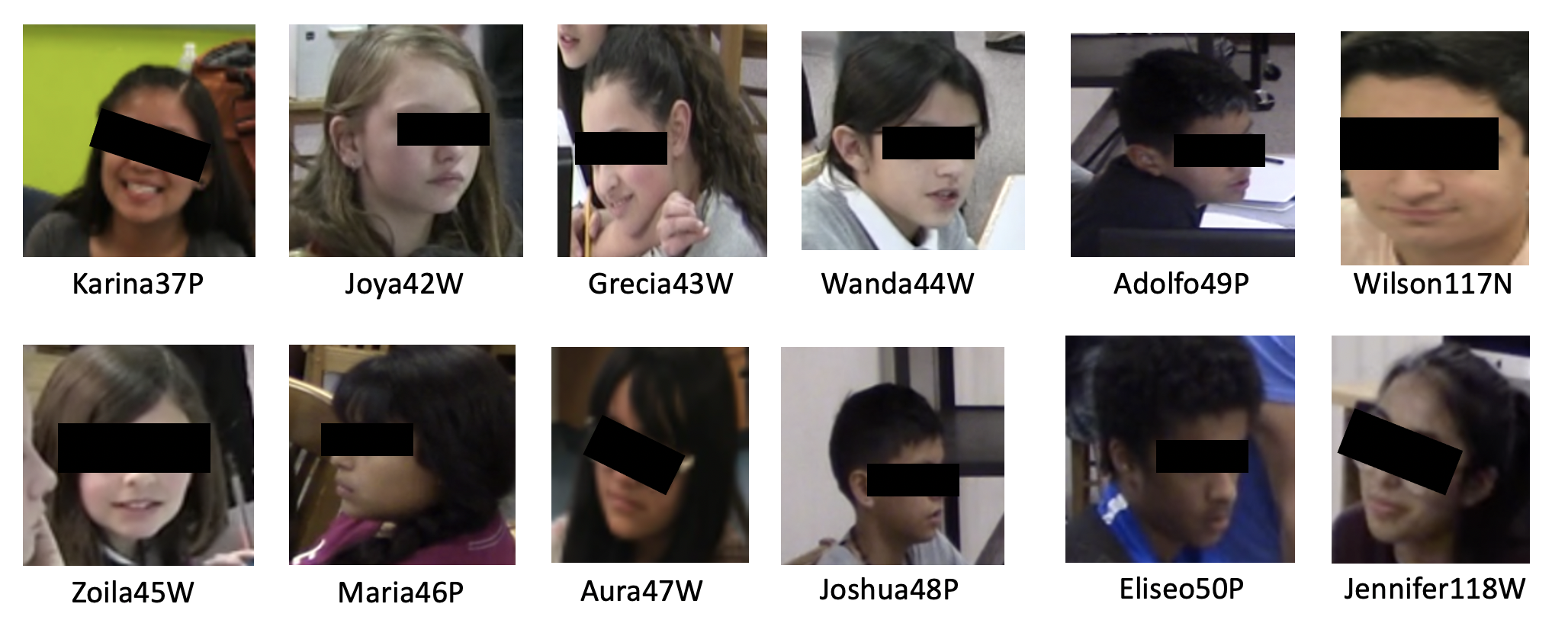}~\\[0.05 true in]
	
	\caption{\label{noKids}AOLME83 dataset (I of III).}
\end{figure*}

\begin{figure*}[!t]
	\centering
	\includegraphics[width=1\textwidth]{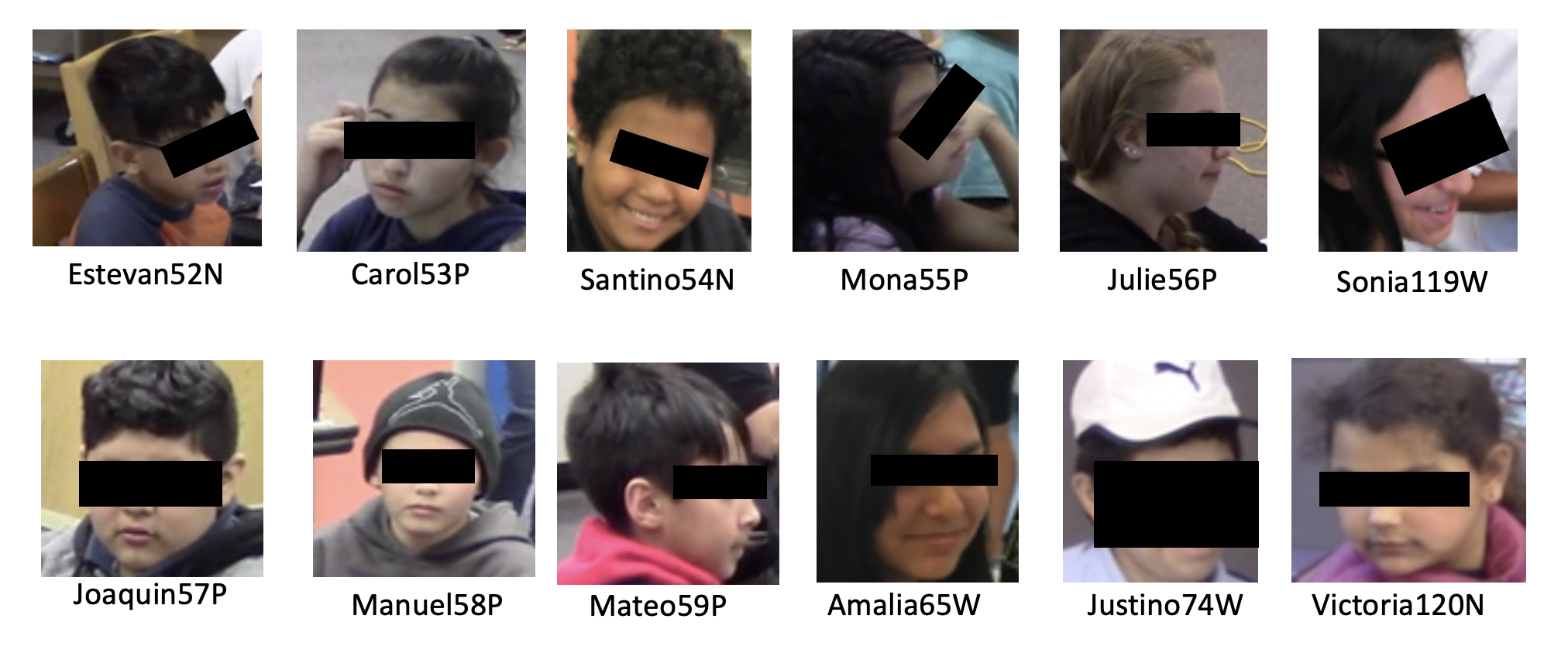}~\\[0.05 true in]
	\includegraphics[width=1\textwidth]{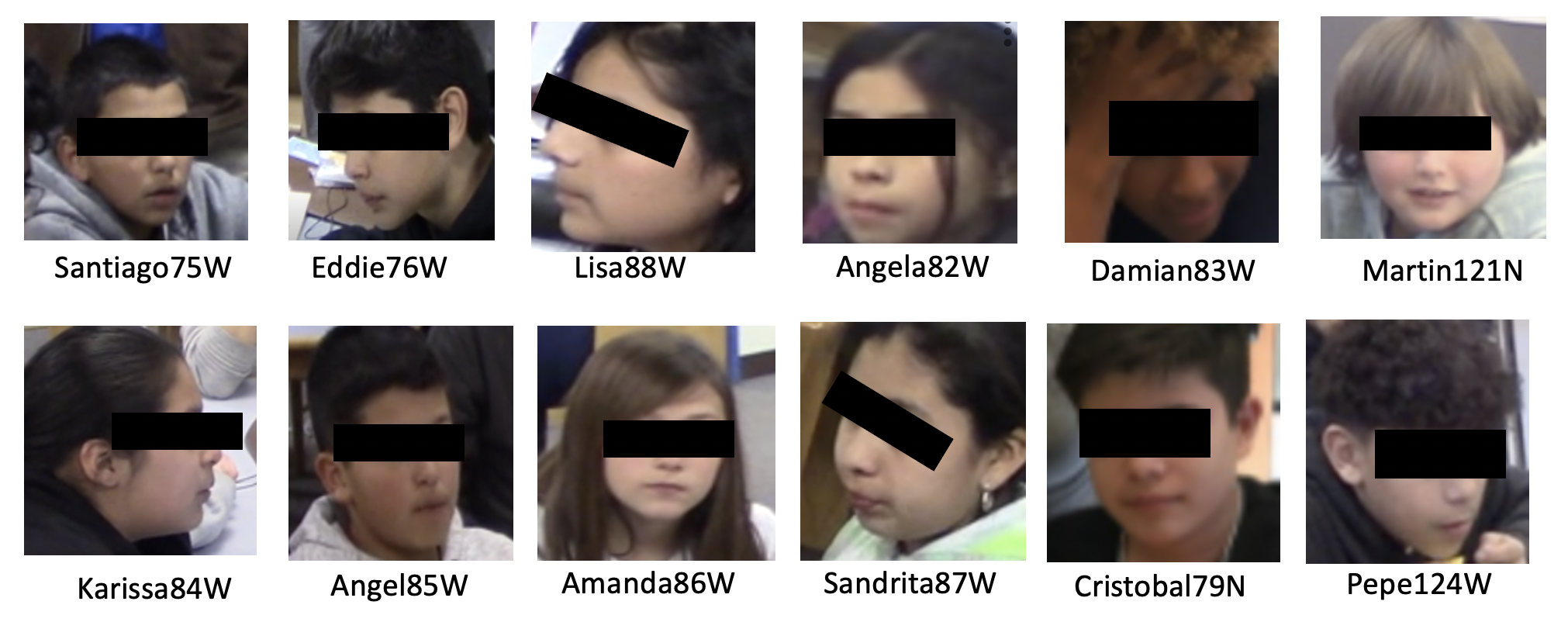}~\\[0.05 true in]
	\includegraphics[width=1\textwidth]{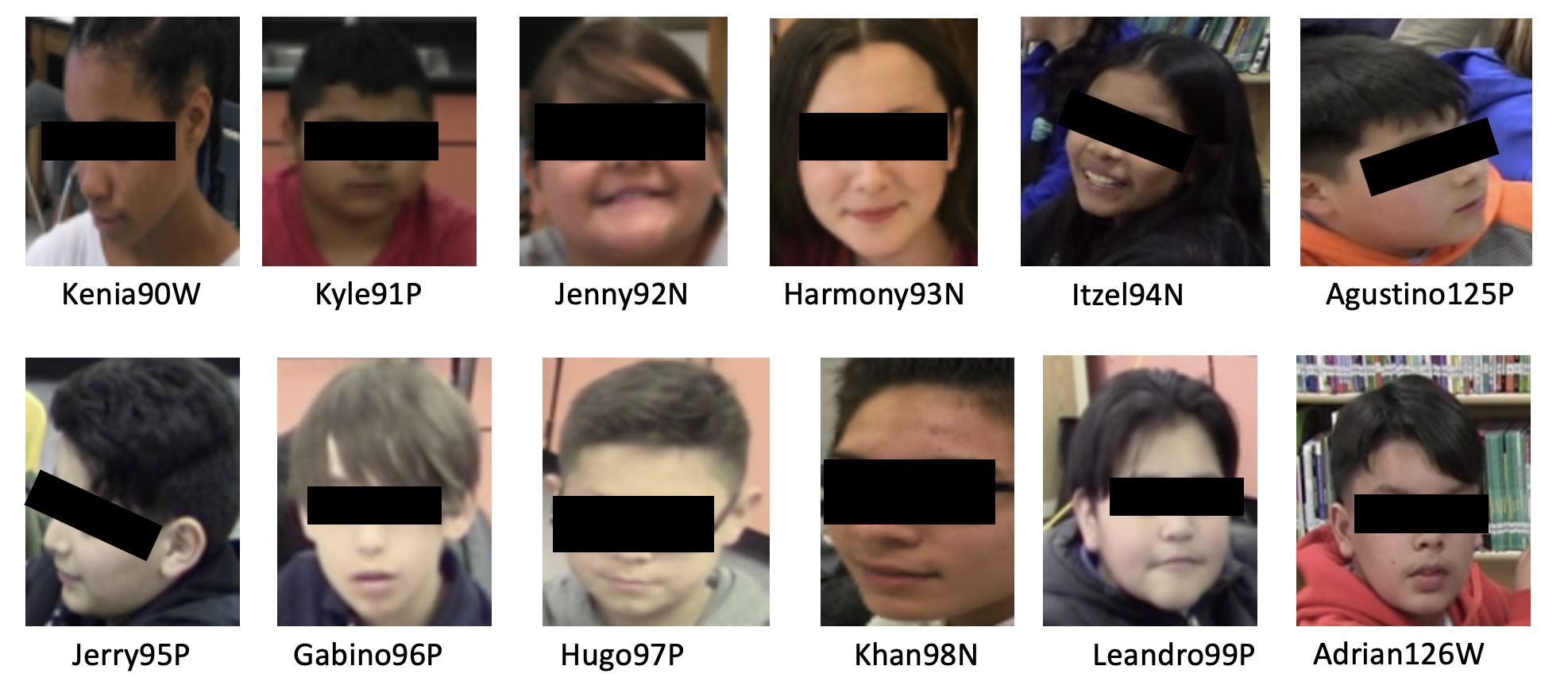}~\\[0.05 true in]
	
	\caption{\label{noCont}AOLME83 dataset (II of III).}
\end{figure*}

\begin{figure*}[!t]
	\centering
	\includegraphics[width=1\textwidth]{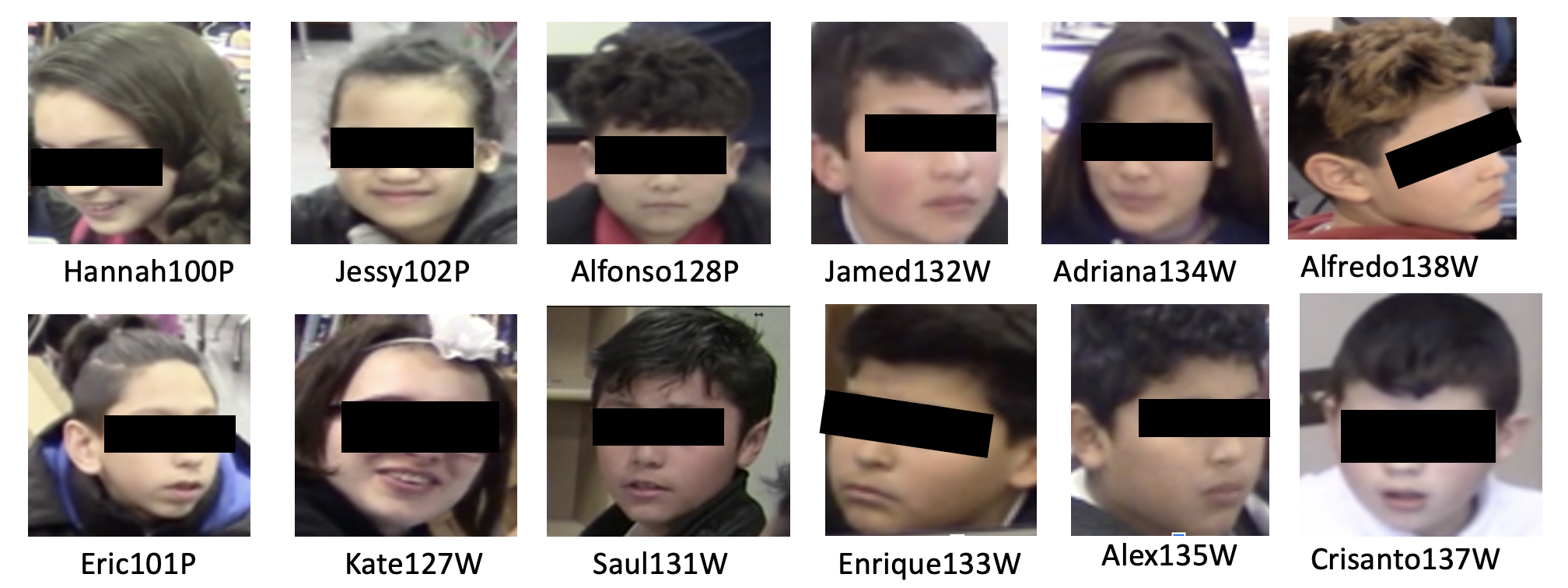}~\\[0.05 true in]
	\caption{\label{noCont2}AOLME83 dataset (III of III).}
\end{figure*}

This appendix summarizes the entire AOLME dataset, which is given in Figures \ref{fig:trainedDatabase} to \ref{noCont2}. With three Cohorts and three Levels across two years (2017 to 2019), the AOLME dataset consists of 138 students. Table \ref{aolmeCount} displays more detailed information on the program's participants. The dataset includes the teachers from both schools, the professors in charge of the program, the undergraduate students, and the graduate researchers that aided the program with facilitating and programming in addition to the student participants.


\begin{table}[]
	\caption[Summary of the people who appear in the AOLME program]{\label{aolmeCount} Summary of the people who appear in the AOLME program. Data collected from 2 schools in different areas that are referred to as Urban and Rural middle schools. Time range varies from C1L1 (2017) to C3L1 (2019). Neither represents students who were not from either schools. These students can be the schools' teachers' children or had heard about the program through other current or former participants.}  
	\centering
	\begin{tabular}{cccc}
		\hline
		\textbf{}      & \textbf{Rural} & \textbf{Urban} & \textbf{Neither} \\ \hline
		\textbf{Boys}  & 44             & 28             & 9                \\
		\textbf{Girls} & 28             & 24             & 5                \\
		\textbf{Total} & 72             & 52             & 14               \\ \hline
	\end{tabular}
\end{table}

\clearpage
\bibliographystyle{amsplain}
\bibliography{references.bib}
	
\end{document}